\documentclass[superscriptaddress,secnumarabic,
amssymb,amsmath,nobibnotes,aps,prd,showkeys,showpacs,nofootinbib]{revtex4-2}%
\usepackage{graphicx}
\usepackage{epsf}
\usepackage{bm}
\usepackage{amsmath}
\usepackage{amsfonts}
\usepackage{amssymb}
\usepackage{epstopdf}
\usepackage{subfigure} 
\usepackage{natbib}
\usepackage{color}%
\usepackage{hyperref}
\hypersetup{dvips,dvipdfm,linktoc=page,colorlinks=true,linkcolor=blue,citecolor=red,filecolor=magenta,urlcolor=magenta,bookmarks=true}
\providecommand{\U}[1]{\protect\rule{.1in}{.1in}}
\newcommand{\be}{\begin{equation}}
	\newcommand{\ee}{\end{equation}}

\newcommand{\mincir}{\raise
	-3.truept\hbox{\rlap{\hbox{$\sim$}}\raise4.truept\hbox{$<$}\ }}
\newcommand{\magcir}{\raise
	-3.truept\hbox{\rlap{\hbox{$\sim$}}\raise4.truept\hbox{$>$}\ }}

\begin{document}
\title{Dynamical systems analysis of a cosmological model with interacting Umami Chaplygin fluid in adiabatic particle creation mechanism: Some bouncing features} 
\author{Goutam Mandal}
\email{gmandal243@gmail.com; 
	rs\_goutamm@nbu.ac.in}
\affiliation{Department of Mathematics, University of North Bengal, Raja Rammohunpur, Darjeeling-734013, West Bengal, India.}
\author{Sujay Kr. Biswas\footnote{corresponding author}}
\email{sujaymathju@gmail.com; sujay.math@nbu.ac.in}
\affiliation{Department of Mathematics, University of North Bengal, Raja Rammohunpur, Darjeeling-734013, West Bengal, India.}
\begin{abstract}
	
The present work aims to investigate an interacting Umami Chaplygin gas in the background dynamics of a spatially flat Friedmann-Lemaitre-Robertson-Walker (FLRW) universe when adiabatic particle creation is allowed. Here, the universe is taken to be an open thermodynamical model where the particle is created irreversibly and consequently, the creation pressure comes into the energy-momentum tensor of the material content. The particle creation rate is assumed to have a linear relationship with the Hubble parameter ($\Gamma \propto H$) and the created particle is dark matter (pressureless). With this creation rate a single fluid model studied and found no phase transition. Then, we studied an interacting two-fluid model where second fluid is taken as perfect fluid equation of state and late-time acceleration is obtained.
Next, interacting Umami chaplygin gas is studied in context of particle creation. Dynamical stability of the model is performed by perturbing the autonomous system around critical points upto first order. Classical stability of the model is also studied at each critical point. This study explores some cosmologically viable scenarios when we constrain the model parameters. Some critical points exhibit the accelerated de Sitter expansion of the universe at both the early phase as well as the late phase of evolution which is characterized by completely Umami Chaplygin fluid equation of state. Scaling solutions are also described by some other critical points showing late-time accelerated attractors in phase space satisfying present observational data, and solving the coincidence problem. In a specific region of parameters, a sequence of critical points is achieved exhibiting a unified cosmic evolution of the universe starting from early inflation (represented by source point), which is followed by a decelerated intermediate phase (described by saddle solution), and finally goes through the late-time dark energy dominated universe (represented by stable point). Finally, non-singular bouncing behavior of the universe is also investigated for this model numerically.
	  
\end{abstract}
\maketitle
\section{Introduction}
One of the main crucial issues in modern cosmology is the present acceleration of the universe which is favored by numerous observational data obtained from Supernovae Type Ia (SNe Ia) \cite{Riess:1998cb,Perlmutter:1998np}, Cosmic Microwave Background Radiation (CMBR) \cite{Komatsu:2011}, the Baryon Acoustic Oscillations (BAOs) \cite{Sanchez:2012}, Plank data \cite{Ade:2014,Ade:2016} etc. Cosmologists have addressed these in theory by introducing a new matter source in the energy-momentum tensor in the Einstein Field Equation. The matter source is dubbed as Dark Energy (DE), and being an exotic type fluid having a huge negative pressure, it has significant role in making gravitational repulsive effect to drive the present acceleration. 
Although, the DE can successfully explain the present acceleration of the universe, the nature of the DE is completely unknown to us. In addition, it violates the strong energy condition (i.e., $\rho+3p<0$) and the equation of state of the DE satisfies $\gamma_d=\frac{p}{\rho}<-\frac{1}{3}$. In this premise, `cosmological constant' ($\Lambda$) is the most preferred candidate for the DE having its constant equation of state ($\gamma_{\Lambda}=-1$). The cosmological constant along with cold dark matter constitutes the famous $\Lambda$CDM model \cite{Carroll1992,Sahni2000,Peebles2003} which provides the best fitted result to the observational data. However, the model has two serious theoretical issues: one is the `cosmological constant problem' arising due to the disagreement between observed small value and the theoretical large value of vacuum predicted by the quantum field theory. Another is the `cosmic coincidence problem' which refers to as `why the energy densities of two dark fluids are of the same order today, though they scale differently in their cosmic evolution?'. Motivated from this, people are looking for new cosmological models which are free from the above two problems.  \\
In order to address such issues, cosmologists have proposed several models based on perfect fluid equation of state. In this context, a newly proposed fluid model having perfect fluid equation of state turns out to be very promising. The equation of state of this model is expected to satisfy two limits which follow the expansion rate at high redshift as well as at low redshift values. That is, this type of fluid can mimic the matter(dust) dominated epoch at early stage of its evolution and an accelerated expansion of universe's evolution at late-times. This class of models belongs to the Chaplygin gas type fluid. The model belonging to Chaplygin gas is popular in cosmology due to its ability to provide a mechanism of unified description of DE and DM. The original Chaplygin gas was first introduced by Kamenschik, Moschella, and Pasquier \cite{Kamenschick2001} 
and characterized by the following EoS:
$$p=-\frac{A}{\rho}$$
where $A$ is positive constant 
and this was introduced as an alternative to quintessence. After that a plenty generalizations have been reported in the literature (see for instance in Ref. \cite{Pourhassan2014}). One of such generalizations, known as Generalized Chaplygin gas (GCG) which was introduced by Bento et. al.\cite{Bento2002} and was characterized by EoS:
$$p=-\frac{A}{\rho^{\alpha}}$$
where $\alpha$ is positive constant and $\alpha \in [0,~1]$.
Latter, a modified Chaplygin gas (MCG) was also proposed with the equation of state
$$p=B\rho-\frac{A}{\rho^{\alpha}}$$
where $B$ is positive constant. For $B=0$ the EoS reduces to the generalized Chaplygin gas (GCG) EoS while for $B=0,~\alpha=1$, it recovers the original Chaplygin gas. Specifically, the conditions: $A=0$ and $B=\frac{1}{3}$ will imply that the EoS leads to radiation dominated evolution. In cosmology, he models of modified Chaplygin gas and Generalized Chaplygin gas have been extensively studied in the literature (see in Refs. \cite{Gorini}).\\
Recently, Lazkoz et al. \cite{Lazkoz2019} further generalized the fluid model and introduced a new type of Chaplygin fluid in the following way:
\begin{equation}\label{Umami Chaplygin}
	p=-\frac{\rho}{\frac{1}{|\omega|}+\frac{\rho^{2}}{|A|}}
\end{equation}
where $\rho$ is the energy density and $p$ is the pressure of the fluid. The parameters $A$ and $\omega$ are the real constants. This fluid can describe different phases of evolution depending on restrictions of parameters $A$ and $\omega$. In particular, at high energy limit, the equation of state of the fluid mimics as the equation of state of usual Chaplygin gas
$p\longrightarrow -\frac{|A|}{\rho}$. On the other hand, at the low energy density limit one obtains the relation $p\longrightarrow -|\omega|\rho$ that describes the negative effective equation of state. Unlike the other models (previously described) of Chaplygin fluid, the Umami Chaplygin gas describes not only unified model of DE and DM, but also it can represent the unified description of dark fluid and baryonic matter \cite{Lazkoz2019}. Therefore, the Umami Chaplygin fluid has a significant role in contributing the total matter distribution of the universe. Therefore, it is interesting to study the Umami Chaplygin gas as an alternative to $\Lambda$CDM. The Umami Chaplygin fluid has been studied further by Biswas and Biswas in Ref. \cite{Biswas2021} in the context of dynamical and thermodynamical point of view and many interesting results have been reported therein.\\

From several astrophysical data, it is speculated that the main material content of the universe is constituted with pressureless dark matter (DM) contributing $28 \%$ of total energy density and dark energy (DE) contributing around $68 \%$ of the total energy budget of the universe. Hence, it is assumed that the dark sectors (DM and DE) are the dominant source of the universe  \cite{Planck}. In spite of plenty of these components, a very little is known about their nature. So, an interaction between them cannot be neglected. Interestingly, an interacting two-fluid model can also provide the possible mechanism to address the coincidence problem,  see for instance in Refs. \cite{Amendola,Tocchini,Amendola1,del,Campo}. 
But, there is no guiding principle for choosing the interaction term in dark sectors. In fact, it can be taken on purely phenomenological ground. Various phenomenological interacting scenarios have been investigated in the literature (see the Refs. \cite{Yuri.L.Bolotin2014,Andre A.Costa2014,M.Khurshudyan2015,N.Tamanini2015,T.Harko2013,Nunes2016,Wang2016,Copeland1,Aguirregabiria1,Lazkoz1,Fang1,Leon1,Leon2,Odintsov2018a,Odintsov2018b,Bahamonde2018}).
Moreover, many studies recently indicate that introduction of an interaction in the dark sectors can alleviate the $H_{0}$ tension \cite{Di,Kumar,Yang,Y,Pan,Kumar1,E,Akarsu,Lucca,Valentino,L,Renzi,The,Vagnozzi2020,Visinelli2019,Valentino2020,Valentino2020a,Vagnozzi2020a,Pan2020,Yang2021,Gao2021,Lucca2021,Nunes2021,Guo2021,Mancini2022,Ferlito2022,Nunes2022,Pan2022} appearing due to disagreement between the CMB measurements by Planck satellite within $\Lambda CDM$ cosmology \cite{Planck} and SH$0$ES \cite{Riess} and $S_{8}$ tension \cite{Pour,Rui,Kumar2,Araujo,Avila} arising due to discordance (i.e. many sigmas gap) between the Planck and Weak Lensing measurements \cite{Valentino1}. Despite the above, it should be noted that an interaction can be able to explain the phantom phase of the DE without any need of a scalar field with negative kinetic \cite{Pan1,Bonilla,Bonilla1} . Therefore, recent research has a lot of interests to study the dynamics of interacting scenario in cosmology. In this context, one may follow \cite{Yuri.L.Bolotin2014,Wang2016} for a comprehensive study on interacting dark energy.\\

Another explanation for the present cosmic acceleration of the expanding universe can be found in the study of the gravitationally induced matter creation process. The study is started from the year 1939, when Schrodinger gave a microscopic description of matter creation process in his seminal paper \cite{Schrodinger1939}. Latter on, Parker and his collaborators \cite{Parker Colleraboration} developed this idea and applied it to curved space-time based on the quantum field theory (QFT). They proposed that the gravitational field of the expanding universe acts on the quantum vacuum and produces radiation, matter particles acquiring mass, momentum and energy from the time evolving gravitational background in the sense that gravity pumps on the curvature to create particles by a pumping mechanism. The idea of matter creation is thus becoming an interesting issue in the present cosmology. In the matter creation process, there is another puzzle. That is the prediction for baryon to entropy ratio which is $\frac{\eta_B}{s}\simeq 9.2\times 10^{-11}$ \cite{Oikonomou2016} and the existence of such baryon to entropy ratio in universe still remains a mystery and an unresolved problem. However, a work in Ref. \cite{Sakharov1967} indicates that the baryon asymmetry may occur due to nonequilibrium thermodynamical process in an expanding universe and this may refers to the fact that particle creation is happening.  
The above facts motivate to the cosmology communities  to include the particle creation mechanism as a part of cosmological studies. First, Prigogine et. al. \cite{Prigogine1988,Prigogine1989} included the macroscopic description of matter creation mechanism  in the Eienstein Field equations to describe the evolution of the universe. In this case, thermodynamics of an open system was first introduced in the cosmology. Calvao et al. in Ref. \cite{Calvao1992} then extended and generalized the approach in a covariant formulation allowing specific entropy variation as usually expected for nonequilibrium process in fluid. Since then, people have investigated successfully this particle creation mechanism in context of cosmology. In the present thermodynamic formalism, the dissipative phenomenon in FLRW model may arise in form of bulk viscous pressure due to non-conservation of particle numbers (i.e., $N^{\mu}_{;\mu}\neq0$).  There is no clear explanation about what type of particles are created by the gravitational field. It is well known that the problems of dark matter may be one of the main puzzling issues in the present day cosmology and the dark matters consist of weakly interacting massive particles (known as WIMPs). Also, it can only be probed through gravitational effects and it may interact very weakly with baryonic matter and gravitational force. As we know that the dark matter is one of the dominant source in the dark sectors and in addition, radiation has no effect/ impact in the late time evolution of the universe. We may therefore, assume that the cold dark matter particles existing in the post inflationary era, and presently these are created in expanding universe by energy flow from gravitational field to matter \cite{Herring2020}. 
Motivated from the above, many cosmological studies have been performed in the literature \cite{Steigman2009,Lima2010,Lima2014,Fabris2014,Chakraborty2015,Nunes2016} where particle creation formalism is taken into account for closely mimicking the $\Lambda$CDM model. For these cases, the matter is created from time-varying gravitational field mimicking as cold dark matter \cite{Prigogine1988,Prigogine1989,Calvao1992,Lima1991,Lima1992,Lima1996,Lima2008,Basilakos2010}.  
\\
The cosmological models with possibility of gravitationally induced matter creation are  studied in different way. Modified gravity theories are also studied (see in Refs. \cite{Harko2014,Harko2015,Harko2021,Pinto2022,Cipriano2024}) where it was assumed that the irreversible energy flow occurred from the gravitational field to created particles and the energy-momentum tensor of matter content is not covariantly conserved due to nonminimal geometry (curvature, torsion)-matter coupling in an open thermodynamical system. Here, the cosmological evolution is analyzed by deriving the particle creation rate, the creation pressure and the entropy production rate. A new cosmological model, called Created Cold Dark Matter (CCDM) scenario  is proposed in literature \cite{Steigman2009,Lima2010,Basilakos2010} via particle creation mechanism. This can be a viable alternative to $\Lambda$CDM model at the background as well as perturbative level \cite{Jesus2011}. Recently, the authors in Ref. \cite{Cardenas2024} have shown that the analogy between the CCDM and diverse dark energy models at the background level. Gravitationally induced particle creation is also studied in Energy-Momentum-Squared gravity recently in Ref. \cite{Cipriano2024}. 
There is another way to solve cosmological models of matter creation by adopting the creation rate $\Gamma$. In this formalism, the matter creation rate can play a crucial role to describe the evolution of the universe. It is assumed to be a unknown function of time t, and it can be determined through quantum field theory in curved space-time which is yet to be developed. Also, there is no guiding principle to choose the form of creation rate $\Gamma$. Therefore, some phenomenological choices of $\Gamma$ have been made where the creation rate is either constant or has explicit $H$ dependent expressions. Cosmological dynamics of matter creation models have been extensively studied in literature by assuming different forms of creation rates that can successfully explain different evolutionary epochs of the universe which will be discussed in the next section.  \\
  
In view of all above, we would like to approach systematically. First, we consider the universe is filled with created matter (pressureless dust) only and examine whether there is any physically interesting result. Then, we consider a two-fluid model with particle creation where the second fluid is assumed to be perfect fluid with positive non-linear equation of state and finally, we study an interacting umami chaplygin fluid in FLRW universe with the possibility of particle creation. In this formalism, matter is created irreversibly in an open thermodynamic model of universe. The created particles are taken to be cold dark matter and a phenomenological choice of creation rate $\Gamma=\Gamma_{0}H$ is considered.
Our main motivation for considering such a complex system is to examine whether the weird model can explain the overall evolution of the universe.  

From first model, we observe that there is no phase transition in the evolution. In the second model, late time acceleration is possible only when interaction is included.
From the dynamical analysis of the third model we obtain some cosmological interesting solutions. The de Sitter expansion of the universe is found by some critical points that show the early evolution as well as the late-time evolution. Scaling solutions are also achieved by some critical points that are important at late-time dynamics of the model. It worthy to mention that these late-time scaling attractors are capable of supporting the observational phenomena and the solutions can also solve the coincidence problem. Finally, we obtain a parameter region, in which a sequence of critical points depict the unified description of evolution from early inflation to late time DE dominated solution connecting through a matter-dominated transient era. The numerical investigations of $\omega_{eff}$ and $q$ also support the claim. 
It is interesting to note that a non-singular bouncing universe in early evolution is also achieved. Since it is not possible to find the analytical solution of Hubble function, we investigate the behavior of Hubble function $H$ and scale factor $a$ numerically and obtain a nonsingular bounce at early evolution of universe to avoid the big-bang singularity.
\\

Organization of this work is as follows: In the next sect. \ref{Model and autonomous system}, we discuss the basics of adiabatic particle creation mechanism in cosmological model of Umami Chaplygin gas interacting with pressureless dust. Then, sect. \ref{Analytic sol} refers to the study of the single fluid cosmological model by analytical method. A two-fluid interacting model with creation rate $\Gamma=\Gamma_{0} H$ is described in sect. \ref{Two fluid model PF} where a detailed phase plane analysis is executed. Sect. \ref{Int Umami model} comprises the study of interacting Umami chaplyging fluid in perspective of particle creation with rate $\Gamma=\Gamma_{0} H$. In sub-section \ref{phase plane analysis}, phase plane analysis is performed and classical stability of the model is investigated in sub-section \ref{Classical stability}. Sect. \ref{Cosmological implications} contains the cosmological implications of the critical points and the bouncing scenario is studied in the sect. \ref{Bouncing scenario}. Finally, the overall conclusion is presented in the sect. \ref{Conclusion}. We use the natural units  8$\pi G = c = 1$ throughout the paper.


\section{Basic equations of cosmological model with particle creation mechanism}\label{Model and autonomous system}
In this section, we shall derive the basic governing equations in context of adiabatic particle creation mechanism in the framework of spatially flat, homogeneous and isotropic FLRW universe. We consider irreversible matter creation in an open thermodynamical system and the process of matter creation is applied to the FLRW Universe which is assumed to be an open thermodynamical system.
In agreement with inflation and cosmic microwave background, geometry of the universe is well described by the following metric:
\begin{equation}\label{FLRW metric}
	ds^2=g_{\mu \nu}dx^{\mu} dx^{\nu}=-dt^2 +a^{2}(t) \left(dr^2 +r^2 d\Omega^2\right)
\end{equation}  
where $d\Omega^{2}=d\theta^{2}+sin^{2}\theta d\phi^{2}$ is spherical line-element on the unit 2-sphere and $a(t)$ is the scale factor of the universe. The matter content of the universe is described by the perfect fluid with the energy-momentum tensor as follows: 
\begin{equation}\label{E-M tensor}
	T_{\mu \nu}=(p+\rho)u_\mu u_\nu +p g_{\mu \nu},
\end{equation} 
where $\rho=\rho(t)$ is the energy density and $p=p(t)$ is the thermodynamics pressure of the material content depending only on time $t$ for an isotropic universe and $u^{\mu}=(1,0,0,0)$ is the four-velocity vector having magnitude $u^{\mu} u_{\mu}=-1$.
Thus, the conservation equation ($T^{\mu \nu};\nu=0$) for the matter content will follow as below:

\begin{equation}
	\dot{\rho}+3H(\rho+p)=0.
\end{equation} 
Let us consider an open thermodynamical system of volume $V (=a^{3}(t))$ which contains $N$ number of particles where $N$ is non-constant because of irreversible matter creation process. Hence, the non-conservation of particle number ( $N^{\mu}_{;\mu}\neq0$) will follow the following
balance equation for particle number density:
\begin{equation}\label{balance particle no}
	N^{\mu}_{;\mu}=\dot{n}+\Theta n= n \Gamma,
\end{equation}
where the source term in the right hand side indicates the matter production. The term $\Gamma$ denotes the rate of change of particle number and is the basic quantity that describes the deviation of standard cosmology. The standard cosmology is recovered when $\Gamma=0$. However, a nonzero $\Gamma$ can provide deeper insights in evolution of the cosmological dynamics. The particle number density defined by $n=\frac{N}{V}$, where $V=a^3$ is the co-moving volume. The particle flow vector is referred to as $N^{\mu}=n u^{\mu}$, $u$ is the four-velocity vector related to the expansion of congruence of time-like geodesic by the relation $\Theta=u^{\mu}_{;\mu}$. It is to be mentioned that for FLRW universe, one may obtain $\Theta=u^{\mu}_{;\mu}=3H$ and $\dot{n}=n_{\mu} u^{\mu}$.
Now, considering this case in FLRW universe, the balance equation will take the form in terms of particle number density:
(using Eqn. (\ref{balance particle no})):
\begin{equation}\label{balance eqn}
	\dot{n}+3Hn=\Gamma n
\end{equation}
Also, since the universe is considered as a non-equilibrium thermodynamical model where total particle numbers are not conserved  (in Eqn. \ref{balance particle no}), the Gibb’s equation \cite{Prigogine1989,Zimdahl 1996,Zimdahl 2000} reads as:

\begin{equation}\label{Gibbs equation}
	Tds=d\left( \frac{\rho}{n}\right) +pd\left( \frac{1}{n}\right), 
\end{equation}
where $T$ refers to the fluid temperature and $s$ represents the specific entropy (the entropy per particle) and $\rho$ is the total energy density and $p$ is the total thermodynamic pressure of the system.
Now, taking variation in specific entropy, and by using the balance equation (\ref{balance eqn}), the Eqn. (\ref{Gibbs equation}) will lead to the following form \cite{Prigogine1989,Haro}
\begin{equation}\label{entropy}
	nT\dot{s}=\dot{\rho}+3H\left( 1-\frac{\Gamma}{3H}\right) (\rho+p),
\end{equation}
In order to obtain a clear picture of our cosmological model, we shall restrict ourselves by assuming that our universe is an ideal thermodynamical system and its evolution continues to be adiabatic (or isentropic) \cite{Calvao,Barrow} in the sense that the specific entropy will remain constant
(i.e., $\dot{s}=0$). But, the entropy of fluid of the system changes because of enlargement of the phase space of the system due to increasing number of fluid particles.
Therefore, under the adiabatic condition (i.e., the condition of constancy of specific entropy) the Gibbs relation in Eqn. (\ref{entropy}) will lead to the following  form:
\begin{equation}\label{conservation}
	\dot{\rho}+3H(\rho+p)=\Gamma(\rho+p).
\end{equation}
On the other hand, inclusion of matter creation pressure ($p_c$) inside the energy-momentum tensor can be reinterpreted as 
\begin{equation}\label{creation em}
	T^{(c)}_{\mu \nu}=p_c u_\mu u_\nu +p_c g_{\mu \nu}.
\end{equation}
which after adding with the Eqn.(\ref{E-M tensor}), will give the total energy-momentum tensor as follows:
\begin{equation}\label{E-M tensor modified}
	T_{\mu \nu}+T^{(c)}_{\mu \nu}=(p+\rho+p_c)u_\mu u_\nu +(p+p_c) g_{\mu \nu},
\end{equation} 
Then from Bianchi's identity, the energy conservation equation will take the form \cite{Pan}
\begin{equation}\label{conservation1}
	\dot{\rho}+3H(\rho+p+p_{c})=0,
\end{equation}
where, $p_c$ refers to as the creation pressure.
Comparing equations (\ref{conservation}), and (\ref{conservation1}), the creation pressure gets the following form: \cite{Chakraborty, Haro, Lima, Mimoso}
\begin{equation}\label{creation pressure}
	p_{c}=-\frac{\Gamma}{3H}(\rho+p)
\end{equation}
where  $\Gamma$ is the particle production rate. The condition for particle creation is $\Gamma>0$ and the annihilation condition is $\Gamma<0$. However, $\Gamma=0$ implies that there is no particle production. Here, we assume that particle is created gravitationally in the expanding universe, where the creation pressure is taken into account due to dissipative effect. Infact, created matter is assumed to be dark matter particle which will come into the total material content of the universe. If the universe is filled with the dark matter having energy density $\rho_{m}$ and any other fluid with energy density $\rho_{d}$  then the total matter content of the universe defined as: $\rho_{tot}=\rho_{d}+\rho_{m}$ is the sum of energy densities. Consequently, the total thermodynamic pressure of the material content is described by $p_{tot}=p_d+p_c$, where $p_{d}$ is the pressure of the other fluid and $p_c$ is the particle creation pressure which depends on the creation rate.
Thus, the modified Friedmann's equation and Raychaudhuri's equation can be written as
\begin{equation}\label{Friedmann}
	H^{2}=\frac{1}{3} \rho_{tot}=\frac{1}{3} \sum_{i}(\rho_{i}) =\frac{1}{3}(\rho_{d}+\rho_{m})
\end{equation}
and
\begin{equation}\label{Raychaudhuri equation}
	\dot{H}=-\frac{1}{2}(\rho_{tot}+p_{tot})=-\frac{1}{2} \sum_{i} (\rho_{i}+p_{i}) +p_{c} =-\frac{1}{2}(\rho_{d}+p_{d}+\rho_{m}+p_{c})
\end{equation}
where, $H$ is the Hubble parameter defines the expansion rate
of the universe, and is expressed as $H =\frac{\dot{a}}{a}$ in terms of scale factor $a(t)$. An over dot indicates the derivative with respect to the cosmic time $t$. The subscript $i$ in the summation sign indicates the number of fluids content in the universe.
Now, we constrain the formalism by assuming that the created particles describe the pressure-less dark matter (DM) in the form of dust (i.e., $p_{m}=0$) then the creation pressure $p_{c}$ in (\ref{creation pressure}) will be of the following form \cite{Nunes2015}:
\begin{equation}\label{creation pressure1}
	p_{c}=-\frac{\Gamma}{3H}(\rho_{m}).
\end{equation} 
Then, the equation of state for the created particles will be $\gamma_c =\frac{p_c}{\rho_{m}}=-\frac{\Gamma}{3H}$ and the conservation equation for created particles will follow:
\begin{equation}\label{conservation created particle}
	\dot{\rho_m}+3H\rho_{m}=\Gamma\rho_{m}
\end{equation}
If the universe is also filled with dark energy component with variable equation of state parameter 
$\gamma_{d}=\frac{p_{d}}{\rho_{d}}$.
Then the individual fluids will follow the following conservation law:
\begin{equation}\label{DM}
	\dot{\rho}_{m}+3H(\rho_{m}+p_{c})=0
\end{equation}
or, equivalently the Eqn. (\ref{conservation created particle}).
The energy conservation equation for dark energy is
\begin{equation}\label{DE}
	\dot{\rho}_{d}+3H(\rho_{d}+p_{d})=0
\end{equation}
In order to solve the coincidence problem, one has to consider a non-gravitational interaction $Q$
between these dark species. Thus, the individual energy conservation equations for each component will take the following form:
\begin{equation}\label{DM1}
	\dot{\rho}_{m}+3H\rho_{m}\left( 1-\frac{\Gamma}{3H}\right) =-Q
\end{equation}
and
\begin{equation}\label{DE1}
	\dot{\rho}_{d}+3H(\rho_{d}+p_{d})=Q.
\end{equation}
The interaction term $Q$ has a specific form in literature depending on Hubble function and energy densities of dark components (either DE or DM or both). The interaction term defines rate at which energy exchange occurs between dark sectors. Also, for $Q>0$ the energy flow happens from DM to DE, and for $Q<0$ the energy flows in reverse direction.

The main modification of this model is to consider a non-zero matter creation rate (i.e., $\Gamma\neq0$) which measures the deviation from standard cosmology. In the dynamical systems analysis, the cosmological dynamics of the model cannot be realized from the system of ODEs (Ordinary Differential Equations) until the $\Gamma$ is specified. Also, there is no clear idea about the exact functional form of $\Gamma$. However, one can consider a phenomenological choice of $\Gamma$ and constrain the dynamics with observational data. Several choices of creation rate $\Gamma$ have been made in literature and that describe different phases of evolution of the universe. For example, $\Gamma\propto H$, $\Gamma\propto \frac{1}{H}$ are discussed in Ref. \cite{Pan2015} which predicts the matter dominated intermediate phase and transition from deceleration to late time acceleration of the universe. In Ref. \cite{Abramo1996} $\Gamma\propto H^2$ corresponds to prediction of early evolution. Further, $\Gamma=\Gamma_{0}$, a constant is studied in Ref. \cite{Haro} in which early big-bang singularity to late time de Sitter acceleration is obtained. A generalization is found in Ref. \cite{Chakraborty2014PLB} where the choice $\Gamma=\Gamma_{0}+mH+lH^2+\frac{n}{H}$ predicts early inflation, late time acceleration and  future deceleration. Recently, a two fluid solutions of particle creation model performed in Ref. \cite{Pan2019} where $\Gamma$ has more general expression as $\Gamma=\Gamma_{0}+\Gamma_{1}H^{-1}+\Gamma_{2}H^{-2}+\sum^{n}_{i=3} \Gamma_{i}H^{-i}$.
In Ref. \cite{Singh2020}, creation rate $\Gamma$  considered as $\Gamma=3\alpha H_0 +3\beta H$ is linear combinations of $\Gamma_{0}$ and $H$. In context of particle creation, it has been noticed that the creation rate $\Gamma$ has a crucial role in determining the universe's emergent scenario, see for instance in Refs. \cite{Chakraborty2014PLBa,Dutta2016}. A general creation rate $\Gamma=-\Gamma_{0}+m H +\frac{n}{H}$ is considered in Ref. \cite{Pan} to study the early time acceleration as well as the late-time acceleration of the universe. In Ref. \cite{Chakraborty}, the authors have studied different cosmic scenarios by assuming different choices of creation rates. They considered $\Gamma\propto H^{2}$ to describe early epoch, $\Gamma\propto H$ to describe intermediate phase and the assumed $\Gamma\propto H^{-1}$ for describing the late-phase of evolution. The present acceleration and different phases of evolution driven by gravitationally induced particle production are investigated in various articles (see in Refs. \cite{Nunes2016,Nunes2015,Nunes2016IJMPD}) where the creation rate is parameterized as $\Gamma\propto H$. It is to be noted that an interacting dark energy is investigated in Ref. \cite{Biswas2017} in the framework of particle creation mechanism with a simple form of creation rate $\Gamma=\Gamma_{0}H$ in context of dynamical analysis and late-time acceleration of universe is found in quintessence era, cosmological constant era or in phantom era. Further, a recent article, in Ref. \cite{Bhardwaj2024} studied matter creation cosmology with generalised chaplygin gas where $\Gamma=3\beta H$ was taken as matter creation rate.  Thus, from the discussions above, we can choose $\Gamma$ as the function of Hubble parameter $H$ that can be able to describe different evolutionary epochs either analytically or dynamically (when analytic solution can not be obtained).
	
It is worthy to mention that the authors in Ref. \cite{Nunes2015} showed that $\omega_{eff}<-1$ for $\Gamma=3\beta H$ without invoking the dark energy. From the analysis of observational data from Supernova Type Ia, Gamma ray busts, Baryon Acoustic Oscillations and Hubble rate, it was found that $\omega_{eff}=-1.073^{+0.034}_{-0.035}$ for the particular model $\Gamma=3\beta H$ at $1\sigma$ confidence level (see the Ref. \cite{Nunes2015}) and the present density parameter for dark matter $\Omega_{m0}\approx 0.3$. 
Motivated from the above, and in view of the Eqn. (\ref{DM1}) (for mathematical simplicity), we consider the particle production rate $\Gamma$ as a linear function of the Hubble parameter \cite{Pan2015,Chakraborty,Biswas2017,Bhardwaj2024} $\Gamma=\Gamma_{0}H$. Here, $\Gamma_{0}$ is a dimensionless  constant. We then analyse the model systematically by assuming sinle fluid, two-fluid material content of the universe.

\section{Single fluid model with particle creation rate $\Gamma=\Gamma_{0} H$}\label{Analytic sol}
First, we shall start with a single fluid model in context of particle creation. The created particles are assumed to be pressureless dust. In this case, the universe is considered to be filled with pressureless dust only. The modified Friedmann's equation and Raychaudhuri's equation will take the following form
\begin{equation}\label{Friedmann1}
	H^{2}=\frac{1}{3} \rho_{tot} =\frac{1}{3}\rho_{m}
\end{equation}
and
\begin{equation}\label{Raychaudhuri equation1}
	\dot{H}=-\frac{1}{2}(\rho_{tot}+p_{tot}) =-\frac{1}{2}(\rho_{m}+p_{c}).
\end{equation}
By using $\Gamma=\Gamma_{0} H$, the conservation equation in Eqn. (\ref{conservation created particle}) will provide the solution of $\rho_{m}$ in terms of redshift ($z$) as
\begin{equation}
	\rho_{m}=\rho_{m0}~ (1+z)^{3-\Gamma_{0}}
\end{equation}
where $\rho_{m0}$ is integrating constant (energy density for matter at present time). Now, 
from (\ref{Friedmann1}) we get
\begin{equation}\label{E}
	E(z)=\frac{H(z)}{H_0}=\left\lbrace \Omega_{m0} (1+z)^{3-\Gamma_{0}} \right\rbrace^\frac{1}{2} 
\end{equation}
where $H_0$ is the Hubble parameter at present time and $\Omega_{m0}$ is the density parameter for matter at present time. Then, the decelerating parameter in terms of redshift can be obtained as follows:
 \begin{equation}\label{deceleration single fluid}
	q(z)=-1+(1+z)\frac{E^{'}(z)}{E(z)}=\frac{1-\Gamma_{0}}{2}
\end{equation}
where prime denotes the derivative with respect to redshift z. From the analytic solution given in Eqn. (\ref{deceleration single fluid}) it is obvious that the single fluid model of particle creation with rate $\Gamma=\Gamma_{0} H$ will not be able to provide phase transition of the evolution of the universe. But, this phenomenon is important from observational cosmology. However, $\Gamma_{0}<1$ indicates that there is only deceleration and acceleration is occurred for $\Gamma_{0}>1$. But, $\Gamma_{0}=1$ implies that there is no acceleration or deceleration. According to present observations $\Gamma_{0}=1$ is impossible.	

\section{Interacting two-fluid model with particle creation rate $\Gamma=\Gamma_{0} H$}\label{Two fluid model PF}
Results obtained in previous section demand the need of further studies. We shall now investigate the cosmological dynamics of the particle creation model with the same rate $\Gamma=\Gamma_{0} H$, when the universe is assumed to be filled with an additional fluid. In the first case, we will consider second fluid as a perfect fluid (with positive equation of state) having non-linear equation of state (EOS):
\begin{equation}\label{pressure PF}
	p_{d}=\frac{B\rho_{d}}{\frac{1}{|\omega|}+\frac{\rho_{d}^{2}}{|A|}}
\end{equation}
where $B>0$ (positive constant). The variable equation of state becomes
 \begin{equation}\label{EOS PF}
	\gamma_{d}\equiv\frac{p_{d}}{\rho_{d}}=\frac{B}{\frac{1}{|\omega|}+\frac{\rho_{d}^{2}}{|A|}}>0
\end{equation}
Note that the fluid follows the conservation law according to as in Eqn. (\ref{DE}).
Therefore, the total energy density described by the total matter content of the universe is defined as: $\rho_{tot}=\rho_{d}+\rho_{m}$  and the total thermodynamic pressure of the material content is described by $p_{tot}=p_d+p_c$.
Note that $B=-1$ provides that the fluid to be a generalization of Chaplygin fluid model described in Eqn. (\ref{Umami Chaplygin}). The present section is devoted to study of a two-fluid model in conttext of particle creation. In presence of perfect fluid (having $\gamma_{d}>0$), the governing equations are complicated in nature and hard to solve them analytically. Therefore, dynamical system analysis is performed to get a qualitative idea of the overall evolution. Because, this analysis allows us to bypass the non-linearity and complications of a cosmological model. For dynamical analysis, we first introduce the dimensionless dynamical variables in terms of cosmological parameters which are normalized over Hubble scale. A two-dimensional autonomous system of ordinary differential equations is constructed from cosmological model. The nature of critical points is found by analyzing the linear perturbation of the system around the critical points.

\subsection{Dynamical variables and cosmological parameters}	
Below we consider the dimensionless dynamical variables in terms of cosmological variables \cite{Khurshudyan,Biswas2017,Biswas2021}
\begin{equation}\label{variables PF}
	x=\frac{\rho_{d}}{3H^{2}},~~~~~~~~~~~ y=\frac{p_{d}}{3H^{2}}
\end{equation} 
which are normalized over Hubble scale.
Now, by putting the dimensionless variables (\ref{variables PF}) in the equation of state parameter in
(\ref{pressure PF}), the dimensionless variable $y (y = y(x, H))$  can be written in terms of $H$ and $x$ in the form:
\begin{equation}
	y=\frac{B x}{\frac{1}{|\omega|}+\frac{9x^{2}H^{4}}{|A|}}
\end{equation}
which by inverse operation gives the Hubble parameter $(H =
H(x, y))$ in terms of dynamical variables $x$, $ y$ and the model parameters $A$ and $\omega$ as follows:
\begin{equation} \label{expression H PF}
	H^{4}=\frac{|A|}{9x^{2}y}\left(B x-\frac{y}{|\omega|} \right) 
\end{equation}

By using the dimensionless variables in Eqn.  (\ref{variables PF}), and the expression of $H$ in Eqn.(\ref{expression H PF}) the governing equations (\ref{Raychaudhuri equation}), (\ref{DM1}) and (\ref{DE1}) will give the following 2D system of ordinary differential equations:

\begin{eqnarray}\label{autonomous Gen1 PF}
	\begin{split}
		\frac{dx}{dN}& =\frac{Q}{3H^{3}}-3(x+y)+x\left\lbrace 3(1+y)-\frac{\Gamma}{H}(1-x)\right\rbrace ,& \\
		\frac{dy}{dN}& =\left\lbrace \frac{Q}{3H^{3}}-3(x+y) \right\rbrace \left\lbrace \frac{y}{x}\left(-1+\frac{2y}{B|\omega|x } \right)  \right\rbrace+y\left\lbrace 3(1+y)-\frac{\Gamma}{H }(1-x)\right\rbrace   ,
		&
	\end{split}
\end{eqnarray}
Here, the independent variable is chosen as the lapse time $N = \ln a$, which is called the e-folding number and $B>0$.
\\
Now, physical parameters can be expressed in terms of dynamical variables $x$ and $y$ (the co-ordinates of critical points) as follows:
The equation of state parameter for perfect fluid
reads as:
\begin{equation}\label{EOS  PF}
	\gamma_{d}=\frac{p_{d}}{\rho_{d}}=\frac{y}{x}
\end{equation}
from which one can find the nature of the fluid. In the present case, the fluid always behaves as perfect fluid as $B>0$.
The density parameter for the perfect fluid takes the value:
\begin{equation}\label{density parameter PF}
	\Omega_{d}=\frac{\rho_{d}}{3H^2}=x.
\end{equation}
The density parameter for dark matter can take the form
\begin{equation}\label{density parameter DM PF}
	\Omega_{m}=\frac{\rho_{m}}{3H^2}=1-x.
\end{equation}
The deceleration parameter can be expressed in the form
\begin{equation}\label{deceleration PF}
	q=-1-\frac{\dot{H}}{H^{2}}=\frac{1}{2}+\frac{3}{2}\left\lbrace y-\frac{\Gamma_{0}}{3}(1-x) \right\rbrace. 
\end{equation}
Acceleration of the universe is achieved for $q<0$ and the  decelerating nature is observed when $q>0$.
The effective equation of state parameter for the model is defined as 
	$$\omega_{eff}=\frac{p_{total}}{\rho_{total}}=\frac{p_m +p_c + p_d}{\rho_{m}+\rho_{d}}$$	
	which by using equations (\ref{Friedmann}) and (\ref{creation pressure1})	will take the form:
	$$\omega_{eff}=\frac{-\frac{\Gamma}{3H} + p_d}{3H^{2}}$$
	where $p_m =0$, as the matter is taken as pressureless dust.
	Now, by using $\Gamma=\Gamma_{0} H$ and the dynamical variables in (\ref{variables PF}), we obtain the effective equation of the state parameter for the model as:
	\begin{equation}\label{EOS effective PF}
		\omega_{eff}= y-\frac{\Gamma_{0}}{3}(1-x).
	\end{equation}
\subsubsection{Non-interacting case: $Q=0$}
The autonomous system for this case is:

\begin{eqnarray}\label{autonomous Gen1 Non Int PF}
	\begin{split}
		\frac{dx}{dN}& =-3(x+y)+x\left\lbrace 3(1+y)-\Gamma_{0}(1-x)\right\rbrace ,& \\
		\frac{dy}{dN}& =\left\lbrace -3(x+y) \right\rbrace \left\lbrace \frac{y}{x}\left(-1+\frac{2y}{B|\omega|x } \right)  \right\rbrace+y\left\lbrace 3(1+y)-\Gamma_{0}(1-x)\right\rbrace,
		&
	\end{split}
\end{eqnarray}
We obtain the following critical points for this autonomous system (\ref{autonomous Gen1 Non Int PF}):
{\bf 
	\begin{itemize}
		\item  I.  Critical Point : $M_{1}=(1,0)$
		
		\item  II. Critical Point : $M_{2}=(1,B|\omega|)$
				
	\end{itemize}	
}
The corresponding physical parameters are represented in the Table \ref{parameters Non-int}.
\begin{table}[tbp] \centering
	\fontsize{9.2pt}{5pt}
	\caption{The existence of critical points and the corresponding cosmological parameters for the interaction model $Q=0$ }%
	\setlength{\tabcolsep}{0.0001cm}
	\renewcommand{\arraystretch}{}
	\begin{tabular}
		[c]{cccccccccc}\hline\hline
		\textbf{Critical Points}&~~ $\mathbf{\Omega_{m}}$ &~~ $\mathbf{\Omega_{d}}$ &~~ $\mathbf{\gamma_{d}}$ &~~
		$\mathbf{\omega_{eff}}$ &~~  $q$ &~~ Existence &
		\\\hline
		$M_{1}  $ &~~ $0$ &~~ $1$ &~~
		$0$ &~~ $0$ &~~ $\frac{1}{2}$  &~~ $\forall~  \omega, \Gamma_{0}$ \\ \\
		$M_{2}  $ &~~ $0$ &~~ 
		$1$ &~~ $B|\omega|$ &~~ $B|\omega|$ &~~ $\frac{1}{2}(1+3 B|\omega|)$ &~~ $\forall~  \omega, \Gamma_{0}~\mbox{and}~B>0$ 
		\\\hline\hline
	\end{tabular}
	\label{parameters Non-int} \\
	
\end{table}%
%
\subsubsection{Phase plane analysis for the model $Q=0$}
\begin{itemize}
	\item The point $M_1$ exists always in the phase plane. The point is dominated completely by the perfect fluid ($\Omega_{d}=1,~\Omega_{m}=0$). Here perfect fluid behaves as dust ($\gamma_{d}=0$) and there is always a decelerating phase $q=\frac{1}{2}$ (see in Table \ref{parameters Non-int}). Eigenvalues of the linearised Jacobian matrix are $\lambda_1=6,~\lambda_2=\Gamma_{0}$. Therefore the point is always unstable source for $\Gamma_{0}>0$, otherwise saddle in the phase plane. The point represents early matter (perfect fluid) dominated decelerated phase of the universe.
	
	\item Critical point $M_2$ exists for all parameters. It is also perfect fluid dominated solution ($\Omega_{d}=1$) with equation of state $\gamma_{d}>0$ always (since $B>0$). There is no acceleration since $B>0$ (see Table \ref{parameters Non-int}). The eigenvalues of the critical point are $\lambda_1=-6-6B |\omega|,~~\lambda_2=\Gamma_{0} +3B|\omega|$. This point is also not stable solution for $\Gamma_{0}>0$ in the phase plane. So, when particle creation is allowed ($\Gamma_{0}>0$) the model is unable to provide late-time solution and therefore it is not physically interested.
\end{itemize}
{\bf Main results:} There will not be any late time solution in the case of non-interacting two-fluid cosmological model of particle creation with rate $\Gamma=\Gamma_{0} H$. There are two solutions obtained from this model and both the models are unable to provide the late-time acceleration.

\subsubsection{Interacting case: $Q=\alpha H \rho_{m}$}
Now the interaction is allowed in the two-fluid model.
We consider the simple interaction term $Q=\alpha H \rho_{m}$ and particle creation rate $\Gamma=\Gamma_{0} H$ in the system of ordinary differential equations (\ref{autonomous Gen1 PF}), we obtain the following two-dimensional autonomous system:
\begin{eqnarray}\label{autonomous2 PF}
	\begin{split}
		\frac{dx}{dN}& =(1-x)(\alpha-3y-\Gamma_{0}x) ,& \\
		\frac{dy}{dN}& =\left\lbrace x(\alpha+3)+3y-\alpha \right\rbrace \left\lbrace \frac{y}{x}\left(1-\frac{2y}{B |\omega|x } \right)  \right\rbrace+y\left\lbrace 3(1+y)-\Gamma_{0}(1-x)\right\rbrace   ,
		&
	\end{split}
\end{eqnarray}
The autonomous system (\ref{autonomous2 PF}) extracts the following critical points:
{\bf 
	\begin{itemize}
		\item  I.  Critical Point : $C_{1}=(1,0)$
		
		\item  II. Critical Point : $C_{2}=(1,B|\omega|)$
		
		\item  III. Critical Point : $C_{3}=(\frac{\alpha}{\Gamma_{0}},0)$
		
		\item  IV. Critical Point : $C_{4}=(\frac{\alpha}{\Gamma_{0}+3B|\omega|},\frac{\alpha B |\omega|}{\Gamma_{0}+3B|\omega|})$
		
	\end{itemize}
	
}
Existence of the critical points and their corresponding cosmological parameters are displayed in the table (\ref{physical_parameters1 PF Int}). Stability of the critical points are identified by perturbing the system upto first order around the critical points. Eigenvalues of the linearized perturbed matrix are presented in the table (\ref{eigenvalues1 PF Int}).

	\begin{table}[tbp] \centering
		\fontsize{9.2pt}{5pt}
		\caption{The existence of critical points and the corresponding cosmological parameters for the interaction model $Q=\alpha H \rho_{m}$ }%
		\setlength{\tabcolsep}{0.0001cm}
		\renewcommand{\arraystretch}{}
		\begin{tabular}
			[c]{cccccccccc}\hline\hline
			\textbf{Critical Points}&$\mathbf{\Omega_{m}}$& $\mathbf{\Omega_{d}}$ & $\mathbf{\gamma_{d}}$ &
			$\mathbf{\omega_{eff}}$ &  $q$ & Existence &
			\\\hline
			$C_{1}  $ & $0$ & $1$ &
			$0$ & $0$ & $\frac{1}{2}$  & $\forall \alpha, \omega, \Gamma_{0}$ \\ \\
			$C_{2}  $ & $0$ & 
			$1$ & $B|\omega|$ & $B|\omega|$ & $\frac{1}{2}(1+3 B|\omega|)$ & $\forall \alpha, \omega, \Gamma_{0}~\mbox{and}~B>0$ \\ \\
			$C_{3}  $ & $1-\frac{\alpha }{\Gamma_{0} }$ & $\frac{\alpha }{\Gamma_{0} }$ & $0$ & $-\frac{\Gamma_{0}}{3}   \left(1-\frac{\alpha }{\Gamma_{0} }\right)$& $\frac{1}{2}\left\lbrace1-  \Gamma_{0}  \left(1-\frac{\alpha }{\Gamma_{0} }\right)\right\rbrace $ & $\left\lbrace \alpha <0~\mbox{and}~ \Gamma_{0} \leq \alpha )\right\rbrace ~\mbox{or},$ \\ &&&&&& $\left\lbrace \alpha =0~\mbox{and}~(\Gamma_{0} <0~\mbox{or}~ \Gamma_{0} >0)\right\rbrace ~\mbox{or},$\\ &&&&&&
			$ \left\lbrace \alpha >0~\mbox{and}~ \Gamma_{0} \geq \alpha \right\rbrace $\\ \\
			$C_{4}  $ & $1-\frac{\alpha }{\Gamma_{0} +3B| \omega| }$ & $\frac{\alpha }{\Gamma_{0} +3B |\omega| }$& $B|\omega| $ & $\frac{\alpha -\Gamma_{0} }{3}$ & $\frac{1}{2} (1+\alpha -\Gamma_{0} )$ & $B>0~\mbox{and}~$ \\ &&&&&& $\bigg[ $$\left\lbrace \alpha <0~\mbox{and}~ \Gamma_{0} \leq \alpha -3B |\omega| \right\rbrace ~\mbox{or},$ \\ &&&&&& $ \left\lbrace \alpha =0~\mbox{and}~ (\Gamma_{0} <-3B |\omega| ~\mbox{or}~ \Gamma_{0} >-3B |\omega| )\right\rbrace ~\mbox{or},$ \\ &&&&&& $ \left\lbrace \alpha >0~\mbox{and}~ \Gamma_{0} \geq \alpha -3 B |\omega| \right\rbrace \bigg] $
			
			\\\hline\hline
		\end{tabular}
		\label{physical_parameters1 PF Int} \\
		
\end{table}%
%


\begin{table}[h!] \centering
	\caption{The eigenvalues corresponding to critical points are presented}%
	\begin{tabular}
		[c]{cccccccc}\hline\hline
		\textbf{Critical Point}&$\mathbf{\lambda_1}$& $\mathbf{\lambda_2}$ & 
		\\\hline
		$C_1  $ & $6$ & $\Gamma_{0} -\alpha$ &
		\\ \\
		$C_2  $ & $-6 (1+B|\omega|)$ & $\Gamma_{0}-\alpha  +3B |\omega| $ & 
		\\ \\
		$C_3 $ & $\alpha -\Gamma_{0} $ & $-2 (\Gamma_{0}-\alpha -3)$ &
		\\ \\
		$C_4 $ & $-2 (\alpha -\Gamma_{0} +3)$ & $\alpha -\Gamma_{0} -3B |\omega|$ &
		\\ 
		\\\hline\hline
	\end{tabular}
	\label{eigenvalues1 PF Int} \\
	
\end{table}%
%
\subsection{Phase plane analysis}
We shall now discuss the detailed phase plane analysis of the critical points obtained from the autonomous system (\ref{autonomous2 PF}). 
\begin{itemize}
\item Critical point $C_1$ exists for all free parameters $\alpha$, $ \omega$ and $ \Gamma_{0}$ in the phase plane $x-y$. This point corresponds to a completely perfect fluid dominated solution (${\Omega_{d}}=1$), dark matter is absent (${\Omega_{m}}=0$) for this case where the perfect fluid associated to this critical point behaves as dust (since $\gamma_{d}=0$) always. Evolution of the universe near the critical point is always decelerating nature ($q=\frac{1}{2}$). The point is hyperbolic type critical point since all the eigenvalues have non-zero real part see in the table \ref{eigenvalues1 PF Int}). From the linear stability analysis, we observe that the point is always unstable in nature since one of the eigenvalues is always positive ($\lambda_1=6,~~\lambda_2=\Gamma_{0}-\alpha$). Depending on parameters restrictions, the point can behave as saddle-like solution or source in the phase plane. In particular, for $ \Gamma_{0} <\alpha  $ the point corresponds to a saddle solution representing the transient state of the universe, and for $ \Gamma_{0} >\alpha$ it exhibits unstable source describing the early evolution of the universe. The critical point $C_1$ will always evolve in dust dominated decelerated phase ($\omega_{eff}=0$, $q=\frac{1}{2}$ ) for $ \Gamma_{0} <\alpha $ which represents the transient phase of the universe. 

\item Critical point $C_2$ corresponds to a solution in phase plane $x-y$ which is completely dominated ($\Omega_{d}=1$) by the perfect fluid with EoS: $\gamma_{d}=B|\omega|$. The point always exists for all values of model parameters $\alpha$, $ \omega$, $ \Gamma_{0}$ and $B>0$ in the phase plane.
Eigenvalues of the linearized Jacobian  matrix (displayed in the table  \ref{eigenvalues1 PF Int}) indicate that the point is hyperbolic in nature and the linear stability theory is sufficient to determine the stability of the point. The point $C_2$ is saddle-like solution in phase plane when it satisfies the following conditions: \\

$$B>0~~\mbox{and}~~ [ (\omega \leq 0~~\mbox{and}~~ \Gamma_{0} >\alpha +3 B \omega )~~\mbox{or}~~ (\omega >0~~\mbox{and}~~ \Gamma_{0} >\alpha -3 B \omega )]. $$\\
This scenario represents the intermediate phase of the universe.
Finally, depending on parameter restrictions the critical point $C_{2}$ can be stable attractor if either of the following restrictions holds:
$$B>0~~\mbox{and}~~ [ (\omega \leq 0~~\mbox{and}~~ \Gamma_{0} <\alpha +3 B \omega )~~\mbox{or}~~ (\omega >0~~\mbox{and}~~ \Gamma_{0} <\alpha -3 B \omega )]. $$
However, the point will always show the decelerated nature since $\omega_{eff}=B|\omega|>0$. Therefore the point $C_2$ does not support the present observational data. But, interesting fact is that it can predict the future deceleration of the universe.

\item Critical point $C_{3}$ is combination of both the fluid. Therefore, it has scaling nature ($\Omega_{d}\sim \Omega_{m}$) in the phase plane. It exists for the following conditions:

$
(i)~ [\alpha <0~~\mbox{and}~~ \Gamma_{0} \leq \alpha ],~~\mbox{or}~~\\ 
(ii)~[\alpha =0~~\mbox{and}~~ (\Gamma_{0} <0~~\mbox{or}~~ \Gamma_{0} >0)],~~\mbox{or}~~\\
(iii)~[\alpha >0~~\mbox{and}~~ \Gamma_{0} \geq \alpha]$.
The ratio of two fluids for this case is $r=\frac{\Omega_{d}}{\Omega_{m}}=\frac{\alpha }{\Gamma_{0} -\alpha }.$
That is, for $\Gamma_{0}\longrightarrow\alpha$ the energy density of the perfect fluid in the Friedmann equation (\ref{Friedmann}) dominates over the DM ($\Omega_{m}\longrightarrow0,~\Omega_{d}\longrightarrow 1$ see table \ref{physical_parameters1 PF Int}). However, the perfect fluid equation of state associated to this critical point always behaves as dust (since $\gamma_{d}=0$). Depending on parameters the point will show different phases of evolution. In particular, the point evolves in quintessence era (when $-1<\omega_{eff}<-\frac{1}{3}$) for $\alpha \geq 0~~\mbox{and}~~ \alpha +1<\Gamma_{0} <\alpha +3$. On the other hand, the point evolves in cosmological constant era (i.e., $\omega_{eff}=-1$) for $\alpha \geq 0~~\mbox{and}~~ \Gamma_{0} =\alpha +3$ and the point evolves in phantom regime (i.e., $\omega_{eff}<-1$) for $\alpha \geq 0~~\mbox{and}~~ \Gamma_{0} >\alpha +3$.
It is to be mentioned that the acceleration is possible for the restrictions: $\alpha \geq 0~~\mbox{and}~~ \Gamma_{0} >\alpha +1$.\
The critical point $C_{3}$ is hyperbolic (since all the eigenvalues have non-zero real parts) and corresponds to an unstable source for ($\alpha \leq 0~~\mbox{and}~~ \Gamma_{0} <\alpha$), saddle for ($\alpha \geq 0~~\mbox{and}~~ \alpha <\Gamma_{0} <\alpha +3$), and
stable attractor for ($\alpha \geq 0~~\mbox{and}~~ \Gamma_{0} >\alpha +3$).
In summary, the critical point represents the late-time scaling attractor solution corresponding to accelerated universe attracted only in phantom era (satisfying $\lambda_1<0,~~\lambda_2<0,~~ \mbox{and}~~\omega_{eff}<-1$) when $\alpha \geq 0~~\mbox{and}~~ \Gamma_{0} >\alpha +3$.  
Moreover, in a specific parameter region, the point can behave as DM dominated solution. When $\Gamma_{0}= \alpha$ then the point $C_3$ behave as same as the point $C_1$. Here, it is dust dominated decelerated solution in the phase plane. This solution represents transient state of evolution.
On the other hand, for non-interacting case, i.e., for $\alpha= 0$, DM dominates over the other fluid ($\Omega_{m}=1$) and the dynamics of scaling solution completely depends on the parameter $\Gamma_{0}$. In this case, the effective equation of state parameter reduces to the limit $\omega_{eff}=-\frac{\Gamma_{0}}{3}$. Therefore acceleration is possible for $\Gamma_{0}>1$ which is unphysical.  
This is obvious from the above analysis that the point can be physically relevant solution for $0< \Gamma_{0}<1$, i.e., when it is completely dominated by DM representing the intermediate phase of the universe.  

\begin{figure}
	\centering
	\includegraphics[width=9cm,height=9cm]{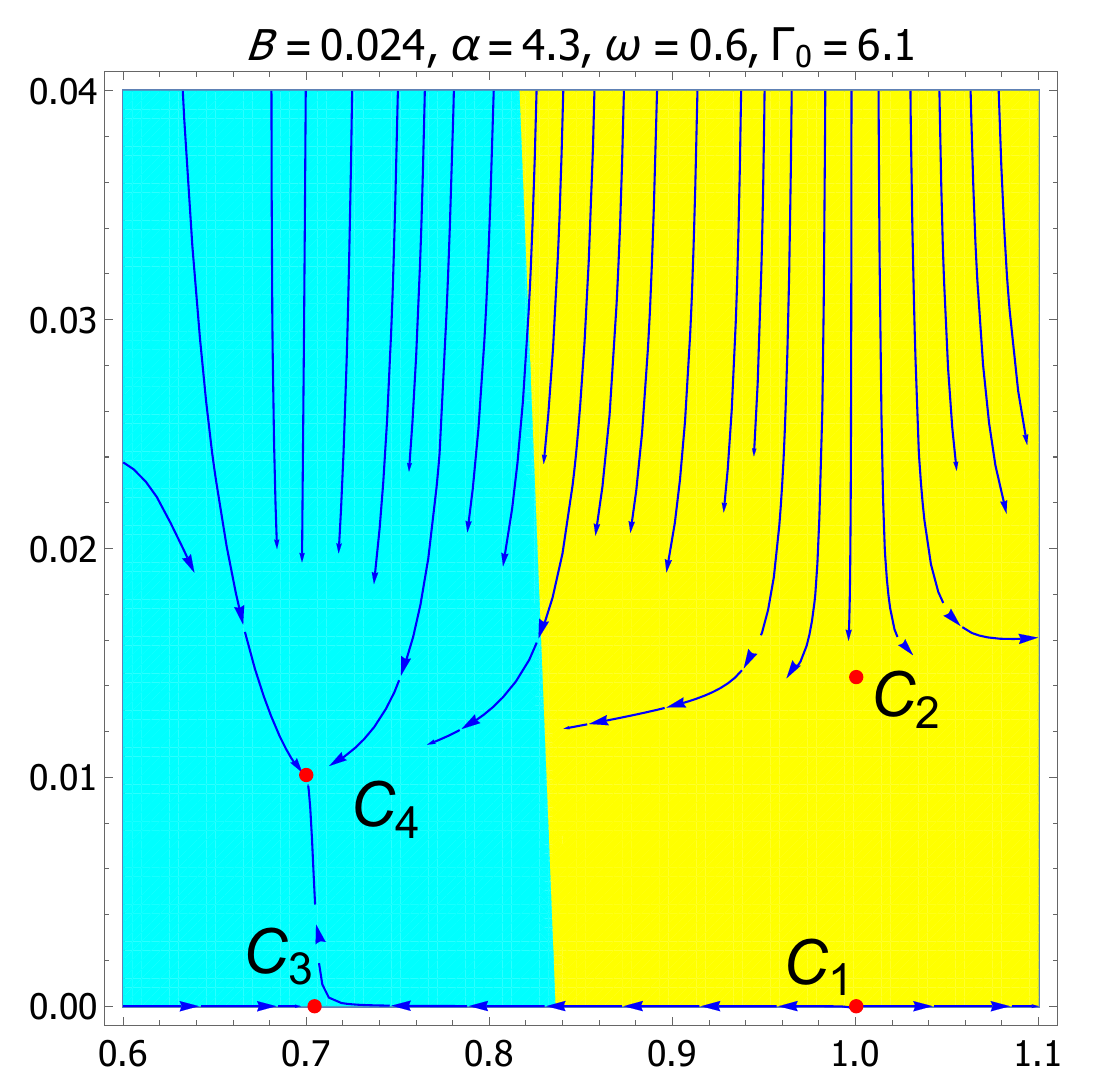}
	
	\caption{The figure shows the phase plane of the autonomous system (\ref{autonomous2 PF}) on the x-y plane for the parameter values: $B=0.024,~~\alpha=4.3,~~\omega=0.6,~~\Gamma_{0}=6.1$. Here, the cyan shaded region represents the accelerated region (i.e. $q < 0$) and the yellow shaded region corresponds to the decelerated region (i.e. $q > 0$).}
	\label{C4}
\end{figure}  
\begin{figure}
	\centering
	\includegraphics[width=9cm,height=9cm]{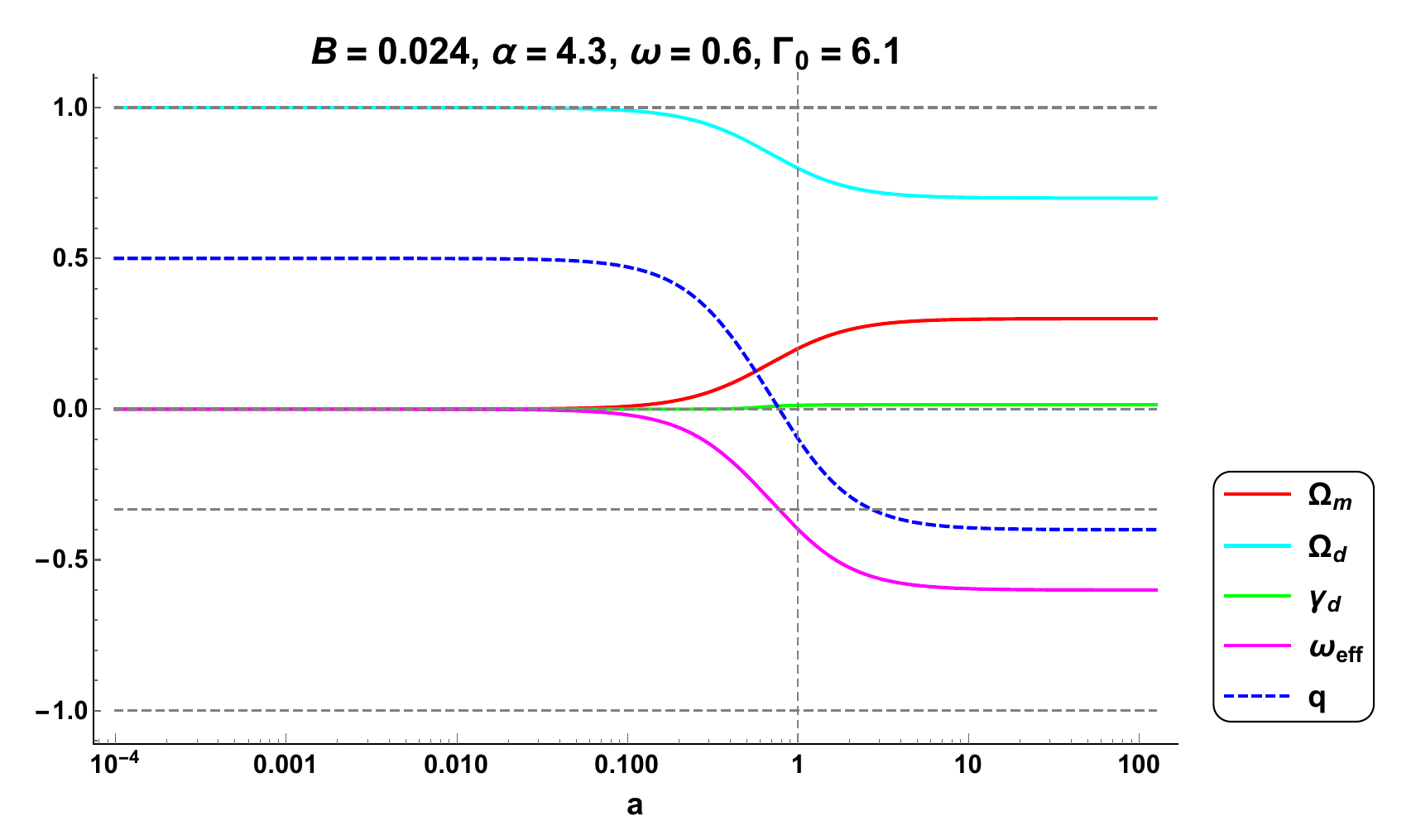}
	
	\caption{The figure shows the evolution of cosmological parameters for the values of free parameters: $B=0.024,~~\alpha=4.3,~~\omega=0.6,~~\Gamma_{0}=6.1$ where the initial conditions are taken as $x(0)=0.7996,y(0)=0.01007$.}
	\label{evolution PF}
\end{figure}  
\item Another scaling solution ($\Omega_{d}\sim \Omega_{m}$) represented by the critical point $C_{4}$ is a combination of both the fluids. The point exists for the following conditions: $B>0$

$
(i)~ (\alpha <0~~\mbox{and}~~ \Gamma_{0} \leq \alpha -3B |\omega| ),~~\mbox{or}~~\\
(ii)~ [\alpha =0~~\mbox{and}~~ (\Gamma_{0} <-3B |\omega| ~~\mbox{or}~~ \Gamma_{0} >-3B |\omega| )],~~\mbox{or}~~\\
(iii)~ (\alpha >0~~\mbox{and}~~ \Gamma_{0} \geq \alpha -3B |\omega| )$.
The ratio of two fluids is $r=\frac{\Omega_{d}}{\Omega_{m}}=\frac{\alpha }{\Gamma_{0}-\alpha +3B |\omega| }.$
So, depending upon parameter restrictions, one fluid can dominate over other. Specifically, the point can be a perfect fluid dominated  ($\Omega_{m}\longrightarrow0,~\Omega_{d}\longrightarrow1$ see table \ref{physical_parameters1 PF Int}) if $ \alpha \neq 0 ~~\mbox{and}~~(\Gamma_{0} -\alpha) \longrightarrow -3B |\omega|  $  while it will be a DM dominated solution for $\alpha\longrightarrow 0$.
Acceleration of the universe near the critical point is possible for the following conditions: ($q<0,~\omega_{eff}<-\frac{1}{3}$):
$$ B>0~~\mbox{and}~~ \alpha \geq 0~~\mbox{and}~~ \Gamma_{0} >\alpha +1 $$
The effective equation of state ($\omega_{eff}$) associated to this critical point determines the universe's evolution in different phases like: quintessence phase, cosmological constant era, or phantom regime or sometimes dust. In particular, evolution of the universe will be in quintessence era ($-1<\omega_{eff}<-\frac{1}{3}$) for:
$$ B>0~~\mbox{and}~~ \alpha \geq 0~~\mbox{and}~~ \alpha +1<\Gamma_{0} <\alpha +3 $$
in cosmological constant era for:
$$ B>0~~\mbox{and}~~ \alpha \geq 0~~\mbox{and}~~ \Gamma_{0} =-3 \left(-\frac{\alpha }{3}-1\right) $$
and in phantom era for:
$$ B>0~~\mbox{and}~~ \alpha \geq 0~~\mbox{and}~~ \Gamma_{0} >\alpha +3 $$
The critical point $C_{4}$ is a hyperbolic in nature since all the eigenvalues have non-zero real parts (presented in table \ref{eigenvalues1 PF Int}). The point can not describe an unstable source ($\lambda_1>0,~\lambda_2>0$).
The point represents the stable attractor ($\lambda_1<0,~\lambda_2<0$) for: $B>0$
\begin{enumerate}
	\item  $\alpha =0 :~~\\
	(i)~(\omega \leq 0~~\mbox{and}~~ 3 B \omega <\Gamma_{0} <3),~~\mbox{or}~~\\
	(ii)~ \left(0<\omega <\frac{1}{B}~~\mbox{and}~~ -3 B \omega <\Gamma_{0} <3\right),~~\mbox{or}~~\\
	(iii)~ \left(\omega =\frac{1}{B}~~\mbox{and}~~ -3 B \omega <\Gamma_{0} <3 B \omega \right),~~\mbox{or}~~\\
	(iv)~ \left(\omega >\frac{1}{B}~~\mbox{and}~~ -3 B \omega <\Gamma_{0} <3\right)$
	
	\item  $\alpha > 0 :~~\\
	(i)~\left(\omega <-\frac{1}{B}~~\mbox{and}~~ \alpha +3 B \omega <\Gamma_{0} <\alpha +3\right),~~\mbox{or}~~\\
	(ii)~ \left(\omega =-\frac{1}{B}~~\mbox{and}~~ \alpha +3 B \omega <\Gamma_{0} <\alpha -3 B \omega \right),~~\mbox{or}~~\\
	(iii)~ \left(-\frac{1}{B}<\omega \leq 0~~\mbox{and}~~ \alpha +3 B \omega <\Gamma_{0} <\alpha +3\right),~~\mbox{or}~~\\
	(iv)~ \left(0<\omega <\frac{1}{B}~~\mbox{and}~~ \alpha -3 B \omega <\Gamma_{0} <\alpha +3\right),~~\mbox{or}~~\\
	(v)~ \left(\omega =\frac{1}{B}~~\mbox{and}~~ \alpha -3 B \omega <\Gamma_{0} <\alpha +3 B \omega \right),~~\mbox{or}~~\\
	(vi)~ \left(\omega >\frac{1}{B}~~\mbox{and}~~ \alpha -3 B \omega <\Gamma_{0} <\alpha +3\right)$.\
	
\end{enumerate}

The point describes the saddle-like solution ($\lambda_1 \lambda_2<0$) for the following conditions :
$$ B>0~~\mbox{and}~~ \alpha \geq 0~~\mbox{and}~~ \Gamma_{0} >\alpha +3 $$

In summary, one can conclude that the critical point $C_4$ can be late-time attractor and it evolves in quintessence era only ($\lambda_1<0,~~\lambda_2<0,~~\mbox{and}~~-1<\omega_{eff}<-\frac{1}{3}$) for :
$$ B>0~~\mbox{and}~~ \alpha \geq 0~~\mbox{and}~~ \alpha +1<\Gamma_{0} <\alpha +3 $$

Further, it should be mentioned that for particular choice of parameters, the critical point corresponds to either perfect fluid or DM dominated solution. In particular, for ($\Gamma_{0}-\alpha)\longrightarrow -3B|\omega|$, perfect fluid dominates the cosmological dynamics $\Omega_{d}\longrightarrow1$, as a result of which the effective equation of state becomes $\omega_{eff}\longrightarrow B|\omega|$.
While, for $\alpha\longrightarrow 0$, DM energy density dominates the cosmological dynamics ($\Omega_{m}\longrightarrow 1$) in Friedmann equation. In this case, $\omega_{eff}\longrightarrow -\frac{\Gamma_{0}}{3}$ which implies that acceleration is possible only when $\Gamma_{0}>1$ and the corresponding eigenvalues are $\lambda_1\longrightarrow -2(-\Gamma_{0}+3)$ and $\lambda_2\longrightarrow -\Gamma_{0}-3B|\omega|$ which results the critical point's stability when $-3B|\omega|<\Gamma_{0}<3$. Therefore, an accelerating DM dominated late-time solution is achieved through this critical point and this result is unphysical according to the present observational data. Therefore, only physically relevant solution in this case will be the matter dominated decelerating transient phase of the universe which is achieved in the parameter region : $\Gamma_{0}<-3B|\omega|.$
Fig. (\ref{C4}) shows that the critical point $C_4$ is a stable solution in the phase plane. The point describes the late time accelerated evolution of the universe. The fig. (\ref{evolution PF}) will also show that the cosmological model with creation rate $\Gamma=\Gamma_{0} H$ can provide the late time cosmic acceleration only when an additional fluid is added with the created matter.\\

{\bf Main results:} Late-time acceleration is possible with the critical points $C_3$ and $C_4$.
Depending on particle creation rate ($\Gamma_{0}$), coupling of interaction ($\alpha$), the critical point $C_3$ represents accelerated universe at late-times attracted only in phantom era (satisfying $\lambda_1<0,~~\lambda_2<0,~~ \mbox{and}~~\omega_{eff}<-1$) when $\alpha \geq 0~~\mbox{and}~~ \Gamma_{0} >\alpha +3$.

On the other hand, the critical point $C_4$ behaves also a late-time accelerated solution depending upon particle creation rate ($\Gamma_{0}$) and coupling of interaction $\alpha$. The critical point $C_4$ can be late-time attractor and it evolves in quintessence era only ($\lambda_1<0,~~\lambda_2<0,~~\mbox{and}~~-1<\omega_{eff}<-\frac{1}{3}$) for :
$$ B>0~~\mbox{and}~~ \alpha \geq 0~~\mbox{and}~~ \alpha +1<\Gamma_{0} <\alpha +3. $$ The stability is shown in 2D plot in the fig. (\ref{C4}). The time evolution of the cosmological parameters for the autonomous system (\ref{autonomous2 PF}) presented in fig. (\ref{evolution PF}) describes the late-time acceleration of the universe.
Therefore, a two-fluid cosmological model with rate $\Gamma=\Gamma_{0} H$ can provide the late time acceleration when second fluid is taken in the form of perfect fluid and the interaction between two fluid is allowed. However, it cannot provide the complete evolution of the universe. For this purpose we need to study by adopting other exotic type of fluid with the created matter.

\end{itemize}

\section{Interacting Umami Chaplygin fluid with particle creation rate $\Gamma=\Gamma_{0} H$}\label{Int Umami model}
Now we consider the universe is filled with pressureless dust and an exotic fluid with negative equation of state $\gamma_{d}<0$. This is the case for $B<0$ in the expression (\ref{EOS PF}). In particular, we consider $B=-1$ in the Eqn. (\ref{pressure PF}) and this is the Umami chaplygin gas as given in Eqn (\ref{Umami Chaplygin}).
we consider another material content of the universe is the Umami Chaplygin gas (UCG) which will be assumed as the dark energy with energy density $\rho_{d}$ and thermodynamic pressure $p_d$. Then, the equation of state in Eqn. (\ref{Umami Chaplygin}) can be rewritten as \cite{Lazkoz2019}
\begin{equation}\label{pressure Umami}
	p_{d}=-\frac{\rho_{d}}{\frac{1}{|\omega|}+\frac{\rho_{d}^{2}}{|A|}}
\end{equation}
The variable equation of state parameter for the Umami Chaplygin gas is denoted by
$\gamma_{d}=\frac{p_{d}}{\rho_{d}}$ and is defined by
\begin{equation}\label{EOS Umami}
	\gamma_{d}\equiv\frac{p_{d}}{\rho_{d}}=-\frac{1}{\frac{1}{|\omega|}+\frac{\rho_{d}^{2}}{|A|}}
\end{equation}

\subsection{Dynamical variables and cosmological parameters}\label{Dynamical variables}

 As the governing equations with irreversible matter creation and with the evolution equation of interacting Umami Chaplygin fluid (given in Eqn.(\ref{pressure Umami})) having the equation of state parameter (in Eqn.(\ref{EOS Umami})) are nonlinear and complicated in form, the exact analytical solution cannot be obtained. However, dynamical system analysis is applied to get a qualitative idea of the overall evolution. By using the dimensionless dynamical variables defined in Eqn. (\ref{variables PF}), cosmological governing equations in (\ref{Raychaudhuri equation}), (\ref{DM1}) and (\ref{DE1}) will give the following 2D system of ordinary differential equations:  
\begin{eqnarray}\label{autonomous Gen Umami}
	\begin{split}
		\frac{dx}{dN}& =\frac{Q}{3H^{3}}-3(x+y)+x\left\lbrace 3(1+y)-\frac{\Gamma}{H}(1-x)\right\rbrace ,& \\
		\frac{dy}{dN}& =\left\lbrace \frac{Q}{3H^{3}}-3(x+y) \right\rbrace \left\lbrace -\frac{y}{x}\left(1+\frac{2y}{|\omega|x } \right)  \right\rbrace+y\left\lbrace 3(1+y)-\frac{\Gamma}{H }(1-x)\right\rbrace   ,
		&
	\end{split}
\end{eqnarray}
Here, the independent variable is chosen as the lapse time $N = \ln a$, which is called the e-folding number.

By assuming the form $\Gamma=\Gamma_{0}H$ in the system (\ref{autonomous Gen Umami}), we obtain a two dimensional system of ODEs:

\begin{eqnarray}\label{autonomous Umami}
	\begin{split}
		\frac{dx}{dN}& =\frac{Q}{3H^{3}}-3(x+y)+x\left\lbrace 3(1+y)-\Gamma_{0}(1-x)\right\rbrace ,& \\
		\frac{dy}{dN}& =\left\lbrace \frac{Q}{3H^{3}}-3(x+y) \right\rbrace \left\lbrace -\frac{y}{x}\left(1+\frac{2y}{|\omega|x } \right)  \right\rbrace+y\left\lbrace 3(1+y)-\Gamma_{0}(1-x)\right\rbrace   ,
		&
	\end{split}
\end{eqnarray}
Note that the above system in explicit presence of $H$ is not an autonomous system. However, the above system of equations is different from that of other works recently done in Ref. \cite{Banerjee2024} where the creation rate $\Gamma$ was considered as $\Gamma=\Gamma_{0}H+\frac{\Gamma_{1}}{H}$. Also, in the present work, we have considered the possibility of matter creation, i.e., $\Gamma_{0}\neq0$ and this makes difference from the system of equations found in Ref.\cite{Biswas2021}, where Umami Chaplygin fluid is taken as DE component. However, for $\Gamma_{0}=0$, the model recovers the standard cosmological solutions as derived in Ref.\cite{Biswas2021} with matter content described by an interacting Umami Chaplygin fluid. Also, the process of gravitational particle creation independently describes the evolutionary history of the Universe and incorporates the observed late-time acceleration quite nicely, for example, see Ref.\cite{Lima}. However, in the present work we consider cosmological model of gravitational matter creation with creation rate $\Gamma=\Gamma_{0} H$, $\Gamma_{0}$, a constant, can not be able to provide the phase transition from deceleration to acceleration epoch. In section \ref{Analytic sol}, we showed that the analytic solution of the model gives the constant value of deceleration parameter $q=\frac{1-\Gamma_{0}}{2}$ (since $\Gamma_{0}$ is a constant). Thus, the model can not provide the phase transition throughout the evolution and the model without having extra fluid does not support the present observation. We, therefore, investigate the particle creation mechanism by considering Umami Chaplygin as an extra fluid with it to fulfill the requirement of providing present acceleration of the universe.  

In view of the above, the present work shows the novelty concerning previously studied models in presence of effect of particle creation rate in interacting Umami Chaplygin fluid which can give a richer dynamics than that of simple interacting model in Ref.\cite{Biswas2021}. Thus, the creation rate can play a vital role in elucidating the evolutionary scheme of the universe. After including a specific choice of $\Gamma$, we have developed the cosmological model in terms of autonomous system and we examine the model qualitatively and numerically when analytical solution is not possible. After finding the sound speed, we can find the stability of the model and also can check the viability by observational data. Now, physical parameters can be expressed in terms of dynamical variables $x$ and $y$ (the co-ordinates of critical points) as in the section \ref{Two fluid model PF}:
the density parameter for dark energy i.e, for Umami Chaplygin gas takes the form in Eqn (\ref{density parameter PF}) whereas the density parameter for dark matter can take the form as in Eqn. (\ref{density parameter DM PF}).
The equation of state parameter for Umami Chaplygin gas
reads similarly as in Eqn.(\ref{EOS  PF}).
From which we obtain the condition for the fluid to be dark energy for $\gamma_{d}=\frac{y}{x}<-\frac{1}{3}.$
The deceleration parameter can be expressed in the form of Eqn.(\ref{deceleration PF}) and the effective equation of state is described in Eqn.(\ref{EOS effective PF}).
Acceleration of the universe is achieved for $q<0$, i.e., for $\omega_{eff}<-\frac{1}{3}$ and the universe is decelerated otherwise. Different epochs of evolution of the universe can be realized by the restrictions of effective equation of state parameter. The universe evolves in quintessence era for $-1<\omega_{eff}<-\frac{1}{3}$, in phantom era for $\omega_{eff}<-1$.
Moreover, we have the evolution equation of the Hubble function as
\begin{equation}
	-\frac{2\dot{H}}{H^{2}}=3(1+y)-\Gamma_{0}(1-x).
\end{equation}
For the energy condition ($0\leq\Omega_{d}\leq 1$), the
phase space becomes bounded in the physical region of dynamical variable as
\begin{equation}
	0\leq x\leq 1.
\end{equation}

\subsection{Formulation of autonomous system with the interaction term $Q=\alpha H \rho_{m}$ and stability analysis }\label{Autonomous system}

In this section, we shall discuss the phase space analysis of the system (\ref{autonomous Umami}) with the phenomenological interaction term $Q=\alpha H \rho_{m}$ \cite{Boehmer2008,Chen2009}, where the coupling parameter $\alpha$ is a dimensionless constant and it measures the strength of interaction. Positive coupling ($\alpha>0$) indicates that the energy transfer occurs from DM to DE and negative is for reverse. The interaction term is popular and is well accepted to solve many DE problems. Local stability of the critical points as well as the classical stability of the model will be performed in this section.
Now, by applying the interaction term in the system of ordinary differential equations (\ref{autonomous Umami}), we obtain the following two-dimensional autonomous system:
\begin{eqnarray}\label{autonomous1}
	\begin{split}
		\frac{dx}{dN}& =(1-x)(\alpha-3y-\Gamma_{0}x) ,& \\
		\frac{dy}{dN}& =\left\lbrace x(\alpha+3)+3y-\alpha \right\rbrace \left\lbrace \frac{y}{x}\left(1+\frac{2y}{|\omega|x } \right)  \right\rbrace+y\left\lbrace 3(1+y)-\Gamma_{0}(1-x)\right\rbrace,
		&
	\end{split}
\end{eqnarray}
To perform dynamical analysis of the system, first we extract all possible critical points from the system above. After that stability (local) analysis will be performed.
The critical points for this system (\ref{autonomous1}) are the following:
{\bf 
	\begin{itemize}
		\item  I.  Critical Point : $P_{1}=(1,0)$
		\item  II. Critical Point : $ P_{2}=(1,-1)$
		
		\item  III. Critical Point : $P_{3}=(1,-|\omega|)$
		
		\item  IV. Critical Point : $P_{4}=(\frac{\alpha}{\Gamma_{0}},0)$
		
		\item  V. Critical Point : $P_{5}=(\frac{\alpha}{\Gamma_{0}-3|\omega|},-\frac{\alpha |\omega|}{\Gamma_{0}-3|\omega|})$
		
		\item  VI. Set of Critical Points : $ P_{6}=(x_{c},\frac{\alpha-\Gamma_{0} x_{c}}{3})$
		
	\end{itemize}
	
}
The existence of critical points and their cosmological parameters are displayed in the table (\ref{physical_parameters}). Local stability and the classical stability will be performed in the following two different subsections. To perform the local stability of the critical points, the eigenvalues of the linearized Jacobian matrix have to be identified for each critical point by perturbing the system up to first order. The eigenvalues of critical points are presented in the table (\ref{eigenvalues}).

	\begin{table}[tbp] \centering
		\fontsize{9.2pt}{5pt}
		\caption{The existence of critical points and the corresponding cosmological parameters for the interaction model $Q=\alpha H \rho_{m}$ }%
		\setlength{\tabcolsep}{0.0001cm}
		\renewcommand{\arraystretch}{}
		\begin{tabular}
			[c]{cccccccccc}\hline\hline
			\textbf{Critical Points}&$\mathbf{\Omega_{m}}$& $\mathbf{\Omega_{d}}$ & $\mathbf{\gamma_{d}}$ &
			$\mathbf{\omega_{eff}}$ &  $q$ & Existence &
			\\\hline
			$P_{1}  $ & $0$ & $1$ &
			$0$ & $0$ & $\frac{1}{2}$  & $\forall \alpha, \omega, \Gamma_{0}$ \\ \\
			$P_{2}  $ & $0$ & $1$ &
			$-1$ & $-1$ & $-1$ &  $\forall \alpha, \omega, \Gamma_{0}$\\ \\
			$P_{3}  $ & $0$ & 
			$1$ & $-|\omega|$ & $-|\omega|$ & $\frac{1}{2}(1-3|\omega|)$ & $\forall \alpha, \omega, \Gamma_{0}$ \\ \\
			$P_{4}  $ & $1-\frac{\alpha }{\Gamma_{0} }$ & $\frac{\alpha }{\Gamma_{0} }$ & $0$ & $-\frac{\Gamma_{0}}{3}   \left(1-\frac{\alpha }{\Gamma_{0} }\right)$& $\frac{1}{2}\left\lbrace1-  \Gamma_{0}  \left(1-\frac{\alpha }{\Gamma_{0} }\right)\right\rbrace $ & $\left\lbrace \alpha <0~\mbox{and}~ \Gamma_{0} \leq \alpha )\right\rbrace ~\mbox{or},$ \\ &&&&&& $\left\lbrace \alpha =0~\mbox{and}~(\Gamma_{0} <0~\mbox{or}~ \Gamma_{0} >0)\right\rbrace ~\mbox{or},$\\ &&&&&&
			$ \left\lbrace \alpha >0~\mbox{and}~ \Gamma_{0} \geq \alpha \right\rbrace $\\ \\
			$P_{5}  $ & $1-\frac{\alpha }{\Gamma_{0} -3| \omega| }$ & $\frac{\alpha }{\Gamma_{0} -3 |\omega| }$& $-|\omega| $ & $\frac{\alpha -\Gamma_{0} }{3}$ & $\frac{1}{2} (1+\alpha -\Gamma_{0} )$ & $\left\lbrace \alpha <0~\mbox{and}~ \Gamma_{0} \leq \alpha +3 |\omega| \right\rbrace ~\mbox{or},$ \\ &&&&&& $ \left\lbrace \alpha =0~\mbox{and}~ (\Gamma_{0} <3 |\omega| ~\mbox{or}~ \Gamma_{0} >3 |\omega| )\right\rbrace ~\mbox{or},$ \\ &&&&&& $ \left\lbrace \alpha >0~\mbox{and}~ \Gamma_{0} \geq \alpha +3 |\omega| \right\rbrace $\\ \\
			$P_{6}  $ & $1-x_{c}$ & $x_{c}$ & $\frac{\alpha -x_{c} (\alpha +3)}{3 x_{c}}$ & $-1$ & $-1$ & $0\leq x_{c}\leq 1~\mbox{and}~ \Gamma_{0} =3+\alpha$
			
			\\\hline\hline
		\end{tabular}
		\label{physical_parameters} \\
		
\end{table}%
%


\begin{table}[h!] \centering
	\caption{The eigenvalues corresponding to critical points are presented}%
	\begin{tabular}
		[c]{cccccccc}\hline\hline
		\textbf{Critical Point}&$\mathbf{\lambda_1}$& $\mathbf{\lambda_2}$ & 
		\\\hline
		$P_1  $ & $6$ & $\Gamma_{0} -\alpha$ &
		\\ \\
		$P_2  $ & $-\frac{6 (|\omega| -1)}{|\omega| }$ & $\Gamma_{0}-\alpha  -3$ &
		\\ \\
		$P_3  $ & $6 (|\omega| -1)$ & $\Gamma_{0}-\alpha  -3 |\omega| $ & 
		\\ \\
		$P_4 $ & $\alpha -\Gamma_{0} $ & $-2 (\Gamma_{0}-\alpha -3)$ &
		\\ \\
		$P_5 $ & $-2 (\alpha -\Gamma_{0} +3)$ & $\alpha -\Gamma_{0} +3 |\omega|$ &
		\\ \\
		$P_6  $ & $0$ & $-6-\alpha+\frac{\alpha }{x_{c}}+\frac{2 \left\lbrace \alpha -x_{c} (\alpha +3)\right\rbrace ^2}{3 x_{c}^2 |\omega| }$ &   \\
		\\\hline\hline
	\end{tabular}
	\label{eigenvalues} \\
	
\end{table}%
%

\subsection{Phase plane analysis  }\label{phase plane analysis}
We shall now discuss the detailed phase space analysis of the critical points obtained from the autonomous system (\ref{autonomous1}). 
\begin{itemize}
\item Critical point $P_1$ exists for all free parameters $\alpha$, $ \omega$ and $ \Gamma_{0}$ in the phase plane $x-y$. This point corresponds to a completely dark energy (Umami Chaplygin fluid) dominated solution (${\Omega_{d}}=1$), dark matter is absent (${\Omega_{m}}=0$) here in the phase plane. The dark energy associated to this critical point behaves as dust (since $\gamma_{d}=0$) always. Evolution of the universe near the critical point is always decelerating ($q=\frac{1}{2}$). This is a hyperbolic type critical point except for $\Gamma_{0}\neq \alpha$ (since all the eigenvalues have non-zero real part see in the table \ref{eigenvalues}). From the linear stability analysis, we observe that the point is always unstable in nature since one of the eigenvalues is always positive ($\lambda_1=6,~~\lambda_2=\Gamma_{0}-\alpha$). Depending on parameters restrictions, the point can behave as saddle-like solution or source in the phase plane. In particular, for $ \Gamma_{0} <\alpha  $ the point corresponds to a saddle solution representing the transient state of the universe, and for $ \Gamma_{0} >\alpha$ it exhibits unstable source describing the early evolution of the universe. The point evolves in dust dominated decelerated era always since $\omega_{eff}=0$, $q=\frac{1}{2}$ and the DE mimics as dust.

\item Critical point $P_2$ also exists for all model parameters $\alpha$, $ \omega$ and $ \Gamma_{0}$ in the phase plane. The point represents completely Umami Chaplygin gas (DE) dominated solution where DM is absent for this case (${\Omega_{d}}=1,~~{\Omega_{m}}=0$). Dark energy mimics as cosmological constant fluid ($\gamma_{d}=-1$). Accelerated universe always exists near the point (since $q=-1$). Eigenvalues of linearized Jacobian matrix (in table \ref{eigenvalues}) ($\lambda_1=-\frac{6(|\omega|-1)}{|\omega|},~~\lambda_2=\Gamma_{0}-\alpha-3$) indicate that the point is hyperbolic type  and hence `linear stability theory' is sufficient to show the nature of critical point. We observe that the critical point $P_2$ describes an unstable source (if all the eigenvalues are positive) for\\
$$[-1<\omega <0,~~ \Gamma_{0} >\alpha +3],~~\mbox{or}~~[ 0<\omega <1,~~ \Gamma_{0} >\alpha +3]. $$\\
Whereas it behaves as a saddle like solution (if the eigenvalues are of opposite sign) in phase plane for\\

$[-1<\omega <0,~~\Gamma_{0} <\alpha +3],~~\mbox{or}~~[0<\omega <1,~~ \Gamma_{0} <\alpha +3],~~\mbox{or}~~[\omega <-1,~~ \Gamma_{0} >\alpha +3],~~\mbox{or}~~[ \omega >1~~\mbox{and}~~ \Gamma_{0} >\alpha +3],$\\

and it corresponds to a stable attractor if the following conditions hold:
$$[\omega <-1,~~ \Gamma_{0} <\alpha +3],~~\mbox{or}~~[ \omega >1,~~ \Gamma_{0} <\alpha +3].$$
The stability of the point $P_2$ is shown in different parametric region in figs. (\ref{Alpha}), (\ref{Gamma}) and in sub-fig. \ref{SROL}.
From the above stability analysis, it is worthy to note that the critical point may describe the de Sitter solution ($\Omega_{d}=1,~~\Omega_{m}=0,~~\omega_{eff}=q=-1$) which corresponds to early evolution (for $|\omega|<1$)  as well as the late evolution (for $|\omega|>1$) of the universe. This type of solution is very much important in cosmological context. The fig. \ref{P2latedS} for $\alpha=0.1,~\omega=1.2,~\Gamma_{0}=0.7$ and the fig. \ref{P2earlydS} for $\alpha=7.7,~\omega=0.6,~\Gamma_{0}=11$ exhibit that the point $P_2$ corresponds to accelerated de Sitter expansion of the universe at late-time and early-time respectively. 

\item Critical point $P_3$ corresponds to a solution in phase plane $x-y$ which is completely dominated by Umami Chaplygin fluid (DE) ($\Omega_{d}=1$). The point always exists for all values of model parameters $\alpha$, $ \omega$ and $ \Gamma_{0}$ in the phase plane. The DE associated to this critical point behaves as an exotic type of fluid with the EoS parameter $\gamma_{d}=-|\omega|$. Therefore, the DE can behave as quintessence, phantom or cosmological constant depending upon some choices of $\omega$. In particular, for $\frac{1}{3}<\omega<1$ or $-1<\omega<-\frac{1}{3}$, the DE evolves as quintessence, while for $\omega>1$ or $\omega<-1$, the DE behaves as phantom fluid and  for $\omega=\pm 1$, the DE describes cosmological constant like fluid. Accelerating universe exists for either $\frac{1}{3}<\omega$ or $\omega<-\frac{1}{3}$ in the phase plane.
Eigenvalues of the linearized Jacobian  matrix (displayed in the table  \ref{eigenvalues}) indicate that the point is hyperbolic in nature and the linear stability theory is sufficient to determine the stability of the point. The point $P_3$ will be an unstable source if all the eigenvalues are positive and the conditions for that is the following:\\
$$[\omega <-1,~~ \Gamma_{0} >\alpha -3 \omega],~~\mbox{or}~~[ \omega >1,~~ \Gamma_{0} >\alpha +3 \omega], $$\\
which indicates the dynamics of early evolution represented by the point.
On the other hand,  the following conditions determine that the point is saddle-like solution in phase plane: \\

$[-1<\omega \leq 0,~~ \Gamma_{0} >\alpha -3 \omega],~~\mbox{or}~~[0<\omega <1,~~ \Gamma_{0} >\alpha +3 \omega], ~~\mbox{or}~~[\omega <-1,~~ \Gamma_{0} <\alpha -3 \omega],~~\mbox{or}~~[\omega >1~~\mbox{and}~~ \Gamma_{0} <\alpha +3 \omega].$\\

In this case, the critical point describes the transient phase of the universe.
Finally, the critical point $P_{3}$ will be stable attractor if either of the following restrictions holds:
$$[-1<\omega \leq 0,~~ \Gamma_{0} <\alpha -3 \omega],~~\mbox{or}~~[0<\omega <1,~~ \Gamma_{0} <\alpha +3 \omega].$$
This criteria describes the late time behavior of the evolution. The stability of the critical point $P_3$ is shown in different regions of parameter space in the figs. (\ref{Alpha}), (\ref{Gamma}) and (\ref{Omega}). It is worthy to note that the critical point $P_{3}$ can be able to show the accelerated late-time stable attractor in quintessence era only for 
$$\left[ -1<\omega<-\frac{1}{3},~~ \Gamma_{0}<\alpha-3\omega\right],~~\mbox{or}~~ \left[  \frac{1}{3}<\omega<1,~~ \Gamma_{0}<\alpha+3 \omega\right] $$
\begin{figure}
	\centering
	\subfigure[]{%
		\includegraphics[width=7.2cm,height=7.2cm]{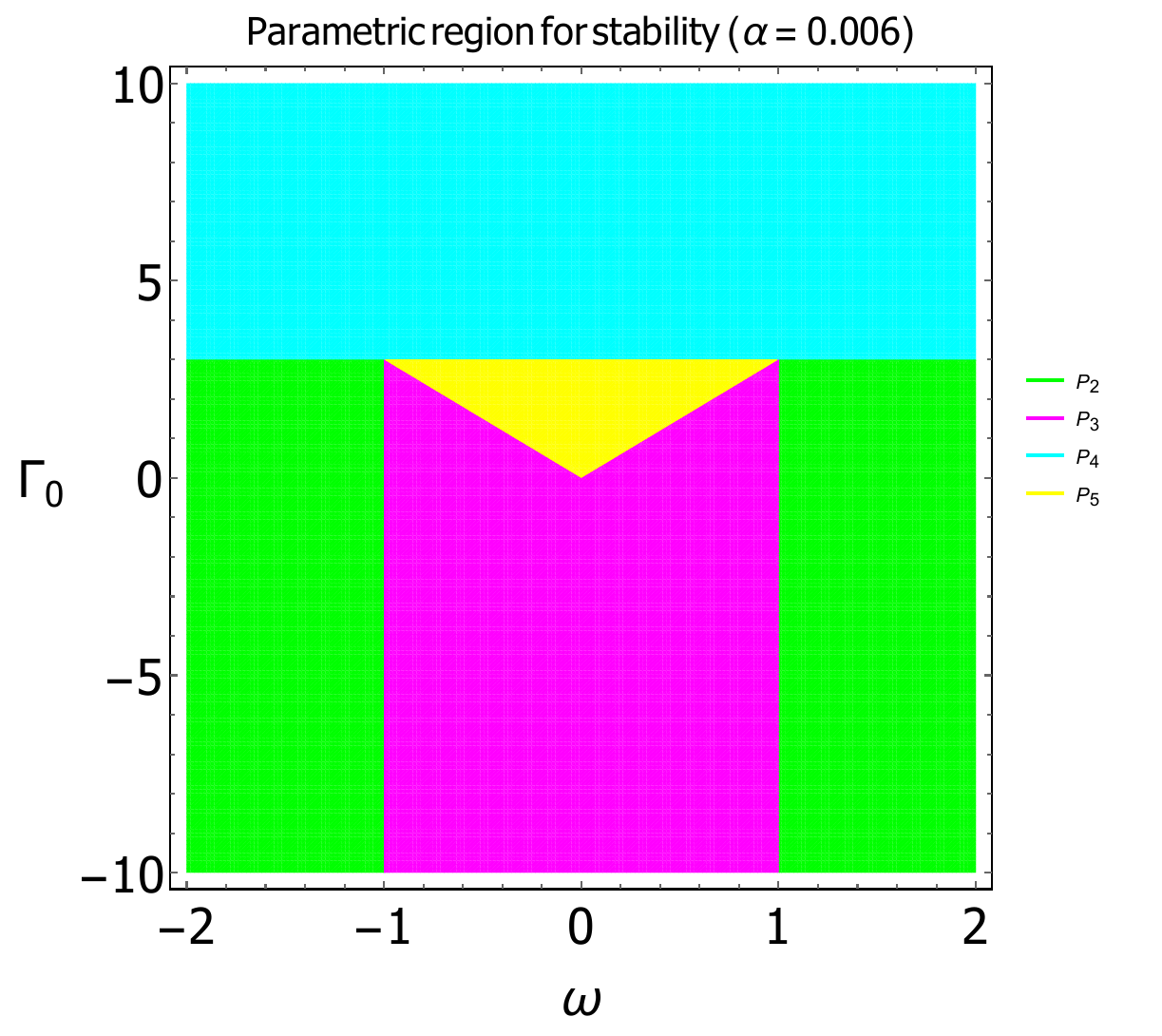}\label{SRA}}
	\qquad
	\subfigure[]{%
		\includegraphics[width=7.2cm,height=7.2cm]{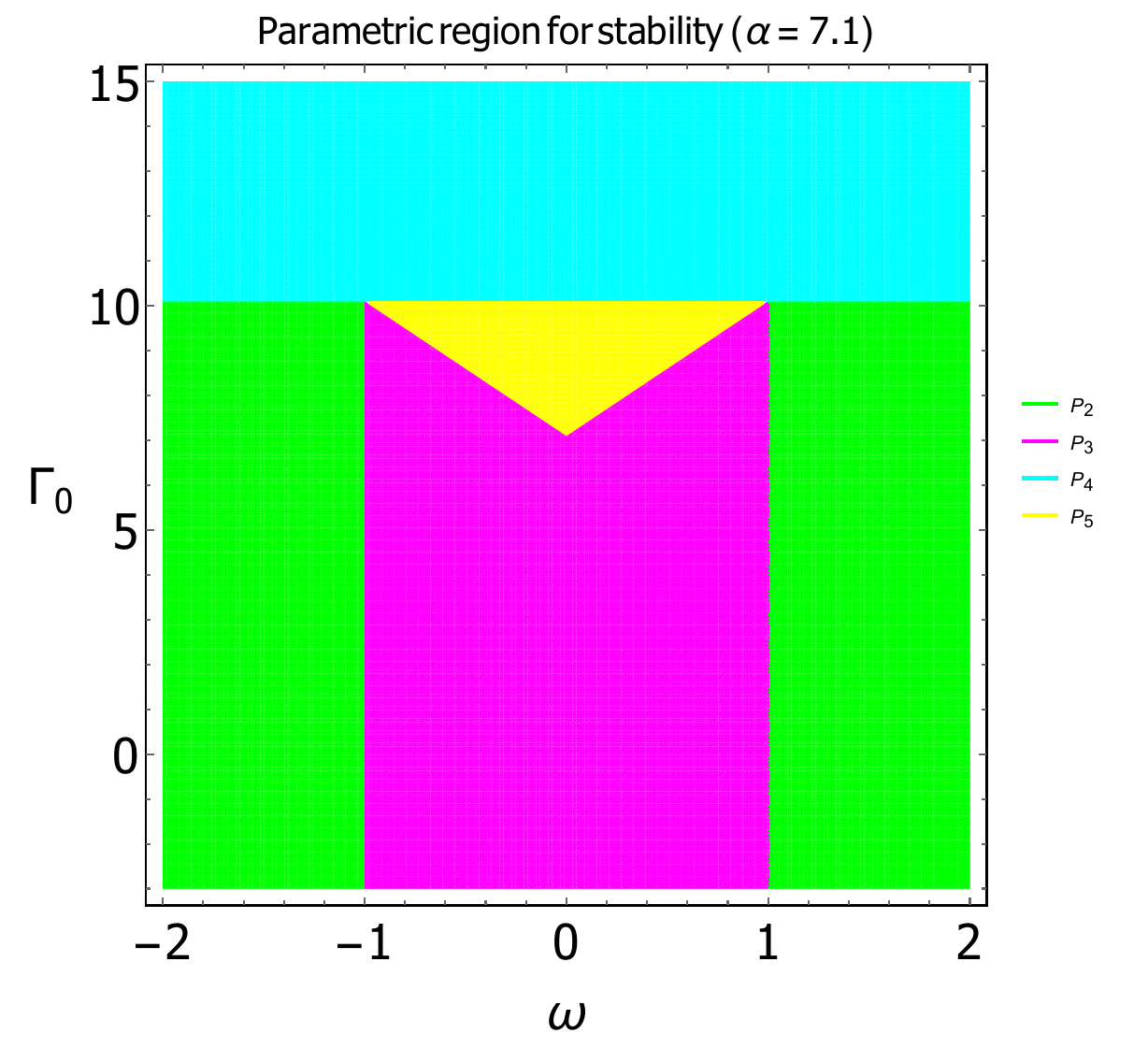}\label{SRAL}}
	
	\caption{The figure shows regions of stability of critical points $P_{2}, P_{3}, P_{4}, P_{5}$ in the $(\omega, \Gamma_{0})$ parameter space for $\alpha=0.006$ in panel (a) and $\alpha=7.1$ in panel (b). Here, green shaded region corresponds to the region of stability of point $P_{2}$, magenta shaded region corresponds to the region of stability of point $P_{3}$, cyan shaded region corresponds to the region of stability of point $P_{4}$ and yellow shaded region corresponds to the region of stability of point $P_{5}$. }
	\label{Alpha}
\end{figure} 

\begin{figure}
	\centering
	\subfigure[]{%
		\includegraphics[width=7.2cm,height=7.2cm]{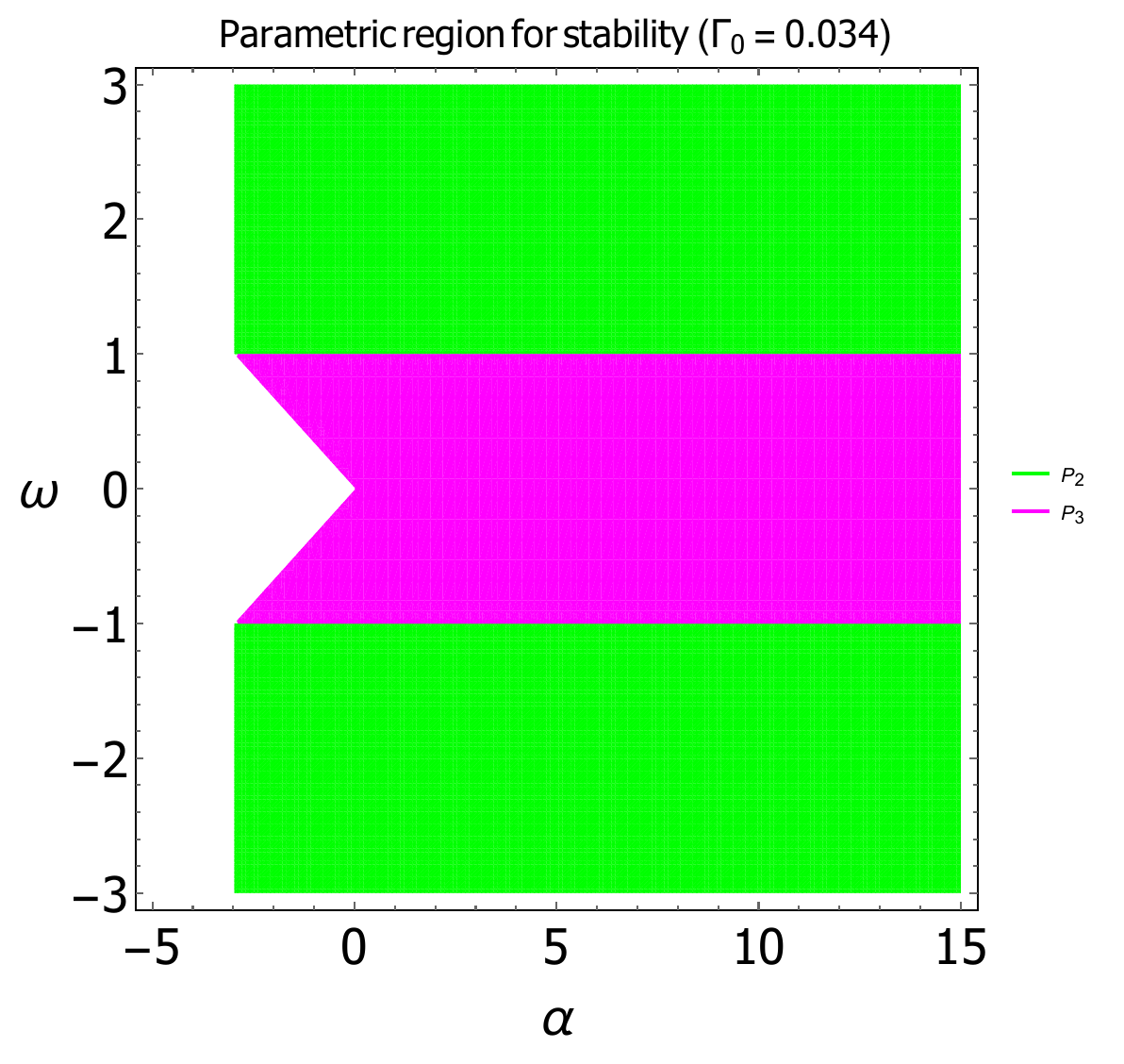}\label{SRG}}
	\qquad
	\subfigure[]{%
		\includegraphics[width=7.2cm,height=7.2cm]{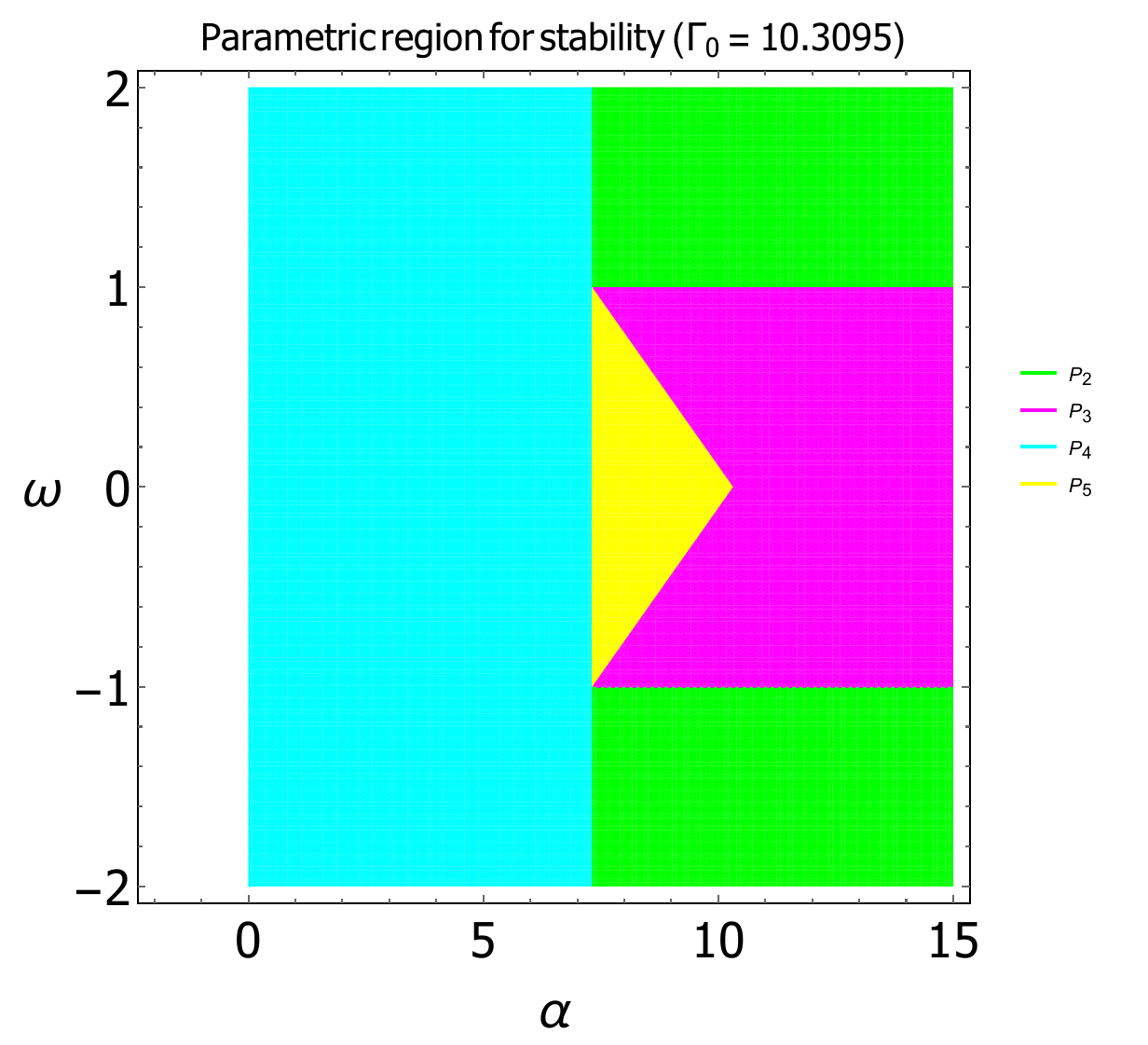}\label{SRGL}}
	
	\caption{The figure shows regions of stability of critical points $P_{2}, P_{3}, P_{4}, P_{5}$ in the $(\alpha,\omega)$ parameter space for $\Gamma_{0}=0.034$ in panel (a) and $\Gamma_{0}=10.3095$ in panel (b). Here, green shaded region corresponds to the region of stability of point $P_{2}$, magenta shaded region corresponds to the region of stability of point $P_{3}$, cyan shaded region corresponds to the region of stability of point $P_{4}$ and yellow shaded region corresponds to the region of stability of point $P_{5}$. }
	\label{Gamma}
\end{figure} 
\begin{figure}
	\centering
	\subfigure[]{%
		\includegraphics[width=7.2cm,height=7.2cm]{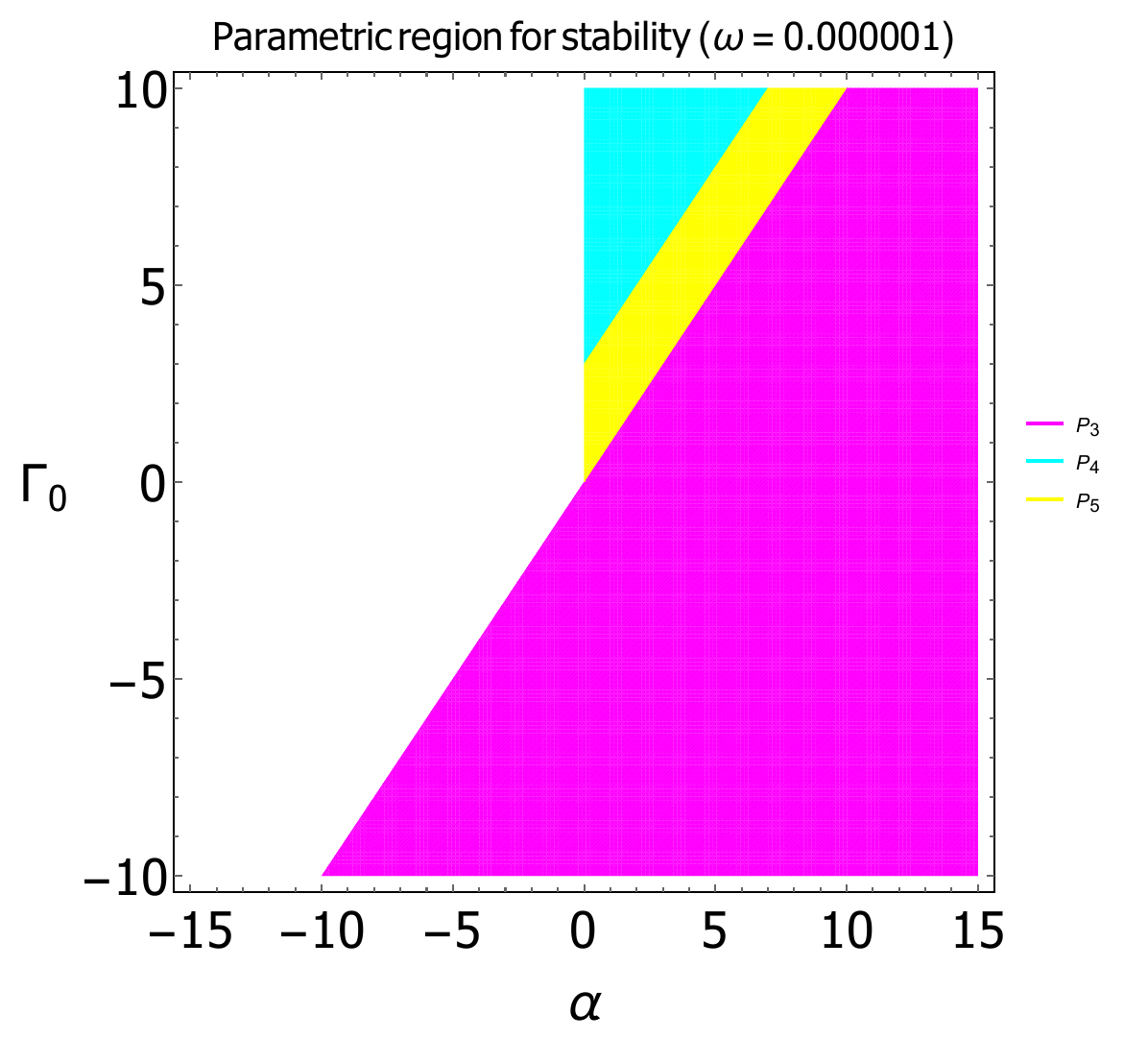}\label{SRO}}
	\qquad
	\subfigure[]{%
		\includegraphics[width=7.2cm,height=7.2cm]{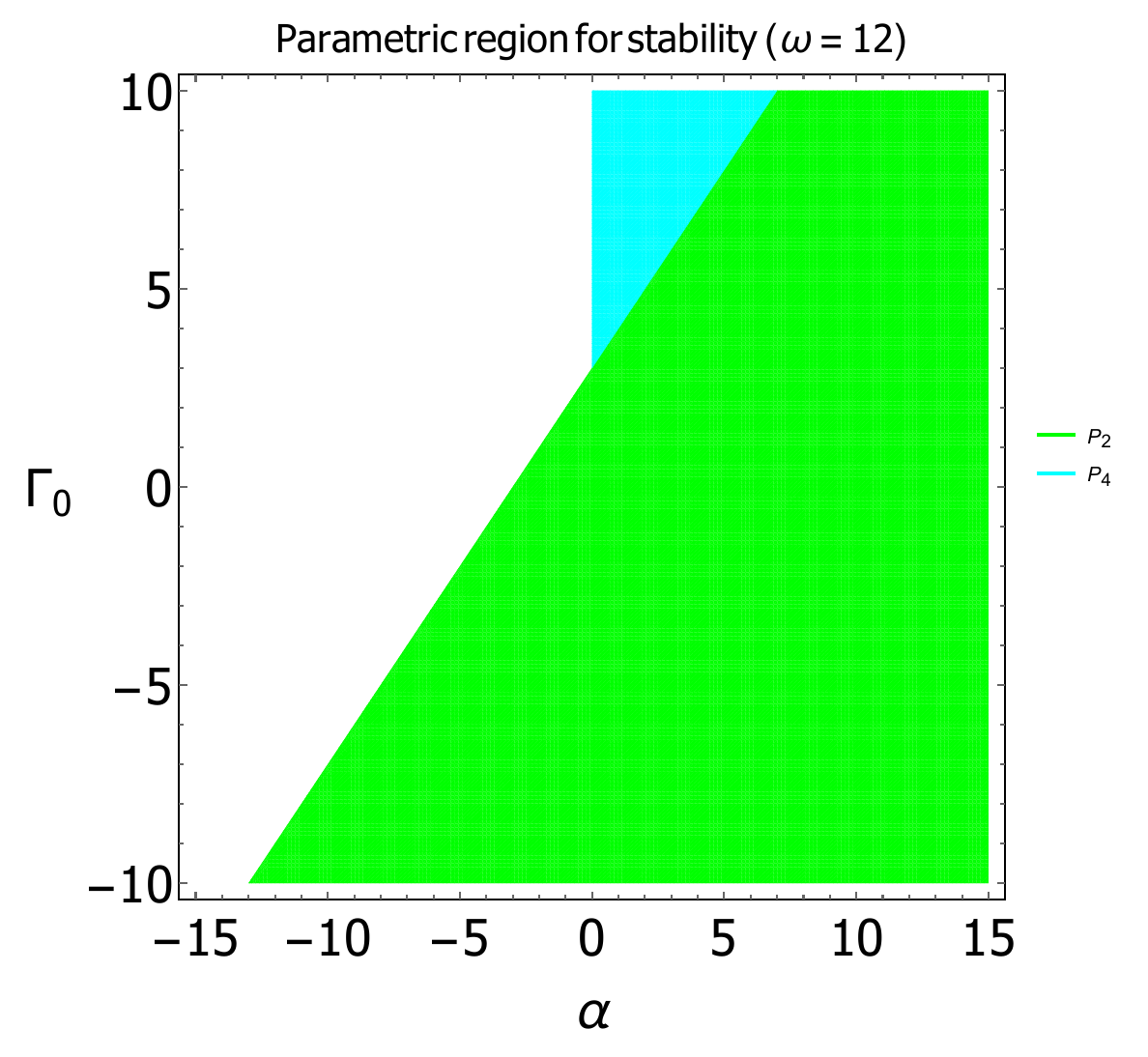}\label{SROL}}
	
	\caption{The figure shows regions of stability of critical points $P_{2}, P_{3}, P_{4}, P_{5}$ in the $(\alpha,\Gamma_{0})$ parameter space for $\omega=0.000001$ in panel (a) and $\omega=12$ in panel (b). Here, green shaded region corresponds to the region of stability of point $P_{2}$, magenta shaded region corresponds to the region of stability of point $P_{3}$, cyan shaded region corresponds to the region of stability of point $P_{4}$ and yellow shaded region corresponds to the region of stability of point $P_{5}$. }
	\label{Omega}
\end{figure} 
\begin{figure}
	\centering
	\subfigure[]{%
		\includegraphics[width=7.2cm,height=7.2cm]{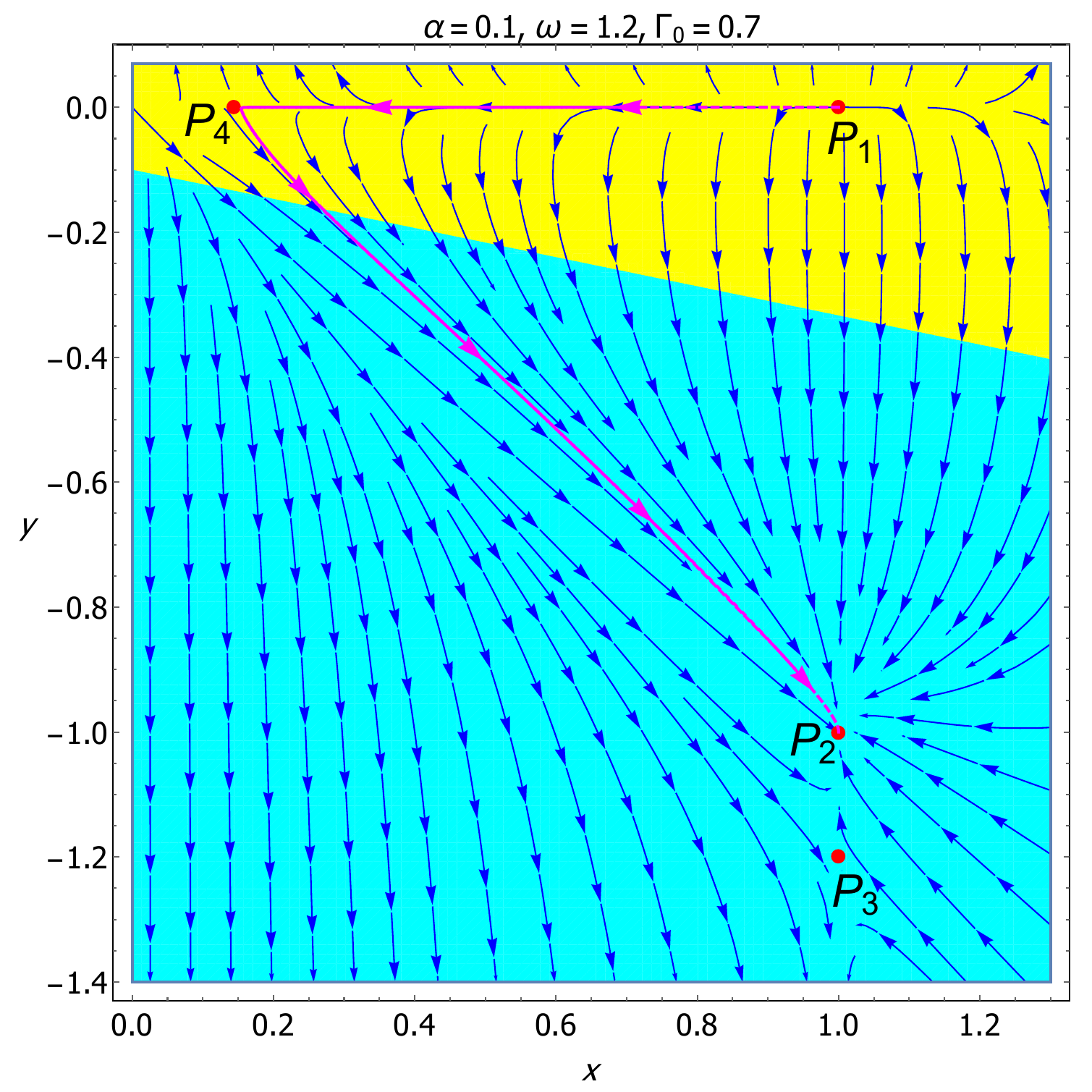}\label{P2latedS}}
	\qquad
	\subfigure[]{%
		\includegraphics[width=7.2cm,height=7.2cm]{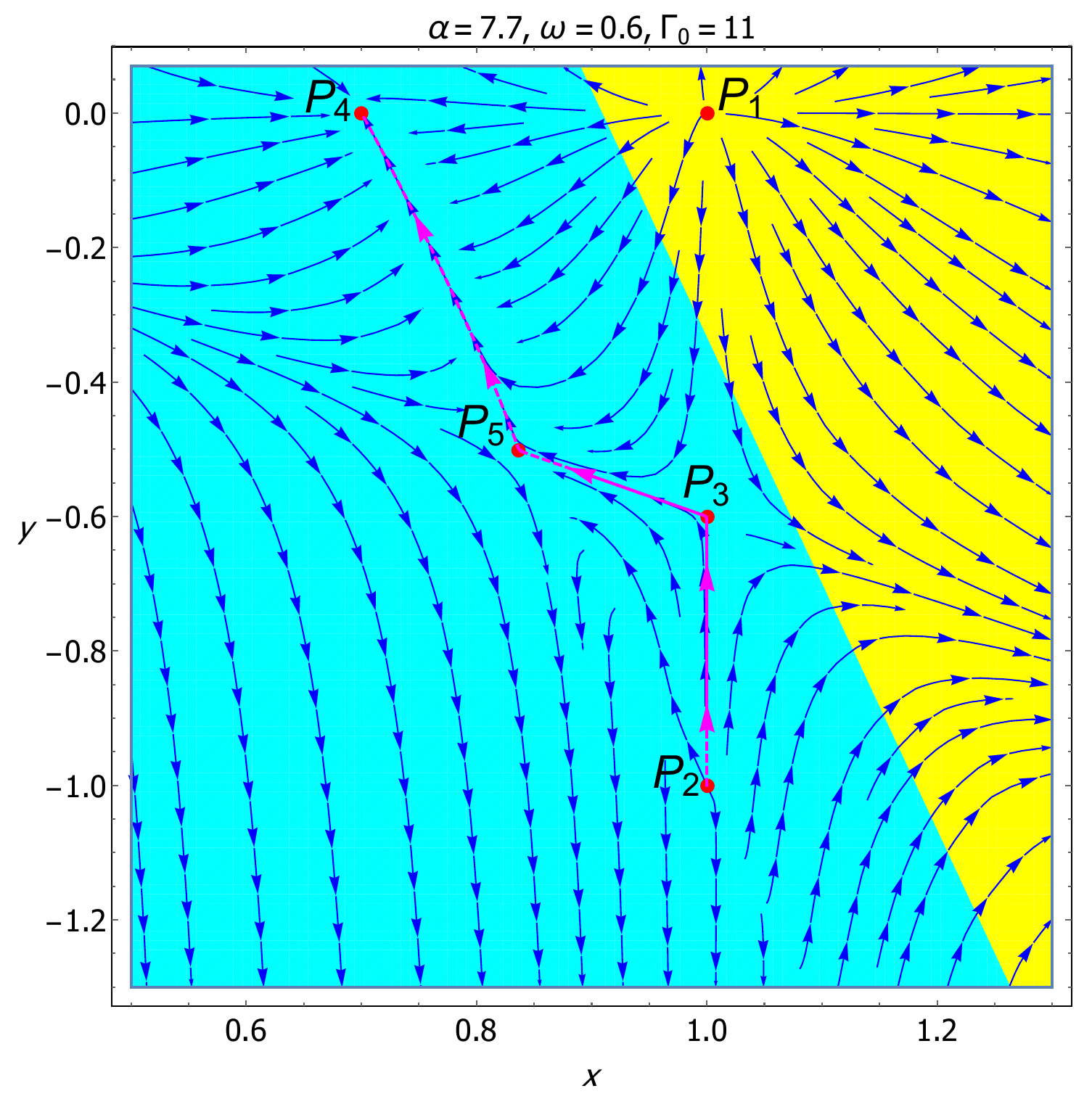}\label{P2earlydS}}
	
	\caption{The phase plot of the autonomous system (\ref{autonomous1}) on the x-y plane for different parameter values:  
		In sub-figure \ref{P2latedS} for $\alpha=0.1,~~\omega=1.2,~~\Gamma_{0}=0.7$ the point $P_2$ exhibits the late time {\it de Sitter} solution, point $P_4$ is matter dominated saddle-like solution and the point $P_1$ is DE dominated decelerated past attractor. Also, the point $P_3$ is DE dominated saddle-like solution. 
		The sub-figure \ref{P2earlydS} shows for $\alpha=7.7,~~\omega=0.6,~~\Gamma_{0}=11$ that the point $P_2$ is the early {\it de Sitter} solution, points $P_3$, $P_5$ are DE dominated saddle-like solutions.
		The point $P_4$ is DE dominated accelerated scaling attractor. The point $P_1$ is DE dominated decelerated past attractor.
		The point $P_3$ is DE dominated accelerated saddle-like solution. Here, the cyan-shaded region represents the accelerated region (i.e. $q < 0$) and the yellow-shaded region corresponds to the decelerated region (i.e. $q > 0$). }
	\label{P2deSitter}
\end{figure} 

\item Critical point $P_{4}$ corresponds to DE-DM scaling solution ($\Omega_{d}\sim \Omega_{m}$) in the phase plane. It exists for the following conditions:

$
(i)~ [\alpha <0~~\mbox{and}~~ \Gamma_{0} \leq \alpha ],~~\mbox{or}~~\\ 
(ii)~[\alpha =0~~\mbox{and}~~ (\Gamma_{0} <0~~\mbox{or}~~ \Gamma_{0} >0)],~~\mbox{or}~~\\
(iii)~[\alpha >0~~\mbox{and}~~ \Gamma_{0} \geq \alpha]$.
The ratio of DE and DM for this case is $r=\frac{\Omega_{d}}{\Omega_{m}}=\frac{\alpha }{\Gamma_{0} -\alpha }.$
That is, for $\Gamma_{0}\longrightarrow\alpha$ the energy density of the DE in the Friedmann equation (\ref{Friedmann}) dominates over the DM ($\Omega_{m}\longrightarrow0,~\Omega_{d}\longrightarrow 1$ see table \ref{physical_parameters}). However, the DE equation of state associated to this critical point always behaves as dust (since $\gamma_{d}=0$). Restrictions on the parameters in the expression of effective equation of state  $\omega_{eff}$($=-\frac{\Gamma_{0}}{3}(1-\frac{\alpha}{\Gamma_{0}})$) (in table \ref{physical_parameters}) will show different dynamic phases of evolution. In what follows, the point evolves in quintessence era (parameters satisfying $-1<\omega_{eff}<-\frac{1}{3}$) when $\alpha \geq 0~~\mbox{and}~~ \alpha +1<\Gamma_{0} <\alpha +3$. On the other hand, the point evolves in cosmological constant era (i.e., $\omega_{eff}=-1$) for $\alpha \geq 0~~\mbox{and}~~ \Gamma_{0} =\alpha +3$ and the point evolves in phantom regime (i.e., $\omega_{eff}<-1$) for $\alpha \geq 0~~\mbox{and}~~ \Gamma_{0} >\alpha +3$.
It is to be mentioned that the acceleration will be occurred for $\alpha \geq 0~~\mbox{and}~~ \Gamma_{0} >\alpha +1$.\
The critical point $P_{4}$ is hyperbolic (since all the eigenvalues have non-zero real parts) and corresponds to an unstable source for ($\alpha \leq 0~~\mbox{and}~~ \Gamma_{0} <\alpha$), saddle for ($\alpha \geq 0~~\mbox{and}~~ \alpha <\Gamma_{0} <\alpha +3$), and
stable attractor for ($\alpha \geq 0~~\mbox{and}~~ \Gamma_{0} >\alpha +3$).
The region of stability for this point is shown in different parameter spaces $(\alpha,~ \Gamma_{0},~ \omega)$ that are exhibited in the figs. (\ref{Alpha}), \ref{SRGL} and (\ref{Omega}).
In summary, the critical point represents the late-time scaling attractor solution corresponding to accelerated universe attracted only in phantom era (satisfying $\lambda_1<0,~~\lambda_2<0,~~ \mbox{and}~~\omega_{eff}<-1$) when $\alpha \geq 0~~\mbox{and}~~ \Gamma_{0} >\alpha +3$, and in this case, DE behaves as dust. This is confirmed numerically by plotting figures. Fig. \ref{Complete Stream}, fig. \ref{Complete Trajectory} and fig. \ref{P4stable} exhibit that the late-time accelerated scaling attractor represented by point $P_4$ is attracted in phantom regime for the parameter values $\alpha=7.1,~\omega=0.005,~\Gamma_{0}=10.3095$. Also, Fig. \ref{P2earlydS} shows that the point $P_4$ is late time accelerated scaling attractor for $\alpha=7.7,~\omega=0.6,~\Gamma_{0}=11.$  
Moreover, in a specific parameter region, the point can behave as DM dominated solution. For example:  in the limit $\Gamma_{0}\longrightarrow \alpha$ the DE (dust) dominates ($\Omega_{d}\longrightarrow1$) the cosmological dynamics over DM and the solution represents decelerated universe in dust era ($\omega_{eff}\longrightarrow 0,~ q\longrightarrow \frac{1}{2}$) when the DE behaves as dust ($\gamma_{d}=0$). In this case, the point will be unstable since one of the eigenvalues will always be positive. This solution represents transient state of evolution.
On the other hand, when $\alpha\longrightarrow 0$, (i.e., for uncoupled case) DM dominates over the DE ($\Omega_{m}\longrightarrow1$) and the dynamics of scaling solution completely depends on the parameter $\Gamma_{0}$. In this case, the effective equation of state parameter reduces to the limit $\omega_{eff}\longrightarrow-\frac{\Gamma_{0}}{3}$, and stability of the point is determined by the eigenvalues: $\lambda_1\longrightarrow -\Gamma_{0}$, $\lambda_2\longrightarrow -2(\Gamma_{0}-3)$. 
This is obvious from the above analysis that the point can be physically relevant solution for $0< \Gamma_{0}<1$, i.e., when it is completely dominated by DM representing the intermediate phase of the universe.  

\item Another DE-DM scaling solution ($\Omega_{d}\sim \Omega_{m}$) represented by the critical point $P_{5}$  exists for the following conditions:

$
(i)~ (\alpha <0~~\mbox{and}~~ \Gamma_{0} \leq \alpha +3 |\omega| ),~~\mbox{or}~~\\
(ii)~ [\alpha =0~~\mbox{and}~~ (\Gamma_{0} <3 |\omega| ~~\mbox{or}~~ \Gamma_{0} >3 |\omega| )],~~\mbox{or}~~\\
(iii)~ (\alpha >0~~\mbox{and}~~ \Gamma_{0} \geq \alpha +3 |\omega| )$.
The ratio of DE and DM is $r=\frac{\Omega_{d}}{\Omega_{m}}=\frac{\alpha }{\Gamma_{0}-\alpha -3 |\omega| }.$
So, depending upon parameter restrictions, either DE can dominate the cosmological dynamics over the DM, or the DM dominates over DE in their evolution. Specifically, the point can be a  DE dominated solution ($\Omega_{m}\longrightarrow0,~\Omega_{d}\longrightarrow1$ see table \ref{physical_parameters}) for $ \alpha \neq 0 ~~\mbox{and}~~(\Gamma_{0} -\alpha) \longrightarrow3 |\omega|  $  while it will be a DM dominated solution for $\alpha\longrightarrow 0$. The DE associated to this critical point behaves as an exotic type fluid with equation of state $\gamma_{d}=-|\omega|$. Therefore, depending on $\omega$, DE can describe the quintessence, cosmological constant, or phantom field, or any other perfect fluid. 
Acceleration of the universe near the critical point is possible for the following conditions: ($q<0,~\omega_{eff}<-\frac{1}{3}$):
\begin{enumerate}
	\item  $\left\lbrace \alpha <0~~\mbox{and}~~ \Gamma_{0} >\alpha +1~~\mbox{and}~~\left( \omega \leq -\frac{\Gamma_{0}-\alpha }{3}~~\mbox{or}~~ \omega \geq \frac{\Gamma_{0} -\alpha }{3}\right)\right\rbrace $
	\item  $\left\lbrace \alpha =0~~\mbox{and}~~ \Gamma_{0} >1~~\mbox{and}~~ \left(\omega <-\frac{\Gamma_{0} }{3}~~\mbox{or}~~ -\frac{\Gamma_{0} }{3}<\omega <\frac{\Gamma_{0} }{3}~~\mbox{or}~~ \omega >\frac{\Gamma_{0} }{3}\right)\right\rbrace$
	\item $ \left(\alpha >0~~\mbox{and}~~ \Gamma_{0} >\alpha +1~~\mbox{and}~~-\frac{\Gamma_{0}-\alpha }{3}\leq \omega \leq \frac{\Gamma_{0} -\alpha }{3}\right)$.\
\end{enumerate}
The effective equation of state ($\omega_{eff}$) associated to this critical point determines the universe's evolution in different phases like: quintessence phase, cosmological constant era, or phantom regime or sometimes dust. In particular, evolution of the universe will be in quintessence era ($-1<\omega_{eff}<-\frac{1}{3}$) for:
\begin{enumerate}
	\item $\alpha <0 :~~\\
	(i)~(\omega \leq -1~~\mbox{and}~~ \alpha +1<\Gamma_{0} <\alpha +3),~~\mbox{or}~~\\
	(ii)~ \left(-1<\omega <-\frac{1}{3}~~\mbox{and}~~ \alpha +1<\Gamma_{0} \leq \alpha -3 \omega \right),~~\mbox{or}~~\\
	(iii)~ \left(\frac{1}{3}<\omega <1~~\mbox{and}~~ \alpha +1<\Gamma_{0} \leq \alpha +3 \omega \right),~~\mbox{or}~~\\
	(iv)~ (\omega \geq 1~~\mbox{and}~~ \alpha +1<\Gamma_{0} <\alpha +3)$
	
	\item $\alpha =0 :~~\\
	(i)~(\omega \leq -1~~\mbox{and}~~ 1<\Gamma_{0} <3),~~\mbox{or}~~\\
	(ii)~ [-1<\omega <-\frac{1}{3}~~\mbox{and}~~ (1<\Gamma_{0} <-3 \omega ~~\mbox{or}~~ -3 \omega <\Gamma_{0} <3)],~~\mbox{or}~~\\
	(iii)~ \left(-\frac{1}{3}\leq\omega \leq \frac{1}{3}~~\mbox{and}~~ 1<\Gamma_{0} <3\right),~~\mbox{or}~~\\
	(iv)~ [\frac{1}{3}<\omega <1~~\mbox{and}~~ (1<\Gamma_{0} <3 \omega ~~\mbox{or}~~ 3 \omega <\Gamma_{0} <3)],~~\mbox{or}~~\\
	(v) (\omega \geq 1~~\mbox{and}~~ 1<\Gamma_{0} <3)$
	
	\item $ \alpha >0 :~~\\
	(i)~(-1<\omega <-\frac{1}{3}~~\mbox{and}~~ \alpha -3 \omega \leq \Gamma_{0} <\alpha +3),~~\mbox{or}~~\\
	(ii)~ \left(-\frac{1}{3}\leq\omega \leq \frac{1}{3}~~\mbox{and}~~ \alpha +1<\Gamma_{0} <\alpha +3\right),~~\mbox{or}~~\\ 
	(iii) \left(\frac{1}{3}<\omega <1~~\mbox{and}~~ \alpha +3 \omega \leq \Gamma_{0} <\alpha +3\right)$,
\end{enumerate}
in cosmological constant era for:
\begin{enumerate}
	\item $ \Gamma_{0} =\alpha+3 :~~\\
	(i)~[\alpha <0~~\mbox{and}~~(\omega \leq -1~~\mbox{or}~~ \omega \geq 1)],~~\mbox{or}~~\\
	(ii)~ [\alpha =0~~\mbox{and}~~ (\omega <-1~~\mbox{or}~~-1<\omega <1~~\mbox{or}~~ \omega >1)],~~\mbox{or}~~\\
	(iii)~ (\alpha >0~~\mbox{and}~~-1\leq \omega \leq 1)$,
\end{enumerate}
and in phantom era for:
\begin{enumerate}
	\item $[\alpha <0~~\mbox{and}~~\left\lbrace(\omega <-1~~\mbox{and}~~ \alpha +3<\Gamma_{0} \leq \alpha -3 \omega)~~\mbox{or}~~ (\omega >1~~\mbox{and}~~ \alpha +3<\Gamma_{0} \leq \alpha +3 \omega)\right\rbrace  ]$
	\item $ \alpha =0 :~~\\
	(i)~ [\omega <-1~~\mbox{and}~~ (3<\Gamma_{0} <-3 \omega ~~\mbox{or}~~ \Gamma_{0} >-3 \omega )],~~\mbox{or}~~\\
	(ii)~ (-1\leq \omega \leq 1~~\mbox{and}~~ \Gamma_{0} >3),~~\mbox{or}~~\\
	(iii)~ [\omega >1~~\mbox{and}~~ (3<\Gamma_{0} <3 \omega ~~\mbox{or}~~ \Gamma_{0} >3 \omega )]$
	\item $ \alpha >0 :~~ \\
	(i)~(\omega <-1~~\mbox{and}~~ \Gamma_{0} \geq \alpha -3 \omega),~~\mbox{or}~~\\
	(i)~ (-1\leq\omega \leq 1~~\mbox{and}~~ \Gamma_{0} >\alpha +3),~~\mbox{or}~~\\
	(ii)~ (\omega >1~~\mbox{and}~~ \Gamma_{0} \geq \alpha +3 \omega )$.
\end{enumerate}
The critical point $P_{5}$ is a hyperbolic in nature since all the eigenvalues have non-zero real parts (presented in table \ref{eigenvalues}). According to linear stability theory, the point describes an unstable source ($\lambda_1>0,~\lambda_2>0$) for :\\
$[\alpha \leq 0~~\mbox{and}~~\left\lbrace (\omega <-1~~\mbox{and}~~ \alpha +3<\Gamma_{0} <\alpha -3 \omega)~~\mbox{or}~~ (\omega >1~~\mbox{and}~~ \alpha +3<\Gamma_{0} <\alpha +3 \omega)\right\rbrace ]. $
On the other hand, the point represents the stable attractor ($\lambda_1<0,~\lambda_2<0$) for:\\
$[\alpha \geq 0~~\mbox{and}~~\left\lbrace (-1<\omega \leq 0~~\mbox{and}~~ \alpha -3 \omega <\Gamma_{0} <\alpha +3)~~\mbox{or}~~ (0<\omega <1~~\mbox{and}~~ \alpha +3 \omega <\Gamma_{0} <\alpha +3)\right\rbrace ]$.\
The regions of stability of the critical point in parameter spaces ($\alpha,~\Gamma_{0},~\omega$) are shown in figs. (\ref{Alpha}), \ref{SRGL} and \ref{SRO}.
The point describes the saddle-like solution ($\lambda_1 \lambda_2<0$) for the following conditions :
\begin{enumerate}
	\item  $\alpha \geq 0 :~~\\
	(i)~(\omega \leq -1~~\mbox{and}~~ \Gamma_{0} >\alpha -3 \omega),~~\mbox{or}~~\\
	(ii)~ (-1<\omega \leq 1~~\mbox{and}~~ \Gamma_{0} >\alpha +3),~~\mbox{or}~~\\
	(iii)~ (\omega >1~~\mbox{and}~~ \Gamma_{0} >\alpha +3 \omega )$.\
	
	\item $\alpha \leq 0 :~~\\
	(i)~(\omega \leq -1~~\mbox{and}~~ \Gamma_{0} <\alpha +3),~~\mbox{or}~~ \\
	(ii)~(-1<\omega \leq 0~~\mbox{and}~~ \Gamma_{0} <\alpha -3 \omega ),~~\mbox{or}~~\\
	(iii)~(0<\omega \leq 1~~\mbox{and}~~ \Gamma_{0} <\alpha +3 \omega ),~~\mbox{or}~~\\
	(iv)~ (\omega >1~~\mbox{and}~~ \Gamma_{0} <\alpha +3)$
\end{enumerate}

Saddle solution arises due to instability in one eigendirection associated with positive eigenvalue and stability in another eigendirection associated with negative eigenvalue.
In summary, one can conclude that the critical point $P_5$ corresponding to the late-time accelerated scaling attractor is achieved only in quintessence era  ($\lambda_1<0,~~\lambda_2<0,~~\mbox{and}~~-1<\omega_{eff}<-\frac{1}{3}$) for :
$\alpha \geq 0 ~~\mbox{and}~~$\\
(i)~ $  (-1<\omega \leq -\frac{1}{3}~~\mbox{and}~~ \alpha -3 \omega <\Gamma_{0} <\alpha +3),~~\mbox{or}~~\\
(ii)~ \left(-\frac{1}{3}<\omega \leq \frac{1}{3}~~\mbox{and}~~ \alpha +1<\Gamma_{0} <\alpha +3\right),~~\mbox{or}~~ \\
(iii)~\left(\frac{1}{3}<\omega <1~~\mbox{and}~~ \alpha +3 \omega <\Gamma_{0} <\alpha +3\right).$
It is interesting to note that the point in this case, can alleviate the coincidence problem due to having $0<\Omega_{d}<1$ where DE behaves as quintessence.
Fig. (\ref{P5}) for $\alpha=0.2,~\omega=0.661,~\Gamma_{0}=2.275$ shows that the scaling solution $P_5$ exhibits the accelerated attractor at late-time attracted in quintessence era.
Further, it should be mentioned that for particular choice of parameters, the critical point corresponds to either DE or DM dominated solution. In particular, for ($\Gamma_{0}-\alpha)\longrightarrow 3|\omega|$, DE dominates the cosmological dynamics $\Omega_{d}\longrightarrow1$, as a result of which the effective equation of state becomes $\omega_{eff}\longrightarrow -|\omega|$ which corresponds to accelerated evolution in quintessence era at late times. 
While, for $\alpha\longrightarrow 0$, DM energy density dominates the cosmological dynamics ($\Omega_{m}\longrightarrow 1$) in Friedmann equation. In this case, $\omega_{eff}\longrightarrow -\frac{\Gamma_{0}}{3}$ which implies that acceleration is possible only when $\Gamma_{0}>1$ and the corresponding eigenvalues are $\lambda_1\longrightarrow -2(-\Gamma_{0}+3)$ and $\lambda_2\longrightarrow -\Gamma_{0}+3|\omega|$ which results the critical point's stability when $3|\omega|<\Gamma_{0}<3$. Therefore, an accelerating DM dominated late-time solution is achieved through this critical point and this result is nonphysical due to present observational data. Therefore, only physically relevant solution in this case will be the matter dominated decelerating transient phase of the universe which is achieved in the parameter region : $\Gamma_{0}<\mbox{min}\{1,~~3|\omega|\}.$

\begin{figure}
	\centering
	\includegraphics[width=9cm,height=9cm]{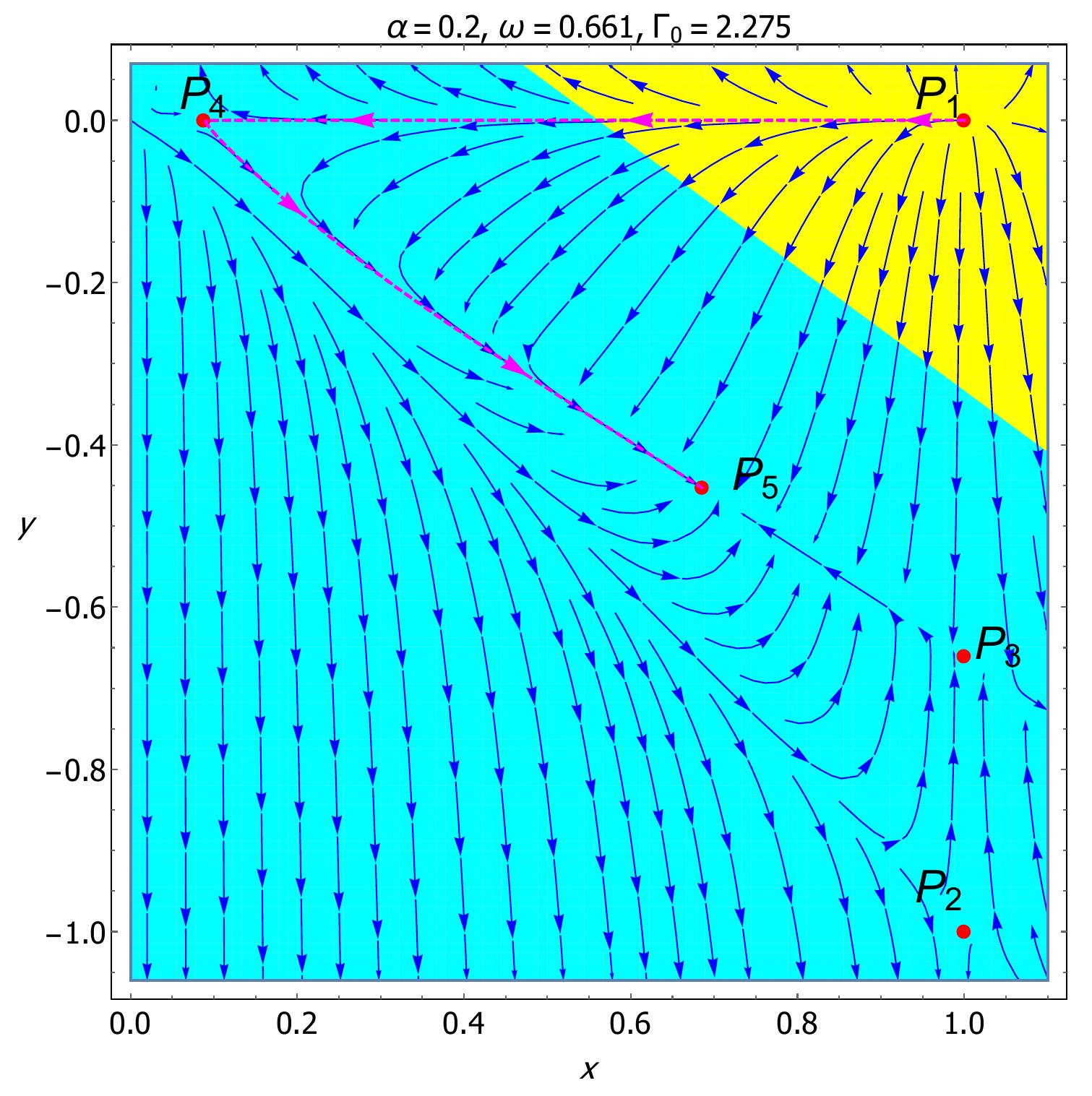}
	
	\caption{The figure shows the phase plane of the autonomous system (\ref{autonomous1}) on the x-y plane for the parameter values:  $\alpha=0.2,~~\omega=0.661,~~\Gamma_{0}=2.275$. The point $P_5$ is accelerated scaling attractor. The point $P_1$ is DE dominated decelerated past attractor where DE behaves as dust. Point $P_4$ is matter dominated accelerated saddle-like solution.
		The point $P_3$ is DE dominated accelerated saddle-like solution. The point $P_2$ is {\it de Sitter} solution describing transient phase. Here, the cyan shaded region represents the
		accelerated region (i.e. $q < 0$) and the yellow shaded region corresponds to the decelerated region (i.e. $q > 0$).  }
	\label{P5}
\end{figure}  



\item The set of critical points $P_{6}$ exists for the following conditions:
($0\leq x_{c}\leq 1~~\mbox{and}~~ \Gamma_{0} =\alpha +3$).
The set corresponds to a solution with combination of both DE and DM. The ratio of DE and DM is $r=\frac{\Omega_{d}}{\Omega_{m}}=\frac{x_{c}}{1-x_{c}}$.
A point on this set will become completely DE dominated for $x_{c}=1$ ($\Omega_{m}=0,~\Omega_{d}=1$ see table \ref{physical_parameters}). The DE behaves as a perfect  fluid with equation of state $\gamma_{d}=\frac{\alpha -x_{c} (\alpha +3)}{3 x_{c}}$. Depending on $x_c$ and the parameter $\alpha$, the DE corresponds to quintessence, cosmological constant or phantom fluid, or any other exotic type fluid. The expansion of the universe will always be accelerating in phase plane (since $\omega_{eff}=-1$, $q=-1$). \\
The set of critical points $P_{6}$ is a non-isolated set of critical points in the phase plane $x-y$. Since, the non-isolated set has exactly one vanishing eigenvalue $\lambda_1 =0$ (see in Table \ref{eigenvalues}), it is called normally hyperbolic set \cite{Coley,Biswas2017,Banerjee2024} and the stability for this type of normally hyperbolic set can be found from the sign of the remaining non-vanishing eigenvalue. Here, the normally hyperbolic set $P_6$ is stable for the following conditions :
\begin{enumerate}
	\item 
	$\Gamma_{0}=\alpha+3~~\mbox{and}~~(0<x_{c}<1) :~~\\
	(i)~~\left\lbrace \omega <0~~\mbox{and}~~-\frac{3x_{c}}{4(x_{c}-1)} \left(\sqrt{(\omega -8) \omega}+(\omega +4)\right)<\alpha <\frac{3x_{c}}{4(x_{c}-1)} \left(\sqrt{(\omega -8) \omega}-(\omega +4)\right)\right\rbrace,~\mbox{or}~\\
	(ii)~~ \left\lbrace \omega >0~~\mbox{and}~~ \frac{3x_{c} }{4(x_{c}-1)} \left(\omega -4-\sqrt{\omega  (\omega +8)}\right)<\alpha <\frac{3x_{c} }{4(x_{c}-1)} \left(\sqrt{\omega  (\omega +8)}+\omega -4\right)\right\rbrace  $
	\item  $\Gamma_{0}=\alpha+3~~\mbox{and}~~\left\lbrace x_{c}=1~~\mbox{and}~~ (\omega <-1~~\mbox{or}~~ \omega >1)\right\rbrace$.
\end{enumerate}
Also, the set $P_{6}$ will be unstable for the following conditions :
\begin{enumerate}
	\item \fontsize{10.4pt}{8pt} $\Gamma_{0}=\alpha+3~~\mbox{and}~~(0<x_{c}<1) :~~\\
	(i)~~ \omega <0~~\mbox{and}~~\left\lbrace\alpha<-\frac{3x_{c}}{4(x_{c}-1)} \left(\sqrt{(\omega -8) \omega}+(\omega +4)\right)  ~~\mbox{or}~~\alpha >\frac{3x_{c}}{4(x_{c}-1)} \left(\sqrt{(\omega -8) \omega}-(\omega +4)\right)  \right\rbrace ,~~\mbox{or}~~\\
	(ii)~~  \omega >0~~\mbox{and}~~\left\lbrace \alpha< \frac{3x_{c} }{4(x_{c}-1)} \left(\omega -4-\sqrt{\omega  (\omega +8)}\right) ~~\mbox{or}~~\alpha >\frac{3x_{c} }{4(x_{c}-1)} \left(\sqrt{\omega  (\omega +8)}+\omega -4\right) \right\rbrace $
	\item  $\Gamma_{0}=\alpha+3~~\mbox{and}~~\left\lbrace x_{c}=1~~\mbox{and}~~ (-1<\omega <0~~\mbox{or}~~ 0<\omega <1)\right\rbrace$.
\end{enumerate}
The fig. (\ref{P6stable}) for $\alpha=7.1,~\omega=1.1,~\Gamma_{0}=10.1$ shows that the set of critical points on the line bounded in $0.609552 < x_{c} \leq 1$ is stable in the phase plane. Interested readers may follow the Refs. \cite{Ghosh2024a,Ghosh2024b} for stability of the non-hyperbolic points and some additional discussions on various dynamical system analysis.\\
It is interesting to note that for $x_c=1$ the set $P_6$ will become a particular isolated critical point. The coordinate of this point is $(1,~\frac{\alpha -\Gamma_{0}}{3})$ which for its existence condition becomes $(1,~-1)$. For this case, the point is non-hyperbolic with eigenvalues $\lambda_1=0,~~\lambda_2=-6+\frac{6}{|\omega|}$. The tools of center manifold theory can be employed to investigate the stability of this non-hyperbolic point and it beyond the scope of our study and it is left for a scope of future study. However, in this work, we will adopt another method to study the stability of this point. We will investigate the stability of this point numerically. In fact, this approach is widely used in cosmological studies in the literature \cite{Roy2014,Roy2015,Roy2017,Roy2018,Zonunmawia2017,Mandal2022,Dutta2019} where dynamical system analysis is applied. We shall now investigate the stability by studying the evolution of perturbations near the critical point. The main mechanism is that the system is perturbed from the critical point and is allowed it to evolve numerically. If the perturbed system comes back to the critical point, then the point is said to be stable, otherwise it is unstable. The evolution of the perturbations are plotted around the critical point (with coordinate $(1,~-1)$). However, the point will have a non-empty stable submanifold for $|\omega|>1$.  The figure (\ref{P6latedS}) shows the phase space projection of perturbation of the system (\ref{autonomous1}) along the x-axis and y-axis for parameter values $\alpha=7.1$, $\omega=1.1$ and $\Gamma_{0}=10.1$. Sub-fig. \ref{Stable_Nx} and sub-fig. \ref{Stable_Ny} exhibit that the perturbations come back to $x=1$ and $y=-1$ axes as $N\longrightarrow \infty$. This indicates that the critical point ($1,~-1$) is stable for parameter values $\alpha=7.1$, $\omega=1.1$ and $\Gamma_{0}=10.1$. Here, cosmological parameters for this point will take the values: $\Omega_{m}=0,~~\Omega_{d}=1,~~\omega_{eff}=q=-1$ and the DE behaves as cosmological constant ($\gamma_{d}=-1$) like fluid. Therefore, the point will describe a late-time de Sitter evolution of the universe.

On the other hand, the point on the set may represent the early accelerated de Sitter solution for $|\omega|<1$, because there is an non-empty unstable sub-manifold in the positive eigendirection. The fig.(\ref{P6earlydS}) with $\alpha=7.1,~\omega=0.6,~\Gamma_{0}=10.1$ exhibits the evolution of perturbations near the critical point (with coordinate ($1,~-1$)) showing that projection of perturbations increase gradually from $x=1$ and $y=-1$ as $N \longrightarrow \infty$. Thus, we conclude that the point is unstable source. In fact, the point describes the accelerated de Sitter evolution at early times.
For $x_c\longrightarrow0$, the set corresponds to the DM dominated solution. But, due to ever accelerating nature, the point cannot be represent the intermediate phase of the universe. So, in this case, it is unrealistic from cosmological point of view.
Finally, for $x_c=\frac{\alpha}{\Gamma_{0}}$, the set will be the critical point $P_4$ which has already been discussed.


\begin{figure}
	\centering
	\subfigure[]{%
		\includegraphics[width=8cm,height=7.2cm]{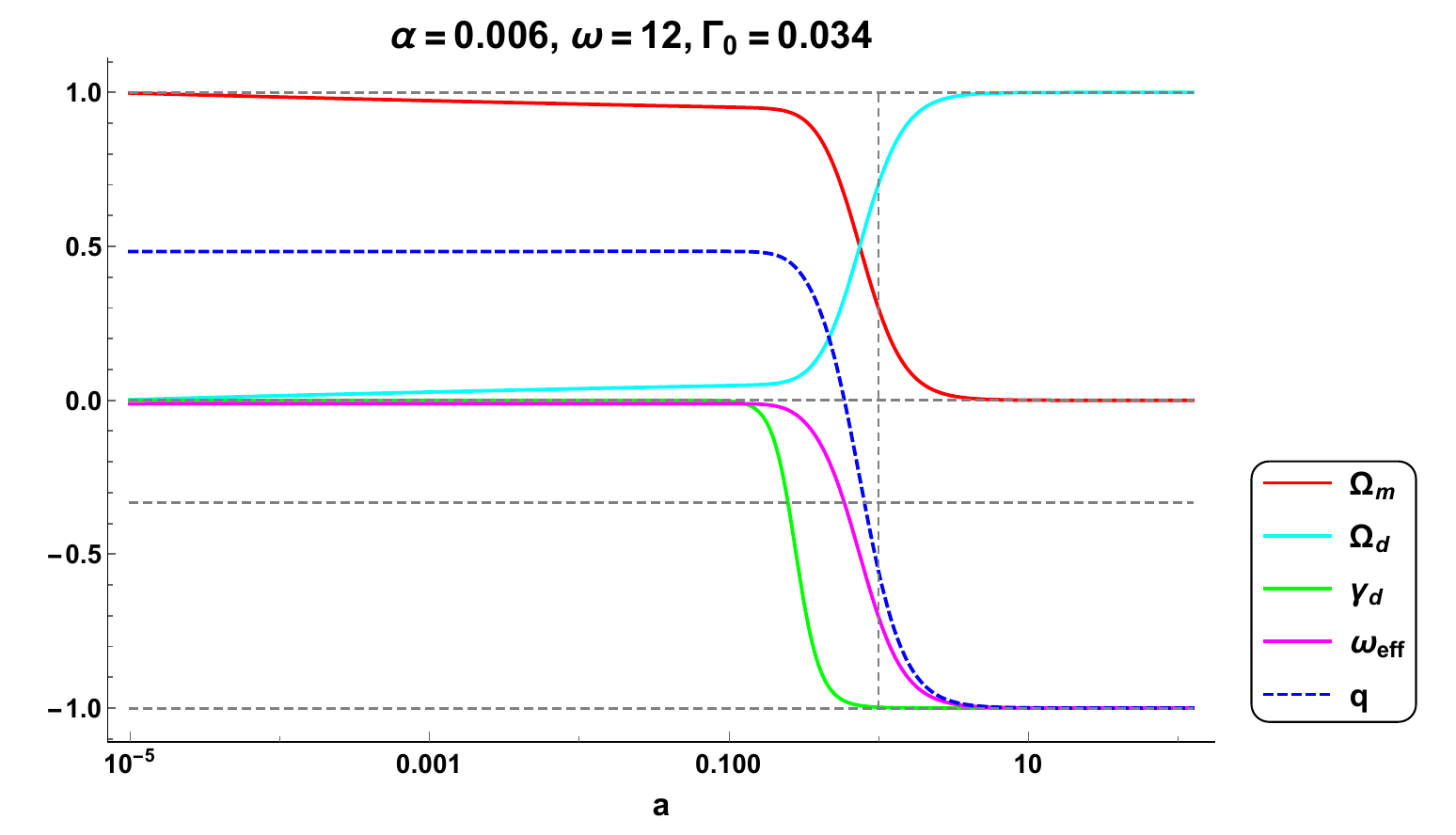}\label{evolutionLCDM1}}
	\qquad
	\subfigure[]{%
		\includegraphics[width=8cm,height=7.2cm]{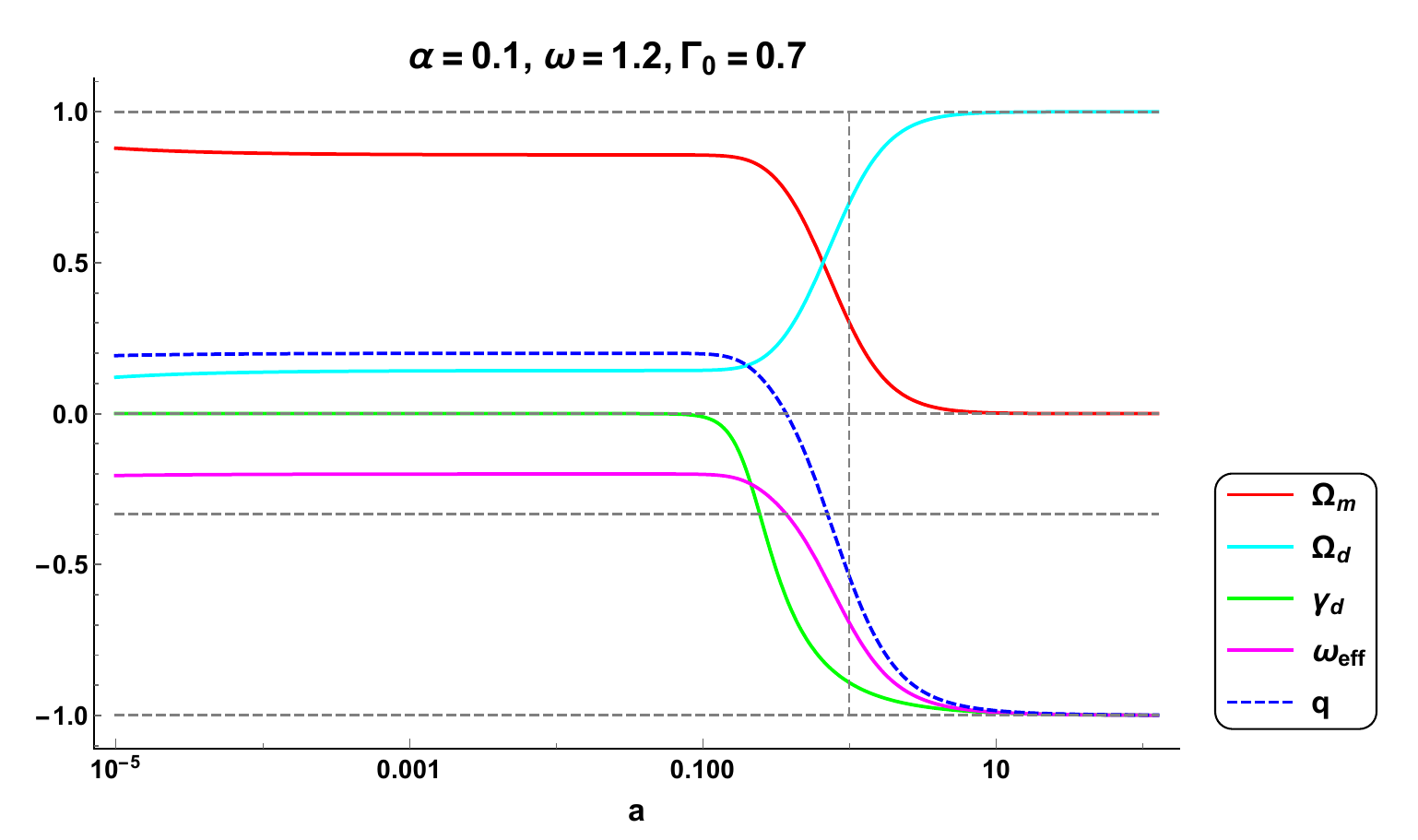}\label{evolutionLCDM2}}
	\caption{ The sub-figure \ref{evolutionLCDM1} shows the evolution of cosmological parameters for the values of free parameters: $\alpha=0.006,~~\omega=12,~~\Gamma_{0}=0.034$ where the initial conditions are taken as $x(0)=0.6995,~y(0)=-0.6976$. Here, the trajectories are attracted by cosmological constant at late-times connecting through a matter-dominated decelerated phase. The sub-fig \ref{evolutionLCDM2} shows the evolution of cosmological parameters for model parameters: $\alpha=0.1,~~\omega=1.2,~~\Gamma_{0}=0.7$ and the initial conditions are $x(0)=0.698994,~y(0)=-0.62304$. This sub-fig also shows that the ultimate evolution of universe is attracted in $\Lambda$CDM.}
	\label{evolutionLCDM}
\end{figure}


\begin{figure}
	\centering
	\includegraphics[width=9cm,height=9cm]{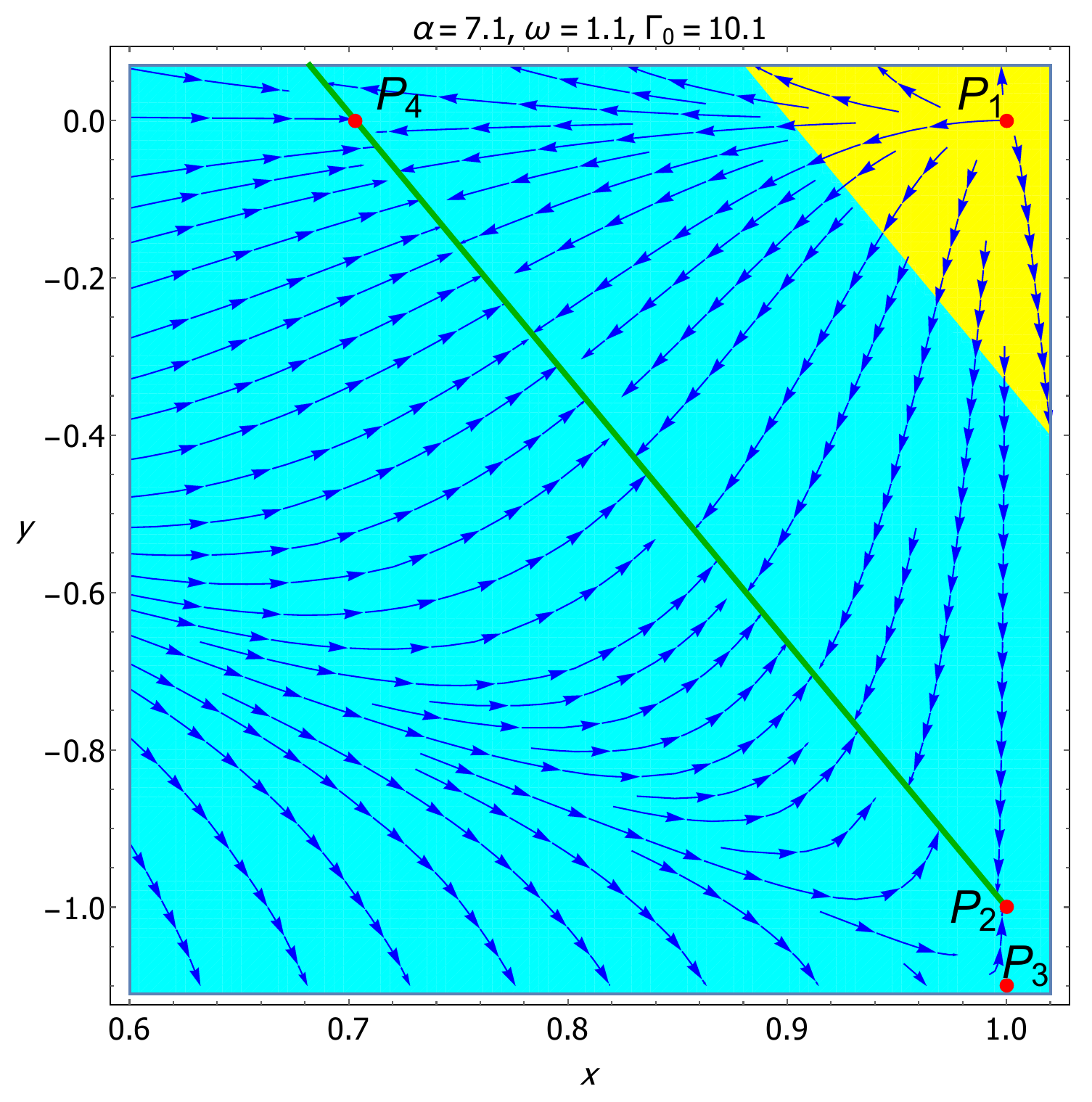}
	
	\caption{The figure shows the vector field of the autonomous system (\ref{autonomous1}) on the x-y plane for the parameter values: $\alpha=7.1,~~\omega=1.1,~~\Gamma_{0}=10.1$.
		The set $P_6$ represented by the green colored line shows the stable attractor. The point $P_1$ is DE dominated decelerated source. But the point $P_3$ gives DE dominated accelerated saddle-like solution. Here, the cyan shaded region represents the accelerated region (i.e. $q < 0$) and the yellow shaded region corresponds to the decelerated region (i.e. $q > 0$). 
	}
	\label{P6stable}
\end{figure}

\begin{figure}
	\centering
	\subfigure[]{%
		\includegraphics[width=7.2cm,height=7.2cm]{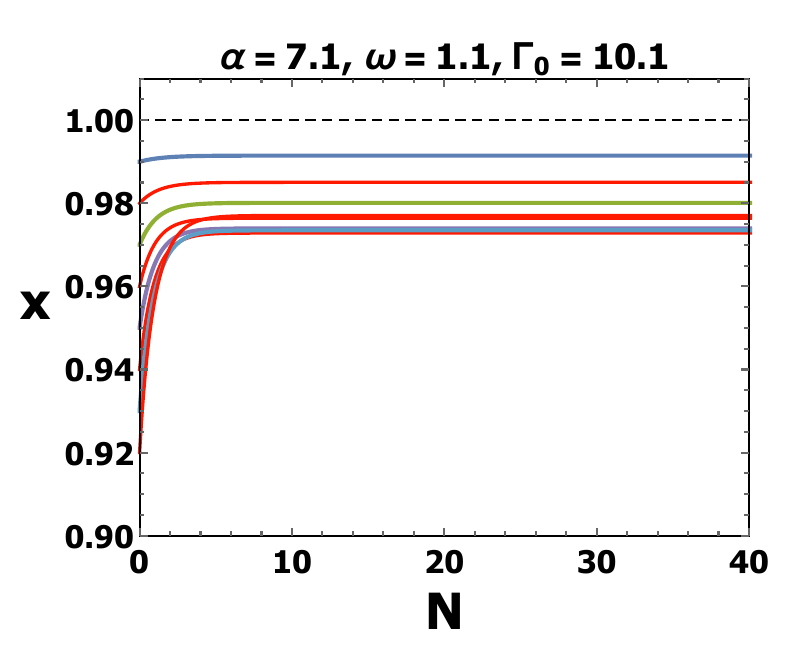}\label{Stable_Nx}}
	\qquad
	\subfigure[]{%
		\includegraphics[width=7.2cm,height=7.2cm]{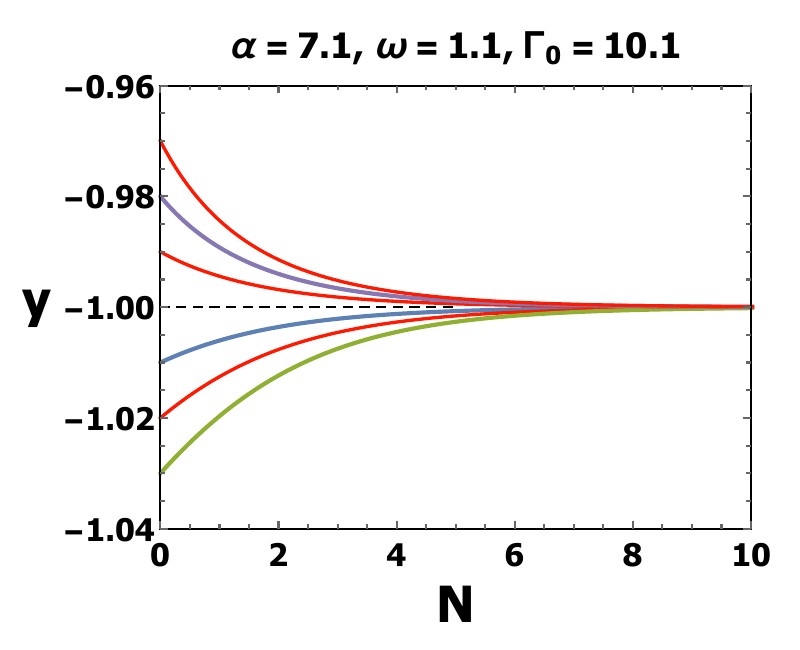}\label{Stable_Ny}}
	\caption{The figures show the phase space projection of perturbation of the autonomous system (\ref{autonomous1}) along the x and y axes for the critical point $P_{6}$ for parameter values $\alpha=7.1$, $\omega=1.1$ and $\Gamma_{0}=10.1$. In panel \ref{Stable_Nx} and \ref{Stable_Ny} perturbations come back. This indicates that the critical point $P_{6}$ is stable for parameter values $\alpha=7.1$, $\omega=1.1$ and $\Gamma_{0}=10.1$.}
	\label{P6latedS}
\end{figure}
\begin{figure}
	\centering
	\subfigure[]{%
		\includegraphics[width=7cm,height=7cm]{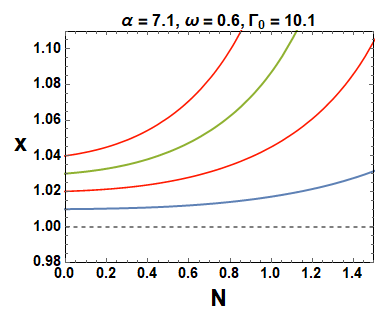}\label{UnStable_Nx}}
	\qquad
	\subfigure[]{%
		\includegraphics[width=7.4cm,height=7.4cm]{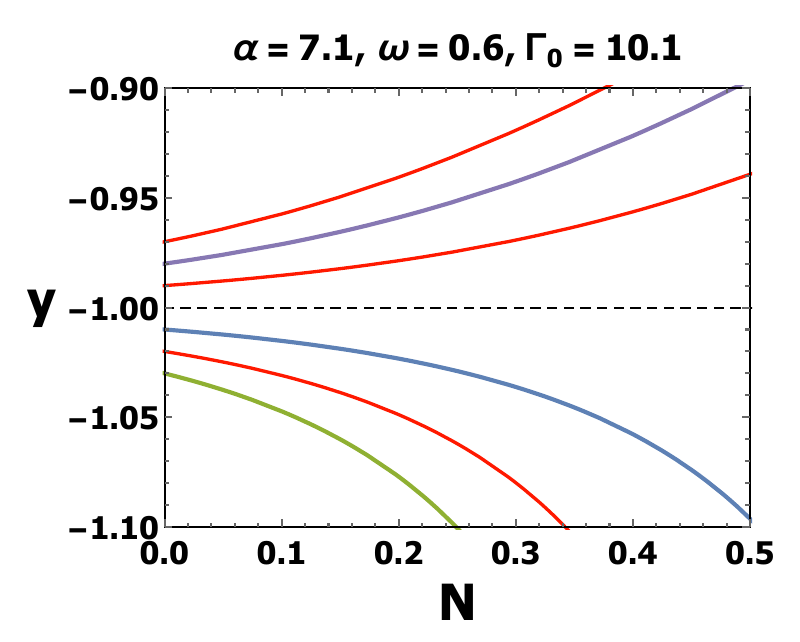}\label{UnStable_Ny}}
	\caption{ The figures show the phase space projection of perturbation of the autonomous system (\ref{autonomous1}) along the x and y axes for the critical point $P_{6}$ for parameter values $\alpha=7.1$, $\omega=0.6$ and $\Gamma_{0}=10.1$. In panel \ref{UnStable_Nx} and \ref{UnStable_Ny} perturbations do not come back. This indicates that the critical point $P_{6}$ is unstable for parameter values $\alpha=7.1$, $\omega=0.6$ and $\Gamma_{0}=10.1$.}
	\label{P6earlydS}
\end{figure}


\end{itemize}

\subsection{Classical stability of the model :}\label{Classical stability}
The local stability of the critical points extracted from the autonomous system is characterized by the eigenvalues of perturbation matrix (presented in table) and the stability criteria are discussed in detail in the previous section. Now, we shall discuss the classical stability of the model. In cosmological perturbation, the sound speed ($C_{s}^{2}$), appearing as coefficient of the term  $\frac{k^{2}}{a(t)^{2}}$ (where $k$ is the co-moving momentum and $a(t)$ is the usual scale factor), plays a vital role to determine the classical stability of the model. 
The sound speed can be defined as \cite{Piazza,Mahata,S.Kr.Biswas2015a,S.Kr.Biswas2015b,Garriga,Sudipta Das,Huang2021,Mandal2022,Bhardwaj2024}
\begin{equation}\label{perturbation}
C_{s}^{2}=\frac{\partial p_{d}  }{\partial \rho_{d}}
\end{equation}
The classical fluctuations of the model is considered to be stable only when the sound speed $C_{s}^{2}$ is positive i.e., $C_{s}^{2}\geq 0$. On the other hand, it violates the causality when $C_{s}^{2}>1$, {\it i.e.}, when the speed of sound diverges. Moreover, ghost instabilities can be realized for $C_{s}^{2}<0$. It is to be noted that the for stability under perturbation, it is needed to satisfy the condition: $0\leq C_{s}^{2}\leq 1$. By applying this method, stability of the model is investigated in many works. For example, Ref. \cite{Huang2021} studied the stability by finding the adiabatic sound speed in Barrow holographic dark energy. In a recent paper \cite{Bhardwaj2024}, the authors have investigated the stability of the model of generalised chaplygin gas in the framework of particle creation mechanism. They have found the speed of sound then have studied the stability of the model in a similar way. In the present work, we shall investigate the classical stability of the model, avoiding the ghost instabilities and we follow the works \cite{Mahata,S.Kr.Biswas2015a,S.Kr.Biswas2015b,Mandal2022} to study the classical stability of the model.

 For that we shall find $\frac{\partial p_{d}  }{\partial \rho_{d}} $ (described in Eqn. (\ref{perturbation})) for the present cosmological model explicitly in terms of dynamical variables and free parameters. 
With the help of  Eqn.(\ref{EOS PF}), we obtain the following
\begin{equation}
C_{s}^{2}=\frac{\partial p_{d}  }{\partial \rho_{d}}=\frac{\frac{\rho_{d}^{2}}{|A|}-\frac{1}{|\omega|}}{\left(\frac{\rho_{d}^{2}}{|A|}+\frac{1}{|\omega|} \right)^{2} }
\end{equation}
After that for simplifying, we use Eqn.(\ref{variables PF}) in the above equation which will give the following:  
\begin{equation}
C_{s}^{2}=\frac{\frac{9x^{2}H^{4}}{|A|}-\frac{1}{|\omega|}}{\left(\frac{9x^{2}H^{4}}{|A|}+\frac{1}{|\omega|} \right)^{2} }
\end{equation}
and finally, from (\ref{expression H PF}) and (\ref{variables PF}) we obtained the expression of sound speed in terms of dynamical variables as follows:
\begin{equation}
C_{s}^{2}=-\left(\frac{y}{x}+\frac{2y^{2}}{|\omega|x^{2}} \right).
\end{equation}
Therefore, one can easily obtain the condition for classical stability from the above as:
\begin{equation}\label{classical_inequality}
-\left(\frac{y}{x}+\frac{2y^{2}}{|\omega|x^{2}} \right) \geq 0.
\end{equation}
The inequality (\ref{classical_inequality}) shows the classical stability criteria for the present model. We shall now discuss the classical stability of the model at each critical point where $x$ and $y$ are the co-ordinates of the corresponding critical point. We observe that
\begin{itemize}
\item
The critical point $P_{1}$ corresponds to classical stability for all free parameters $\alpha$, $ \omega$ and $ \Gamma_{0}$. \
\item
The critical point $P_{2}$ is stable classically for either $ \omega \leq -2 ~~\mbox{or}~~ \omega \geq 2 $.\
\item
The critical point $P_{3}$ corresponds to classically stable for $\omega =0$.\
\item
The critical point $P_{4}$ shows classical stability for all parameters $\alpha$, $ \omega$ and $ \Gamma_{0}$. \
\item
Condition for classical stability of the critical point $P_{5}$ is $\omega =0$.\

\item
Finally, the set of points $P_{6}$ is stable classically for the following parameter restrictions:
\begin{enumerate}
	
	\item  $\Gamma_{0}=\alpha+3~~\mbox{and}~~(0<x_{c}<1) :~~\\
	(i)~~\left\lbrace \omega <0~~\mbox{and}~~\frac{3 x_{c}(\omega+2)}{(1-x_{c})2}\leq \alpha \leq \frac{3 x_{c}}{1-x_{c}}\right\rbrace ,~~\mbox{or}~~\\
	(ii)~~ \left\lbrace \omega >0~~\mbox{and}~~ \frac{3 x_{c}(2-\omega)}{(1-x_{c})2}\leq \alpha \leq \frac{3 x_{c}}{1-x_{c}}\right\rbrace  $
	\item  $\Gamma_{0}=\alpha+3~~\mbox{and}~~\left\lbrace x_{c}=1~~\mbox{and}~~ (\omega \leq -2~~\mbox{or}~~ \omega \geq 2)\right\rbrace$.
\end{enumerate}

\end{itemize}
From the above analysis of classical stability of the model and the local stability of the critical points, the points can be classified into different categories such as: there may be unstable critical points at which the model is stable. On the other hand, there may exist some points which are locally stable critical points at which the model may unstable or stable.
\begin{itemize}
\item
The critical point $P_{1}$ is stable classically for all $\alpha$, $ \omega$ and $ \Gamma_{0}$, but not stable locally for any value of $\alpha$, $ \omega$ and $ \Gamma_{0}$. \
\item
The critical point $P_{2}$ is stable locally as well as classically for :\\ 
$\left\lbrace (\omega \leq -2~~\mbox{and}~~ \Gamma_{0} <\alpha +3)~~\mbox{or}~~ (\omega \geq 2~~\mbox{and}~~ \Gamma_{0} <\alpha +3)\right\rbrace $.\
\item
The critical point $P_{3}$ is stable locally as well as classically for
$(\omega =0~~\mbox{and}~~ \Gamma_{0} <\alpha)$.\
\item
The critical point $P_{4}$ is stable locally as well as classically for
$(\alpha \geq 0~~\mbox{and}~~ \Gamma_{0} >\alpha +3)$.\
\item
The point $P_{5}$ is stable locally as well as classically when the following conditions are satisfied:\\
(i)~$(\alpha =0~~\mbox{and}~~ \omega =0~~\mbox{and}~~ 0<\Gamma_{0} <3)$, ~or\\
(ii)~$ (\alpha >0~~\mbox{and}~~ \omega =0~~\mbox{and}~~ \alpha <\Gamma_{0} <\alpha +3)$.\
\item
Finally, the set of points $P_{6}$ is locally as well as  classically stable for the following parameter restrictions:
\begin{enumerate}
	
	\item  $\Gamma_{0}=\alpha+3~~\mbox{and}~~(0<x_{c}<1) :~~\\
	(i)~~\left\lbrace \omega <0~~\mbox{and}~~\frac{3 x_{c}(\omega+2)}{(1-x_{c})2}\leq \alpha \leq \frac{3 x_{c}}{1-x_{c}}\right\rbrace ,~~\mbox{or}~~\\
	(ii)~~ \left\lbrace \omega >0~~\mbox{and}~~ \frac{3 x_{c}(2-\omega)}{(1-x_{c})2}\leq \alpha \leq \frac{3 x_{c}}{1-x_{c}}\right\rbrace  $
	\item  $\Gamma_{0}=\alpha+3~~\mbox{and}~~\left\lbrace x_{c}=1~~\mbox{and}~~ (\omega \leq -2~~\mbox{or}~~ \omega \geq 2)\right\rbrace$.
\end{enumerate}

\end{itemize} 


\section{Cosmological interpretation of the critical points}\label{Cosmological implications}

We shall now discuss the cosmology behind the critical points arising from the autonomous system (\ref{autonomous1}) in the previous section. The local stability of the critical points are presented in the section \ref{phase plane analysis} where a detailed phase plane analysis is performed. The main findings  from the analysis are as follows:\\
First of all, the autonomous system admits three Umami Chaplygin fluid (DE) dominated solutions, namely the critical points $P_1$, $P_2$ and $P_3$ in the phase plane.
The critical point $P_1$ corresponds to Umami Chaplygin fluid (DE) dominated decelerated universe in dust era ($q=\frac{1}{2},~\omega_{eff}=0$) where the Umami Chaplygin gas behaves as dust ($\gamma_{d}$=0). From the stability analysis, it is observed that the parameter restriction $\Gamma_{0}>\alpha$ describes the early evolution (because the solution is source here) corresponding to dust dominated decelerating phase of the universe. On the other hand, for $\Gamma_{0}<\alpha$, being a saddle-like solution in phase plane, the critical point $P_1$ corresponds to a dust dominated decelerating universe representing intermediate phase of evolution. This is the only physically relevant solution represented by $P_1$. 
The de Sitter expansion of the universe is realized by the critical point $P_2$ which may represent early as well as late phase of the universe. If we constrain the parameters as $|\omega|<1,~(\Gamma_{0}-\alpha)>3$, it is observed that the critical point describes a past attractor in phase plane and the point corresponds to early accelerated de Sitter expansion of the universe  ($\Omega_{d}=1,~\Omega_{m}=0,~\omega_{eff}=-1,~q=-1$), whereas the parameter restriction $|\omega|>1,~(\Gamma_{0}-\alpha)<3$ indicates that the point is a late-time de Sitter solution describing the accelerated expansion of the universe and this acceleration is governed by the Umami Chaplygin gas ($\gamma_{d}=-1$).  These phenomena are also confirmed numerically which are presented in figure (\ref{P2deSitter}). In sub-fig. \ref{P2earlydS}, it is shown for $\alpha=7.7,~\omega=0.6,~\Gamma_{0}=11$ that the universe's evolution starting from early accelerated de Sitter solution ($P_2$) to late-time accelerated scaling attractor (represented by $P_4$). Another sub-fig. \ref{P2latedS} for the parameter values $\alpha=0.1,~\omega=1.2,~\Gamma_{0}=0.7$ shows  that the late-time de Sitter phase (represented by $P_2$) is the ultimate fate of the universe. In this figure, it is also shown that the point $P_4$ is the matter-dominated saddle solution representing the intermediate phase of universe connecting through early (source) Umami Chaplygin fluid dominated decelerated phase of the universe (represented by the point $P_1$) where the Umami Chaplygin fluid mimics as dust ($\gamma_{d}=0$). Evolution of the cosmological parameters are plotted in the parameters region: $|\omega|>1,~(\Gamma_{0}-\alpha)<3$ for different model parameters values in fig.(\ref{evolutionLCDM}) for different initial conditions. In sub-figure \ref{evolutionLCDM1}, the evolution of cosmological parameters ($\Omega_{m},~\Omega_{d},\gamma_{d},~q,~\omega_{eff}$) for the values of free parameters: $\alpha=0.006,~~\omega=12,~~\Gamma_{0}=0.034$ with the initial conditions $x(0)=0.6995,~y(0)=-0.6976$ exhibit that the trajectories are attracted by cosmological constant at late-times connecting through a matter-dominated decelerated phase. On the other hand, the sub-fig \ref{evolutionLCDM2} shows the cosmological evolution of $\Omega_{m},~\Omega_{d},\gamma_{d},~q,~\omega_{eff}$ for other choices of model parameters: $\alpha=0.1,~~\omega=1.2,~~\Gamma_{0}=0.7$ and with the initial conditions: $x(0)=0.698994,~y(0)=-0.62304$. This sub-fig also shows that the ultimate evolution of universe is attracted in $\Lambda$CDM. \\
Another Umami Chaplygin fluid dominated universe is represented by the critical point $P_3$ which will be able to show different cosmic phases for different choices of model parameters. However, a physically relevant solution is achieved only when it evolves in quintessence era at late-times where the DE behaves as quintessence fluid. This is discussed in detail in the previous section \ref{phase plane analysis}.\\

\begin{figure}
\centering
\subfigure[]{%
	\includegraphics[width=7.2cm,height=7.2cm]{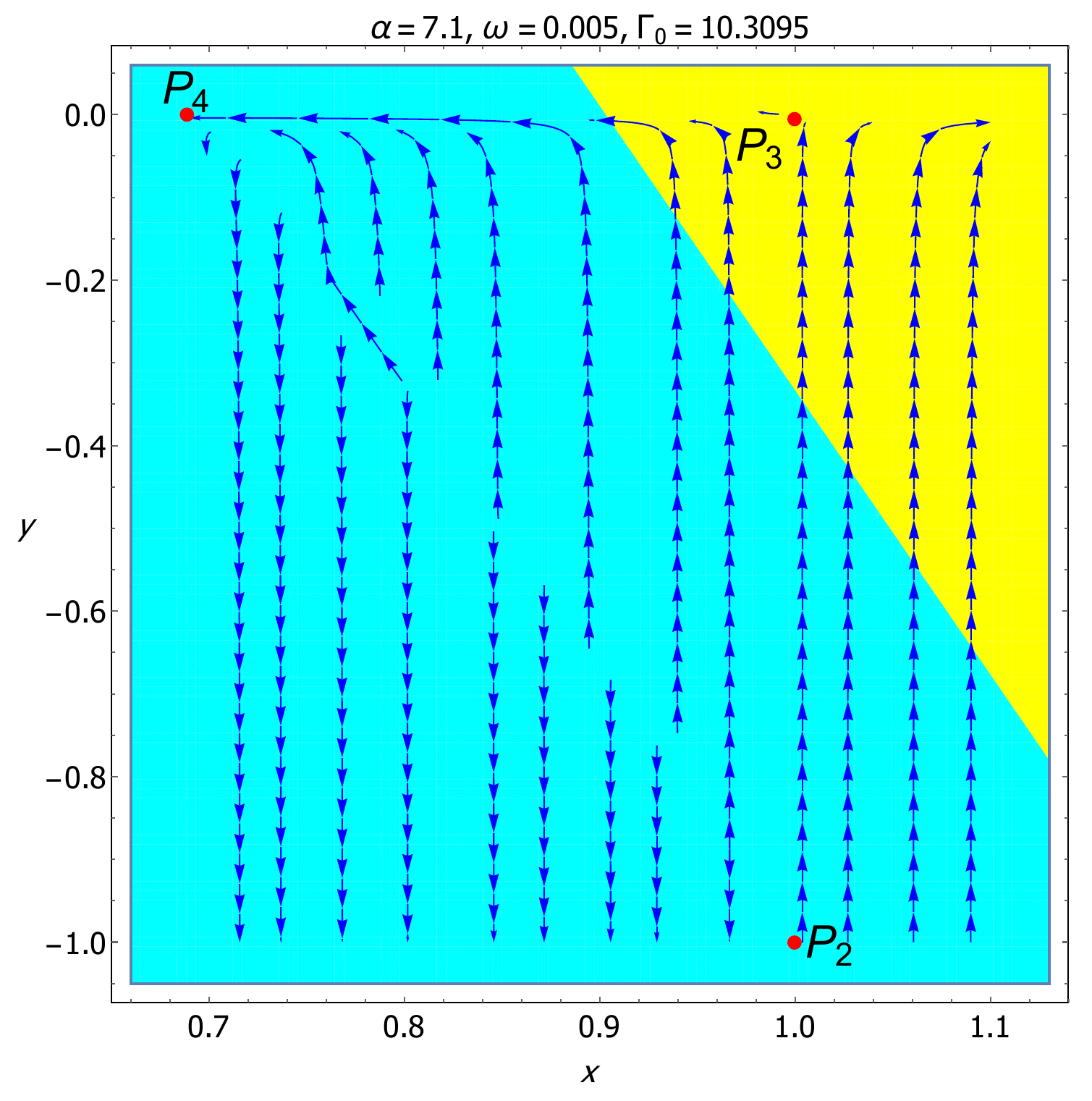}\label{Complete Stream}}
\qquad
\subfigure[]{%
	\includegraphics[width=7.2cm,height=7.2cm]{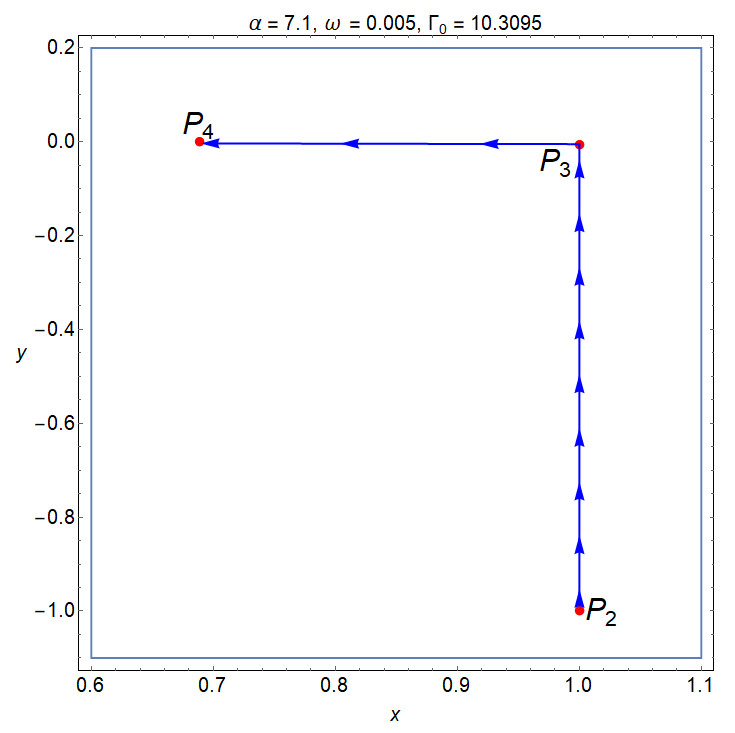}\label{Complete Trajectory}}
\qquad
\subfigure[]{%
	\includegraphics[width=7.2cm,height=7.2cm]{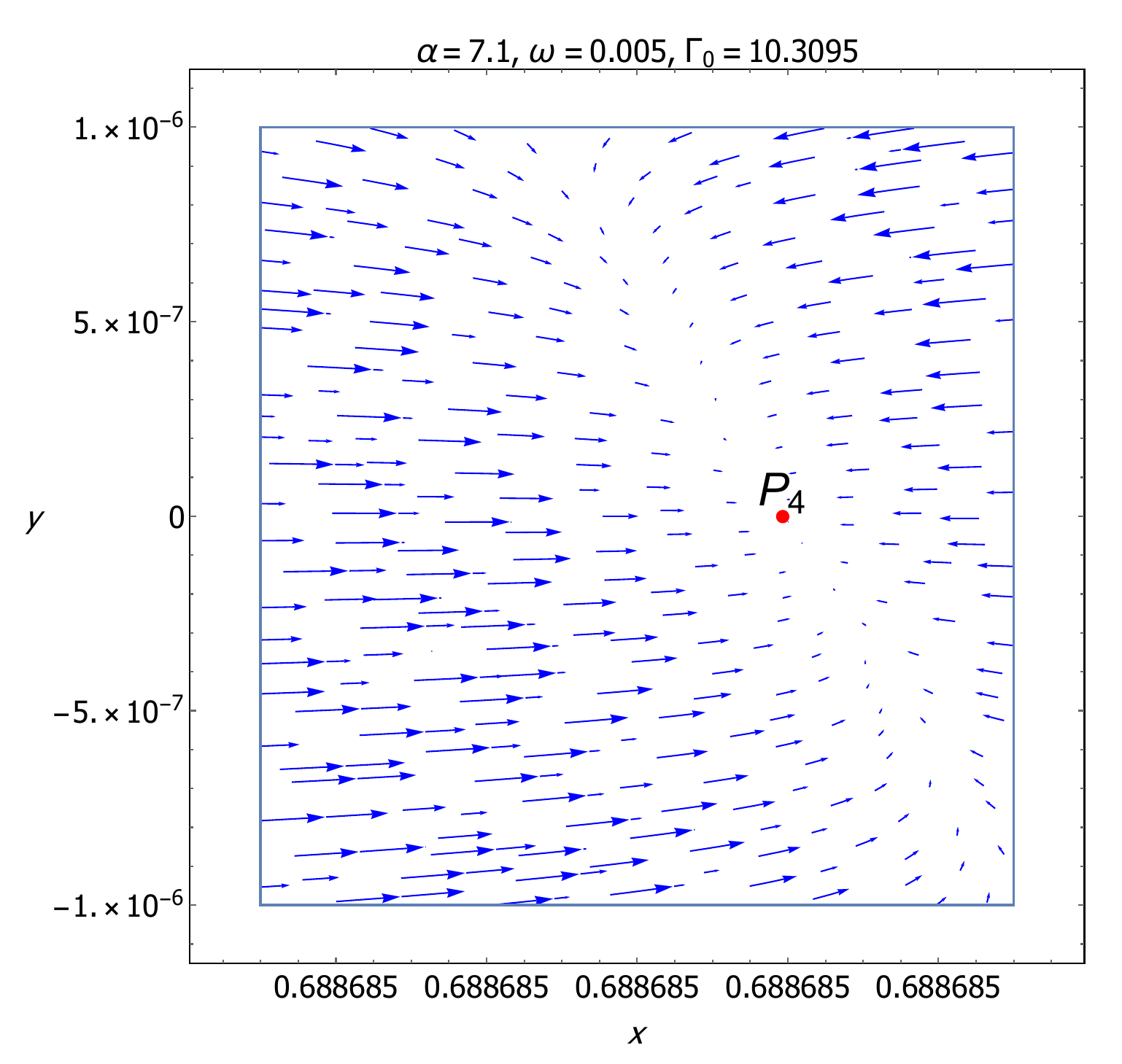}\label{P4stable}}
\qquad
\subfigure[]{%
	\includegraphics[width=7.2cm,height=7.2cm]{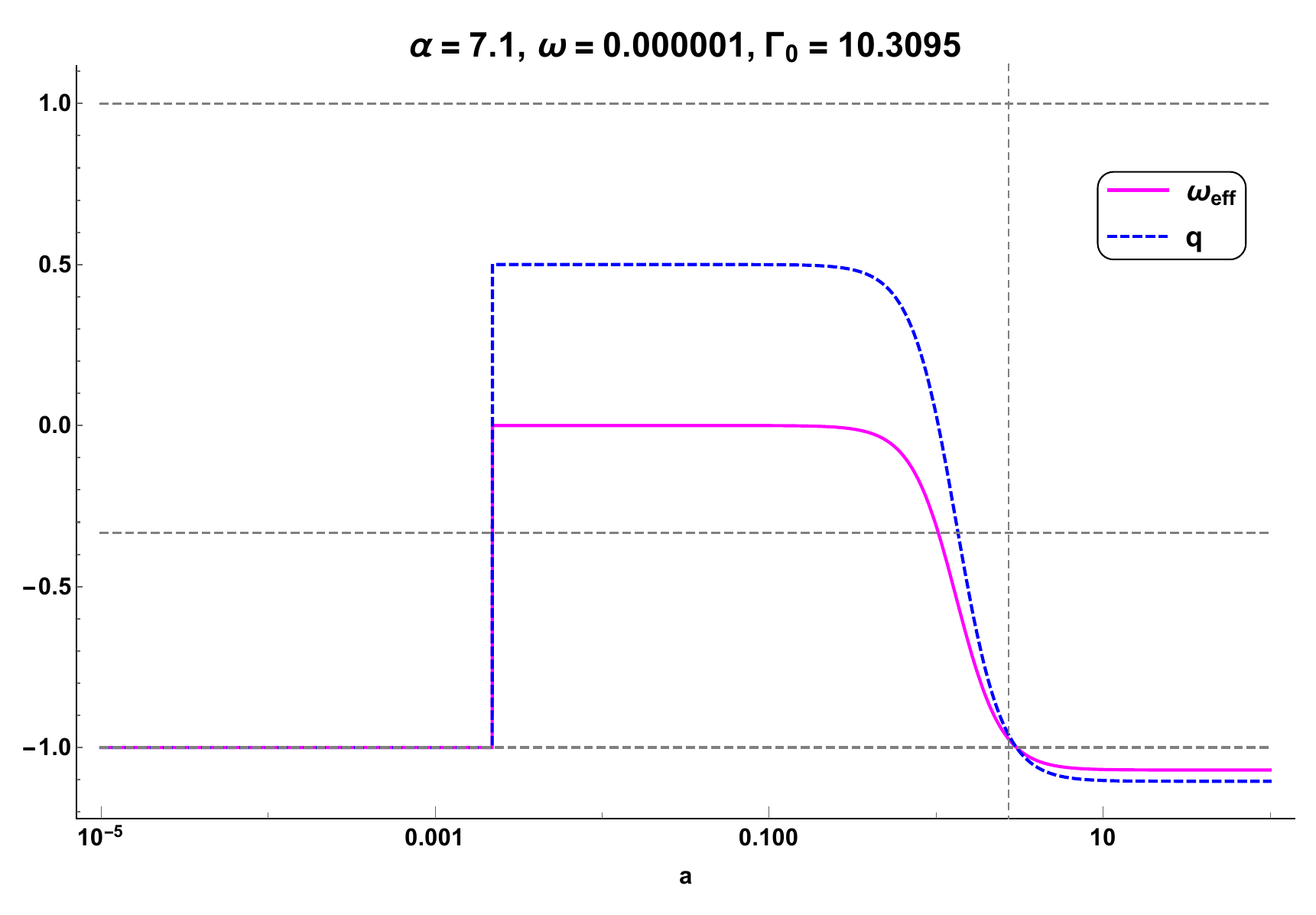}\label{Complete evolution}}
\caption{The sub-figures \ref{Complete Stream}, \ref{Complete Trajectory}  and \ref{P4stable} display the stream plot of the autonomous system (\ref{autonomous1}) on the x-y plane for the parameter values: $\alpha=7.1,~~\omega=0.005,~~\Gamma_{0}=10.3095$. The sub-figure \ref{Complete Stream} shows the point $P_2$ is the early {\it de Sitter} solution, point $P_3$ decelerated saddle-like solution and point $P_4$ gives scaling solution which is a stable attractor. Here, the cyan shaded region represents the accelerated region (i.e. $q < 0$) and the yellow shaded region corresponds to the decelerated region (i.e. $q > 0$). 
	The sub-figure \ref{Complete Trajectory} shows  a sequence of critical points showing complete evolution of the universe where sub-figure \ref{P4stable} exhibits the late-time attractor $P_{4}$ with $(\Omega_{m}=0.311, \Omega_{d}=0.688, \gamma_{d}=0, \omega_{eff}=-1.069, q=-1.105)$.
	Sub-figure \ref{Complete evolution} for $\alpha=7.1,~~\omega=0.000001,~~\Gamma_{0}=10.3095$ exhibits the evolution of decelerating parameter $q$ and effective equation of state parameter $\omega_{eff}$ with initial conditions: $x(0)=0.9099999$ and $y(0)=-0.0000009099999$. This shows a unified description of universe from early inflation to late time acceleration connecting through intermediate decelerated phase.}
\label{coupled complete evolution}
\end{figure}

Next, the autonomous system (\ref{autonomous1}) extracts three scaling solutions namely, $P_4$, $P_5$ and $P_6$ in the phase plane and they exhibit some interesting scenarios which are important in cosmological context. The main findings of these points are following:
imposing some restrictions on parameters, the scaling solution represented by the point $P_4$ can describe the accelerated evolution of the universe attracted in phantom regime at late times. In particular, for parameter values: $\alpha=7.1,~\omega=0.005,~\Gamma_{0}=10.3095$ the sub-figs. \ref{Complete Stream}, \ref{Complete Trajectory}, and \ref{P4stable} exhibit that the scaling solution is stable attractor in phase plane satisfying $0<\Omega_{d}<1$ which indicates that this solution can alleviate the coincidence problem. 
The corresponding physical parameters for this point are: ($\Omega_{m}\approx 0.3,~\Omega_{d}\approx 0.7,~ \omega_{eff}\approx -1.07,~q\approx -1.10$). It is interesting to note that the sub-fig. \ref{P2earlydS} for model parameter values: $\alpha=7.7,~\omega=0.6,~\Gamma_{0}=11$ also refers to the scaling solution $P_4$ which corresponds to accelerated attractor in phantom era at late-times with associated cosmological parameters ($\Omega_{m}= 0.3,~\Omega_{d}= 0.7,~ \omega_{eff}= -1.1,~q= -1.15$). The numerical values of the cosmological parameters support the present observational data which have also been obtained in Ref.\cite{Nunes2015} where the authors studied the model $\Gamma=3\beta H$. It should be mentioned that for both the cases associated DE represented by the Umami Chaplygin fluid  behaves as dust ($\gamma_{d}=0$).\\

It is worthy to mention that the model also support the DESI (Dark Energy Spectroscopic Instrument) DR1 (Data Release 1) \cite{Adame2024} for some restrictions of model parameters. The cosmological parameters result according to DESI: $\Omega_{m}=0.293 \pm 0.015,~ \gamma_{d}=-0.99_{-0.13}^{+0.15}$. It is concluded that for fine tuning of model parameters, the present values of the cosmological parameters can be obtained. Another study in Ref.\cite{Roy2024} suggests that the current EoS of DE is phantom $\gamma_{d}=-1.113_{-0.21}^{+0.22}$. For values of model parameters: $\alpha=0.05,~\omega=0.99,~\Gamma_{0}=3.04072$ the critical point $P_5$ evolves at late-times (since eigenvalues for this case are $\lambda_1=-0.01856,~\lambda_2=-0.02072.$) with corresponding cosmological parameters: ($\Omega_{m}= 0.292986,~\Omega_{d}= 0.707014,~\gamma_{d}=-0.99 ,~ \omega_{eff}=-0.996907,~q=-0.99536$). Also, the model parameter values: $\alpha=-0.791,~\omega=1.2,~\Gamma_{0}=2.209,~x_c=0.7$ exhibit that the critical point $P_6$ supports the observational data at late-times with cosmological parameters: ($\Omega_{m}= 0.3,~\Omega_{d}= 0.7,~\gamma_{d}=-1.113 ,~ \omega_{eff}=-1,~q=-1$).

Another interesting scaling solution represented by critical point $P_5$ can provide the features of late-time cosmological dynamics of the universe. For some parameter restrictions, the solution corresponds to a late-time accelerated scaling attractor in quintessence era satisfying $0<\Omega_{d}<1$ which can give the possible mechanism to alleviate the coincidence problem.
The fig. (\ref{P5}) with the model parameters: $\alpha=0.2,~\omega=0.661,~\Gamma_{0}=2.275$ indicates that the solution $P_5$ describes late time evolution of the universe in agreement with the observation. In this case the numerical values of the cosmological parameters are: ($\Omega_{m}\approx0.31,~\Omega_{d}\approx0.69,~\omega_{eff}\approx-0.692,~q\approx-0.537$ and $\gamma_{d}=-0.661$.) This phenomenon is well agreement with the observational data \cite{Planck,Harko2021}. Note that the Umami Chaplygin gas (as the DE candidate)  for this case mimics as quintessence ($\gamma_{d}=-0.661$).
\\

Finally, we observe that the scaling solution described by the set of critical points $P_6$ can mimic the $\Lambda$CDM model of universe. Fig. (\ref{P6stable}) with parameter values: $\alpha=7.1,~\omega=1.1,~\Gamma_{0}=10.1$ shows that the set is stable solution in phase plane with $\omega_{eff}=-1,~q=-1$.\\

For a specific choice of $x_c=1$, the set represents early time (for $|\omega|<1$) as well as the late phase (for $|\omega|>1$) of de Sitter solution. This is shown by numerical investigation. For $\alpha=7.1,~\omega=1.1,~\Gamma_{0}=10.1$, the set $P_6$ will  become a point with coordinate ($1,~-1$) in the phase plane. Here, the fig. (\ref{P6latedS}) describes the accelerating late-time de Sitter universe (as in sub-fig. \ref{Stable_Nx}, the trajectories approach in $x=1$ line, as well as in sub-fig.\ref{Stable_Ny}, the perturbations coming back to $y=-1$ line). On the other hand, early accelerated de Sitter expansion of the universe is realized in fig. (\ref{P6earlydS}) for $\alpha=7.1,~\omega=0.6,~\Gamma_{0}=10.1$ (as in sub-fig. \ref{UnStable_Nx} refers to unstable along $x=1$ line and sub-fig. \ref{UnStable_Ny} corresponds to instability in $y=-1$ line).\\

\begin{figure}
\centering
\includegraphics[width=15cm,height=15cm]{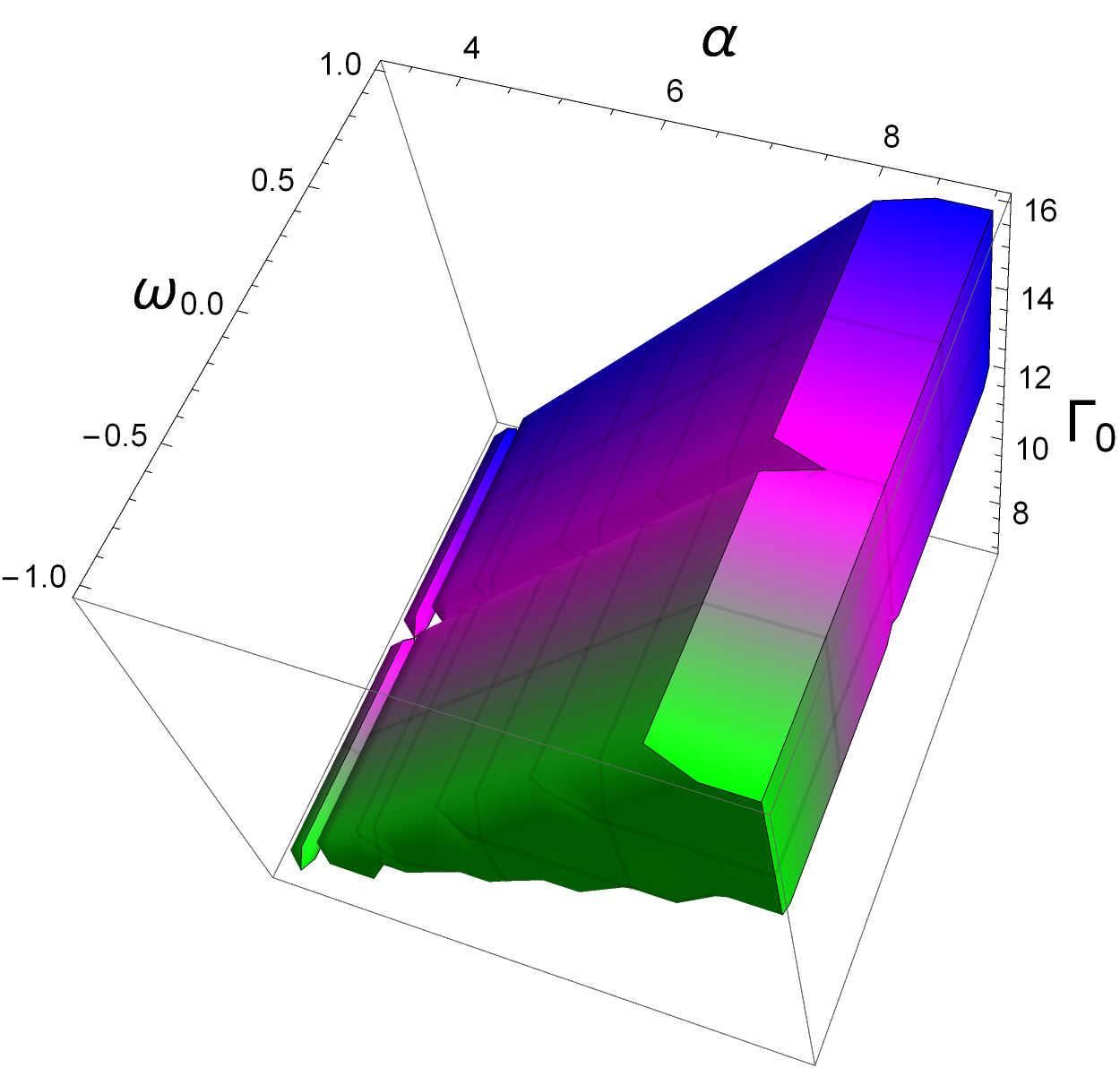}

\caption{ The figure shows the region of parametric space for complete scenario of the model parameters $\alpha$, $\omega$ and $\Gamma_{0}$. 
}
\label{pr}
\end{figure}

In summary, it is worth mentioning that a region of parameter space ($\alpha,~ \Gamma_{0},~ \omega$) can be obtained (see the plot in fig.(\ref{pr})) in which some critical points exhibit a complete description of evolutionary scheme of universe. For example, we obtained the restrictions: $\omega \in (-1,~1),~\omega\neq0$ and $3+\alpha<\Gamma_{0}<2\alpha$, where $\alpha> 3$ such that the critical point $P_2$ corresponds to Umami Chaplygin fluid dominated early accelerated de Sitter solution representing the early inflation of universe, where the Umami Chaplygin gas behaves as cosmological constant like fluid. Next, the critical point $P_3$ describes the decelerated intermediate phase of the universe. In this case, the point $P_3$ is actually Umami Chaplygin (DE) dominated solution, however, the universe appears here as if it is DM dominated solution  ($\omega_{eff}\longrightarrow 0,$ as $|\omega|\longrightarrow0$), the matter mimics the perfect fluid nature here and the Umami Chaplygin fluid behaves as perfect fluid. Finally, a late-time scaling attractor is obtained by the point $P_4$ which is in a good agreement with the observational data. The 
sub-figs. \ref{Complete Stream}, and \ref{Complete Trajectory} display a sequence of critical points:\\

$P_2$ (source) $\longrightarrow$ $P_3$ (saddle) $\longrightarrow$  $P_4$ (stable)\\

which corresponds to a complete description of evolutionary scheme starting from early inflation (represented by $P_2$) to late-time acceleration of the universe (described by $P_4$) connecting through a sufficient amount of matter dominated decelerated phase (represented by $P_3$).
The evolution plot of effective equation of state $\omega_{eff}$ and the deceleration parameter $q$ also confirm the above fact. The fig.\ref{Complete evolution} for $\alpha=7.1,~\omega=0.000001,~\Gamma_{0}=10.3095$ exhibits that the universe's evolution  starting from $\omega_{eff}=q=-1$ (represented by the point $P_2$ with $\Omega_{m}=0,~\Omega_{d}=1,~\gamma_{d}=-1$, i.e, the early de Sitter universe) will go through the phase where $\omega_{eff}\approx0,~q\approx0.5$ (described by the point $P_3$ with $\Omega_{m}=0,~\Omega_{d}=1,~\gamma_{d}\approx0$, i.e., the Umami Chaplygin gas mimicking the dust here) and finally will approach to the state where $\omega_{eff}\approx-1.07,~q\approx-1.05$ (realized by the point $P_4$ with $\Omega_{m}\approx0.3,~\Omega_{d}\approx 0.7$ and $\gamma_{d}=0$.) It is also observed from the plot \ref{Complete evolution} that there is a possibility of crossing the phantom barrier which is also favored by present observation. 
It should also be noted that there is a sudden change in the evolution of $\omega_{eff}$ and $q$ from  $-1$ to $0$ and from $-1$ to $\frac{1}{2}$ respectively at the same time. This refers to the possibility of bouncing universe and this needs further investigation which we shall discuss in the next section. To realize the early evolution of the universe, we perform a numerical investigation of this model in the next section.


\clearpage

\section{Bouncing Scenario}\label{Bouncing scenario}
In this section we shall study the bouncing scenario of interacting Umami Chaplygin fluid with pressureless dust via a phenomenological interaction term in the framework of particle creation mechanism. 
The study of bounce cosmology can be an alternative to big-bang cosmology, i.e., initial singularity. It is well known fact that inflation scenario can resolve many cosmological issues like `horizon problem', `flatness problem' etc.   within the standard big-bang cosmology. But, it faces some phenomenological conceptual obstacles. The singularity problem is one of them, it cannot be addressed by inflationary scenario. In fact, the inflationary scenario cannot address the entire history of early evolution of the universe. For this reason, a successful theory is needed to account related to the central issues of big-bang cosmology. Bouncing cosmology is one in which the fundamental issues in cosmology like problems of big-bang singularities are addressed. In order to address the initial singularity, a huge no. of investigations have been carried out in different gravity theories \cite{Carroll2004 and others}. In recent work, Chakraborty and Chakraborty in Ref. \cite{Chakraborty2023} have studied different kind of bounce in respect of Raychaudhuri's equation. For a detail study of bounce scenario in different modified gravity theories in different aspects, one may refer to the PhD thesis in Ref. \cite{Agarwal2024} and references there in. In context of dynamical systems analysis, bounce scenario is also studied for some scalar field models (see in Ref. \cite{Santiago2013}). A recent paper in Ref.\cite{Agarwal2024a} investigated a global phase space analysis for scalar fields bouncing solutions. Interested readers may also follow the Ref. \cite{Gadbail2023} in this regard.

\begin{figure}
\centering
\subfigure[]{%
	\includegraphics[width=7.2cm,height=7.2cm]{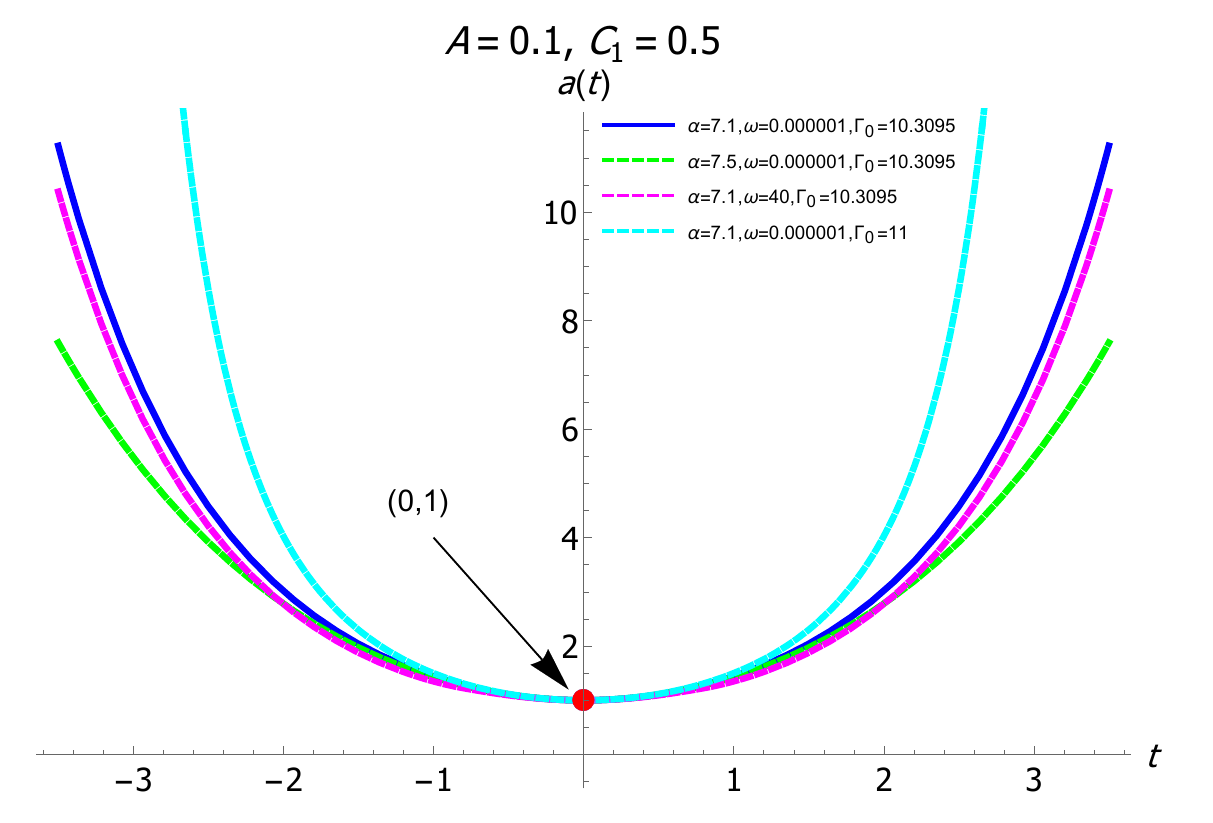}\label{BU1}}
\qquad
\subfigure[]{%
	\includegraphics[width=7.2cm,height=7.2cm]{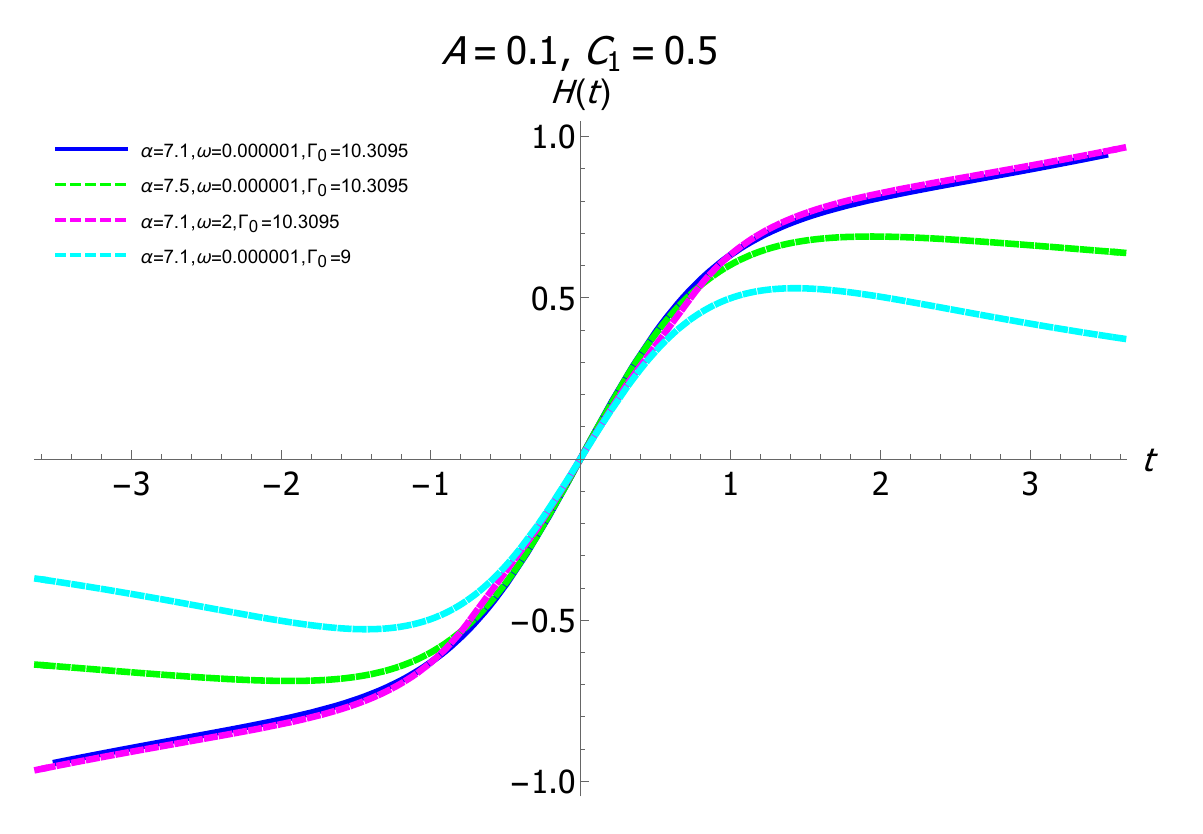}\label{BU2}}

\caption{The figure shows the bouncing behavior of evolution of scale factor $a(t)$ (left panel) and Hubble parameter $H(t)$ (right panel) with respect to the cosmic time $t$ for different parameters values. $C_1$ is integrating constant. In sub-figure \ref{BU1} the point with red dot refers to the bouncing point. }
\label{bouncing_scenario}
\end{figure}

\begin{figure}
	\centering
	\subfigure[]{%
		\includegraphics[width=8.5cm,height=7.2cm]{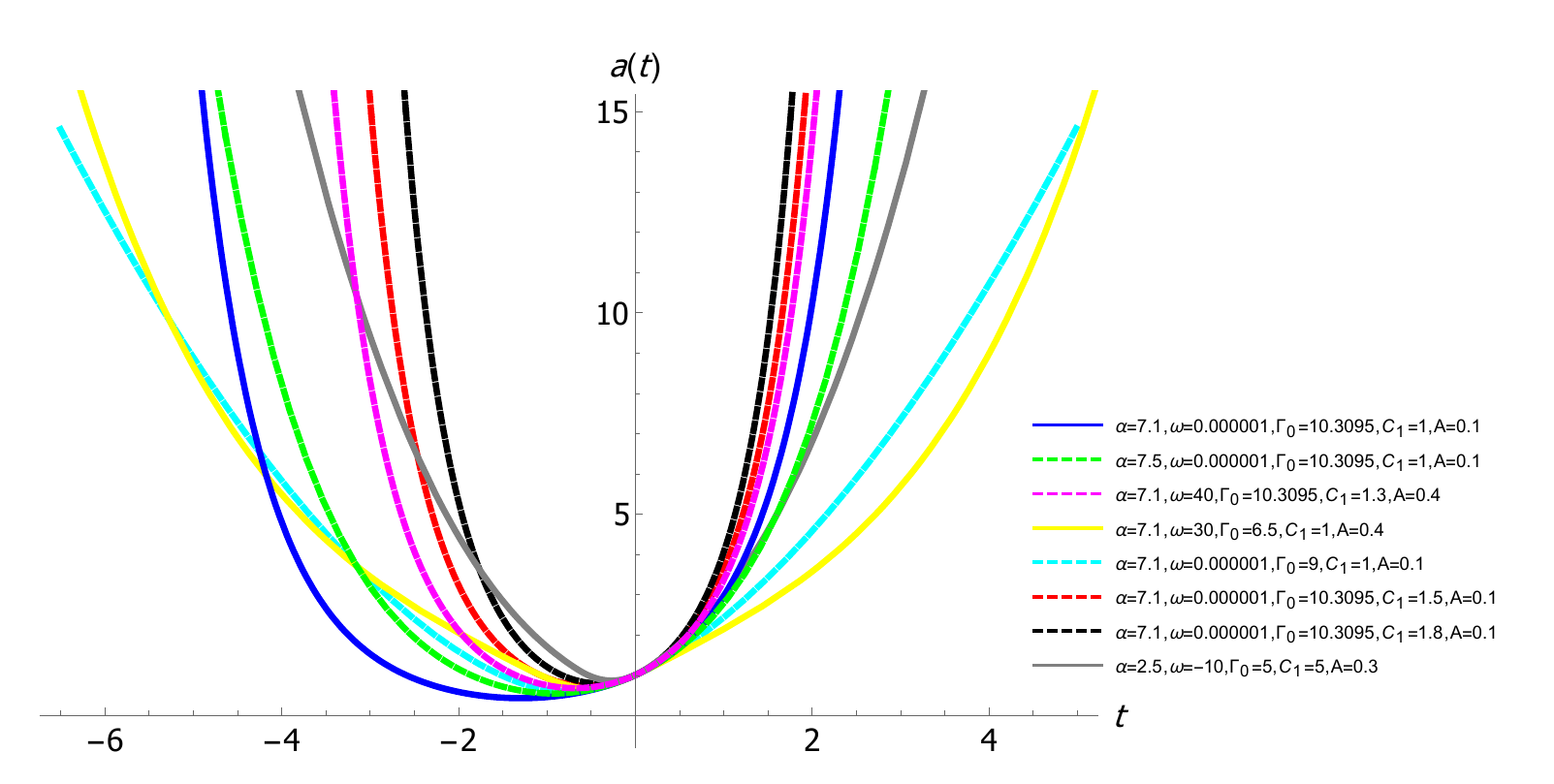}\label{BU3}}
	\qquad
	\subfigure[]{%
		\includegraphics[width=8.5cm,height=7.2cm]{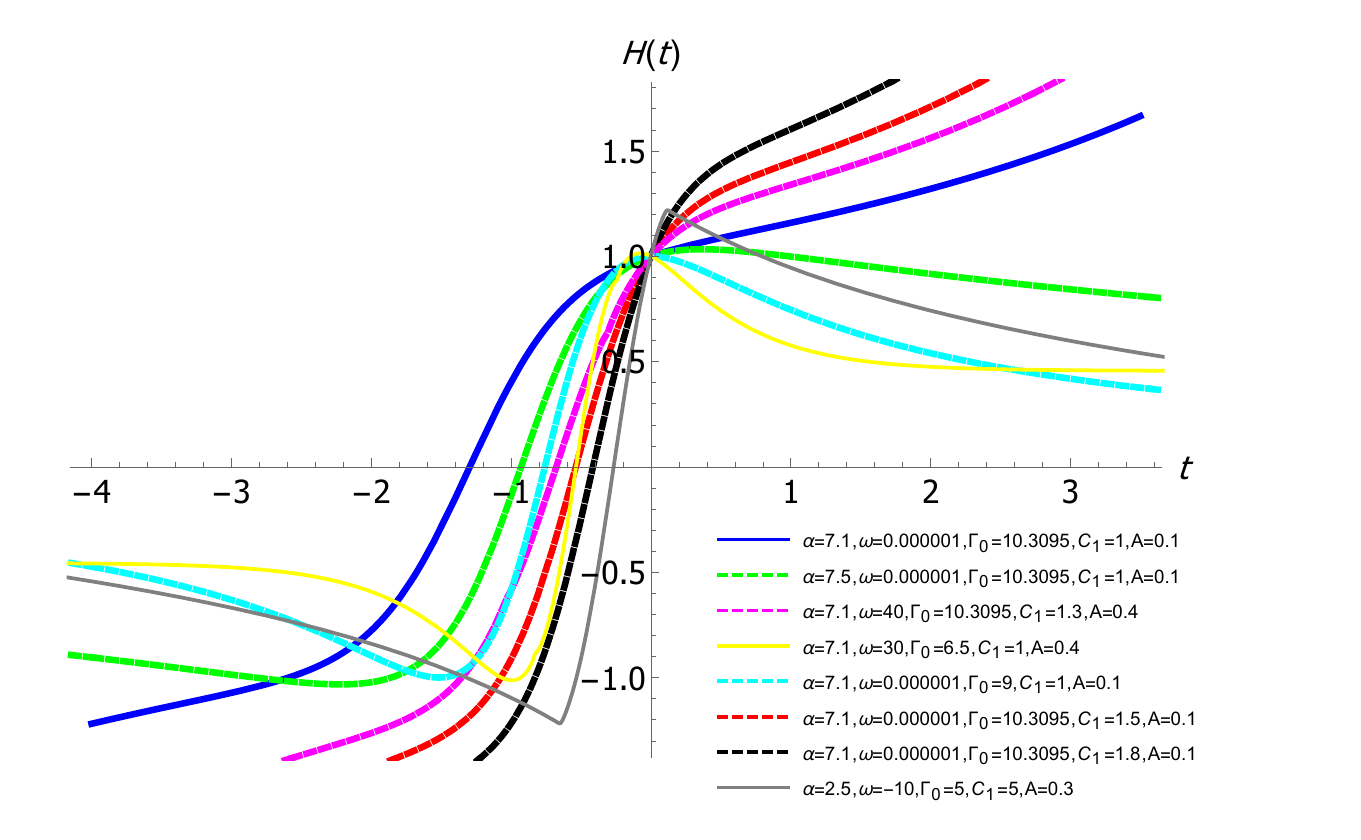}\label{BU4}}
	
	\caption{The figure shows the bouncing behavior of evolution of scale factor $a(t)$ (left panel) and Hubble parameter $H(t)$ (right panel) with respect to the cosmic time $t$ for different parameters values.}
	\label{bouncing_scenario1}
\end{figure}

A cosmological model has no big-bang singularity initially at $t=0$ if it has the scale factor $a(t)\neq 0$ at $t=0$. Here, the universe contracts in the period $t \in (-\infty,~0]$ while it expands for $t\geq 0$. The following are important for the study of bouncing universe:
In non-singular bounce, first, the scale factor $a(t)$ decreases with time, i.e, contracting case ($\dot{a}(t)< 0$) which implies that $H=\frac{\dot{a}}{a}<0$.
Also, after some times the decreasing scale factor reaches to a critical value and finally the Hubble parameter becomes zero at the bounce point.
Now the scale factor increases with time t as a result of which the Hubble parameter becomes $H=\frac{\dot{a}}{a}>0$ after the bounce. In this situation the Hubble parameter $H$ will follow the path of transition from $H<0$ to $H>0$. So, in summary, one can obtain the bounce universe in avoidance of initial singularity in the context of General Relativity after fulfilling the following requirement:
$$a(t=0)\neq0,~~\dot{a}(t=0)=0,~~H(t=0)=0,~~\dot{H}(t=0)>0$$
Therefore, our main aim is to study the evolution of scale factor $a(t)$ and the Hubble parameter $H(t)$ for our model. In this model, particle production rate $\Gamma$ is chosen as $\Gamma=\Gamma_{0}H$ ($\Gamma_{0}$, a constant). This particular choice of production rate is assumed from the phenomenological ground and for mathematical simplicity. Actually, different choices of creation rates can describe different cosmic phases starting from early inflation to late-time cosmology. The rate can only be realized through quantum field theory and it is yet to be developed. Since there is no guiding principle in assuming the form of $\Gamma$, we can choose it as a linear function of Hubble parameter $H$ which can be able to describe evolution of the universe. The cosmological model with this type of creation rate has been studied extensively in the literature \cite{Pan2015,Chakraborty,Biswas2017,Nunes2016,Nunes2015,Nunes2016IJMPD,Bhardwaj2024}. The interaction term is taken as $Q=\alpha H \rho_{m}$, where the coupling parameter $\alpha$ is a dimensionless constant. Now, considering the above, we solve the Eqn.(\ref{DM1}) for $\rho_{m}$ as follows:
\begin{equation}\label{density DM}
\rho_{m}=C_{1} a^{-(\alpha-\Gamma_{0}+3)} ~~~~;~~C_{1}~~ is~~ an~~ integrating~~ constant.
\end{equation}
Using the above explicit solution of $\rho_{m}$ in Eqn.(\ref{density DM}), the Friedmann constraint (\ref{Friedmann}) gives the solution of $\rho_{d}$ as :
\begin{equation}\label{density DE}
\rho_{d}=	3 \left(\frac{\dot{a}}{a}\right)^2- C_{1} a^{-(\alpha -\Gamma_{0} +3)}
\end{equation}
and the Raychaudhuri Equation in (\ref{Raychaudhuri equation}) will immediately provide the solution for $p_{d}$ as
\begin{equation}\label{pressure DE}
p_{d}=\frac{\Gamma_{0}}{3} C_{1}  a^{-(\alpha -\Gamma_{0} +3)}-\left\lbrace \frac{2 \ddot{a}}{a}+\left(\frac{\dot{a}}{a}\right)^2\right\rbrace 
\end{equation}
After putting the explicit form of energy density (in Eqn. (\ref{density DE}))  and the pressure (in Eqn. (\ref{pressure DE})) of Umami Chaplygin fluid, the equation of state of the fluid given in Eqn. (\ref{pressure Umami}) will give the second order differential equation for the scale factor $a(t)$ as follows:
\begin{equation}\label{differential equation}
\frac{2 a \ddot{a}+\left(\dot{a}\right)^2+\frac{a^{\alpha +5} \left| A\right|  \left| \omega \right|  \left(C_{1} a^{\Gamma_{0} }-3 a^{\alpha +1} \left(\dot{a}\right)^2\right)}{a^{2 \alpha +6} \left| A\right| +\left| \omega \right|  \left(C_{1} a^{\Gamma_{0} }-3 a^{\alpha +1} \left(\dot{a}\right)^2\right)^2}}{a^2}-\frac{1}{3} C_{1} \Gamma_{0}  a^{-\alpha +\Gamma_{0} -3}=0
\end{equation}
where the over-dot stands for derivative with respect to cosmic time $t$. Due to complicated form of the non-linear differential equation (\ref{differential equation}), it is very hard to solve analytically. Therefore, we may proceed to carry out a method of numerical computation which may allow us to address such complicated and realistic model that cannot be studied analytically. For this purpose, we use numerical analysis based on Runge-Kutta fourth order method to solve by using MATHEMATICA software. By proper choice of initial condition, we shall solve the differential equation numerically for $a(t)$ and $H(t)$ showing in the fig. (\ref{bouncing_scenario}). From the figure we obtain a non-singular bounce scenario. This scenario consists of the following behavior of scale factor $a(t)$ and the Hubble parameter $H(t)$. A contracting phase is occurred (see in sub-fig. (\ref{BU1})) in pre bounce phase when $\dot{a}(t)<0$ and $H(t)=\frac{\dot a}{a}<0$ which is followed by a smoothly non-singular bounce point where $\dot{a}=0$, i.e., $H=0$ and the co-ordinate of the bounce point is $(a=1,~t=0)$ shown in the sub-fig.(\ref{BU1}). Finally, it goes to the post-bounce expanding phase where $\dot{a}>0$ (see in sub-fig. (\ref{BU1}))  and consequently, $H>0$ (see in sub-fig.(\ref{BU2})). Note that in this computational process, one has to adopt the values of $a(t)$ and $\dot{a}(t)$ at $t=0$ to solve the second order differential equation (\ref{differential equation}). We have used $a(t=0)=1$ and $\dot{a}(t=0)=0$ for plotting fig.(\ref{bouncing_scenario}). Here, we choose the model parameter: $A=0.1$ and integrating constant $C_1 =0.5$. Different trajectories are shown for values of model parameters inside and outside the region: [$\omega \in (-1,~1),~\omega\neq0$ and $3+\alpha<\Gamma_{0}<2\alpha$, where $\alpha> 3$]. Further, we have plotted another figure (\ref{bouncing_scenario1}) for $a(t=0)=1$ and $\dot{a}(t=0)=1$ and have shown trajectories for different values of model parameters and integrating constant $C_1$.

\clearpage


\section{Concluding remarks}\label{Conclusion}
We have investigated cosmological model of particle creation mechanism in the framework of homogeneous, isotropic and spatially flat FLRW universe for a specific form of matter creation rate. In this mechanism, the cosmological matter is assumed to be created irreversibly from the geometry of space-time. We have considered the universe as an open thermodynamical model of a specific volume $V$ containing $N$ number of particles. The dissipative effect in the open system is also considered, as a result of which the number of particles is not conserved in the system. In this formalism, the matter creation rate is included in energy-momentum tensor and reinterpreted the conservation laws. The created matter is assumed to be pressureless dark matter and a suitable choice of production rate ($\Gamma$) is considered as a linear function of Hubble parameter $\Gamma\propto H$ and the matter creation pressure is related to creation rate. 
The particle creation rate can be realized from quantum field theory and the subject is yet to be developed. Also, there is no guiding principle to consider the exact form of the creation rate. However, a particular creation rate has a definite impact in cosmological dynamics of a model. Several creation rates have been studied in context of FLRW cosmology in the literature. Here, in the present work, we considered $\Gamma=\Gamma_{0}H$, where $\Gamma_{0}$ is the dimensionless constant. Note that a non-zero $\Gamma_{0}$ indicates the deviations from standard cosmology and $\Gamma_{0}=0$ recovers the standard cosmology.

As in the literature, it is well established that different forms of the creation rates $\Gamma$ can predict different epochs of evolution. For instance, \\
$(a)~\Gamma \propto H^{2}$ provides the early evolution, $(b)~\Gamma \propto \frac{1}{H}$ predicts the late-time evolution, and $(c)~\Gamma \propto H$ can be able to provide the intermediate phase of the universe. Thus, a particular cosmological model may include either any one of the above froms or a general combination of all the rates. Although, for all the cases, dynamical systems analysis can be performed. However, in the present work, we undertake only $\Gamma \propto H$ to study the dynamical system analysis. Main advantage of considering this rate is mathematical simplicity.  All the rates except $\Gamma \propto H$ cannot provide the closed autonomous system due to the presence of $H$-term in the system of equations. In our study, the systems (\ref{autonomous Gen1 PF}) and (\ref{autonomous Gen Umami}) include $\Gamma$ (in the last term of both the systems) will produce a closed system only when $\Gamma$ takes the form $\Gamma=\Gamma_{0} H$. On the other hand, for all other choices, an extra $H(t)$-term will appear in systems (\ref{autonomous Gen1 PF}) and (\ref{autonomous Gen Umami}). Now, to close the system, one has to consider $H(N)$ as an additional independent dynamical variable for which the dimension of the autonomous system will be increased by one. So, the analysis will be more complicated. The cosmological model of matter creation process with different rates may be of future scope of research.  \\

Here, we approach in a systematic way.
In the first model, we start with the single fluid model where the universe is filled with pressureless dark mater only and the creation rate is $\Gamma=\Gamma_{0} H$. The model is solved analytically and it shows that the model does not describe the phase transition of the evolution of the universe. Therefore, the model does not support the present observations.

Next, in the second model, a perfect fluid (described in Eqn. (\ref{pressure PF})) with a non-linear equation of state $\gamma_{d}>0$, (i.e.,  for $B>0$) is included as an additional matter content of the universe. In context of particle creation, the two-fluid model will become complicated and can not be solved analytically. Thus, we employed dynamical systems tools to solve qualitatively. The non-interacting two-fluid model does not predict late-time acceleration of the universe. So, the model is not interesting from cosmological point of view. Further,  this interacting two-fluid model, without inclusion of matter creation ($\Gamma=0$), can not provide late-time acceleration. However, an interacting two-fluid model with $\Gamma=\Gamma_{0} H$ can give the late-time acceleration (see the fig. (\ref{evolution PF})). Autonomous system (\ref{autonomous2 PF}) extracts critical points $C_1$, $C_2$, $C_3$ and $C_4$. All the points are hyperbolic type and $C_1$, $C_2$ describe perfect fluid dominated decelerated phase. The points $C_3$ and $C_4$ provide the late-time accelerated evolution of the universe. Here, acceleration is driven by perfect fluid with created particles, not by the dark energy. Depending on particle creation rate ($\Gamma_{0}$), the coupling of interaction ($\alpha$), the critical point $C_3$ represents accelerated universe at late-times attracted only in phantom era (satisfying $\lambda_1<0,~~\lambda_2<0,~~ \mbox{and}~~\omega_{eff}<-1$) when $\alpha \geq 0~~\mbox{and}~~ \Gamma_{0} >\alpha +3$. The effects of rate $\Gamma_{0}$ and coupling $\alpha$ determine that the critical point $C_4$ can be a late-time attractor, evolving in quintessence era only ($\lambda_1<0,~~\lambda_2<0,~~\mbox{and}~~-1<\omega_{eff}<-\frac{1}{3}$) for :
$$ B>0~~\mbox{and}~~ \alpha \geq 0~~\mbox{and}~~ \alpha +1<\Gamma_{0} <\alpha +3 $$ (see the fig. (\ref{C4})). \\

We have observed that if we consider the interaction between the DM and the second fluid (perfect fluid) without any matter creation (i.e., $\Gamma_{0}=0$) then it would not be possible to realize the accelerated scaling attractor. However, because of the inclusion of particle creation into this picture we realize the accelerated scaling attractor. This clearly indicates that the scaling attractors are occurring because of inclusion of particle creation. \\
	
Finally, we consider a two-fluid model where an exotic fluid with equation of state: $\gamma_{d}<0$ as in Eqn. (\ref{pressure Umami}) is taken as second fluid. This is the Umami Chaplygin fluid with non-linear equation of state described in (\ref{EOS Umami}). Then, the cosmological dynamics of this model is studied in the context of dynamical systems analysis.
Using the explicit form of creation rate $\Gamma=\Gamma_{0}H$ and the interaction term $Q=\alpha H \rho_{m}$, the cosmological evolution equations have been converted into a two-dimensional autonomous system of ordinary differential equations in two dimensionless variables defined in (\ref{variables PF}). Three dimensionless model parameters namely, $\alpha$, $\Gamma_{0}$ and $\omega$ come into the autonomous system and they play a vital role in describing the universe's evolution. Here, the free parameter $A$ in the equation of state of Umami Chaplygin gas is absent in autonomous system. So, it does not affect to the cosmological dynamics of the model from dynamical perspective. Critical points have been extracted from the autonomous system. The critical points and the corresponding cosmological parameters are obtained in terms of the free parameters $\alpha$, $\Gamma_{0}$ and $\omega$ which are displayed in the table (\ref{physical_parameters}). The nature of critical points is found by perturbing the system around the critical points. The eigenvalues of linearized Jacobian matrix for the critical points are presented in the table (\ref{eigenvalues}). The conditions for stability of critical points have been discussed in the section \ref{phase plane analysis} and various regions of stability in parameters space ($\alpha,~ \Gamma_{0},~ \omega$) for critical points are displayed in figs. (\ref{Alpha}), (\ref{Gamma}) and (\ref{Omega}). Classical stability of the model has also been studied (see in sect.\ref{Classical stability}) at each critical points by finding the speed of sound and we have identified the region of parameters at which the critical points are stable locally as well as classically.

The viability of the model we considered here is examined by numerically. We investigated early inflation cosmological evolution and non-singular bouncing universe by evaluating the Hubble function and the scale factor numerically. Cosmological parameters for some critical points were investigated to compare with the $\Lambda$CDM and with the observational data. Behavior of cosmological parameters: the effective equation of state parameter $\omega_{eff}$ and the deceleration parameter $q$ showed that our model is consistent with observational data. From the dynamical analysis, it is observed that the critical points $P_1,~P_2,~P_3,~P_4$ and $P_5$ are of hyperbolic types and the linear stability theory was employed to determine their stability in phase plane. The point $P_6$ is normally hyperbolic set of critical points. We determined the stability of this set by considering the remaining non-vanishing eigenvalue and also by numerical investigation.  \\
Some cosmologically viable solutions are found from the analysis. We obtained the Umami Chaplygin fluid dominated solutions represented by the critical points $P_1$, $P_2$ and $P_3$. Depending on some parameters restrictions, the points can describe different phases of cosmic evolution. The point $P_2$ can show the cosmological dynamics at late phase as well as the early time of the universe. In particular,  for restrictions: $|\omega|<1$ and ($\Gamma_{0}-\alpha>3$), the point describes accelerated de Sitter solution corresponding to an early evolution of the universe since it is source in the phase plane. On the other hand, the point corresponds to a late-time accelerated de Sitter solution representing the present evolution of the universe for parameter restrictions: $|\omega|>1$ and ($\Gamma_{0}-\alpha<3$), because the point is stable attractor in phase plane. These features are also clearly exhibited in the figure (\ref{P2deSitter}) and are analyzed in the previous section. The cosmological evolution of the cosmological parameters ($\Omega_{m},~\Omega_{d},\gamma_{d},~q,~\omega_{eff}$) are also shown in the fig.(\ref{evolutionLCDM}) for different choices of model parameters in the region: $|\omega|>1$, ($\Gamma_{0}-\alpha)<3$ and for proper initial conditions. These scenarios mimic the standard model of $\Lambda$CDM successfully at late-times.

Next, the critical point $P_3$ is also relevant in late-time evolution of the universe. Depending on parameters, it can describe late-time accelerated universe attracted in quintessence era. But, it cannot solve the coincidence problem since $\Omega_{d}=1$ in its evolution.
Finally, the point $P_1$ describes the dust dominated decelerating universe representing the transient phase for $\Gamma_{0}<\alpha$. \\

Cosmological viable solutions are also achieved by the DE(Umami Chaplygin gas)-DM scaling solutions represented by the points $P_4$, $P_5$, and $P_6$. Here, DE and DM scale similarly in their evolution. This property ensures the possibility to solve the coincidence problem (since $0<\Omega_{d}<1$). The point $P_4$ has interesting nature in the late-time dynamics of the universe. On some parameters restrictions, it behaves like a late-time accelerated attractor in phantom regime which is in well agreement with observations.  This is confirmed in the figure (\ref{coupled complete evolution}).
For values of model parameters: $\alpha=7.1,~\omega=0.005,~\Gamma_{0}=10.3095$ the sub-figs. \ref{Complete Stream}, \ref{Complete Trajectory}, and \ref{P4stable} support the present observation by exhibiting the corresponding cosmological parameter: ($\Omega_{m}\approx 0.3,~\Omega_{d}\approx 0.7,~ \omega_{eff}\approx -1.07,~q\approx -1.10$). Also, the sub-fig. \ref{P2earlydS} for model parameter values: $\alpha=7.7,~\omega=0.6,~\Gamma_{0}=11$ refers to the  associated cosmological parameters ($\Omega_{m}= 0.3,~\Omega_{d}= 0.7,~ \omega_{eff}= -1.1,~q= -1.15$). Therefore, the numerical values of the cosmological parameters are in well agreement with the present observational data which have also been obtained in Ref.\cite{Nunes2015} where the authors studied the model $\Gamma=3\beta H$. It is to be noted that our model also support the DESI (Dark Energy Spectroscopic Instrument) DR1 (Data Release 1) \cite{Adame2024} data with restricted model parameters. The cosmological parameters result according to DESI: $\Omega_{m}=0.293 \pm 0.015,~ \gamma_{d}=-0.99_{-0.13}^{+0.15}$. It is concluded that for fine tuning of model parameters, we can achieve the values of the cosmological parameters. In another study \cite{Roy2024}, the current EoS of DE is crossing the phantom $\gamma_{d}=-1.113_{-0.21}^{+0.22}$.
For values of model parameters: $\alpha=0.05,~\omega=0.99,~\Gamma_{0}=3.04072$ the critical point $P_5$ evolves at late-times (since eigenvalues for this case are $\lambda_1=-0.01856,~\lambda_2=-0.02072.$) with corresponding cosmological parameters: ($\Omega_{m}= 0.292986,~\Omega_{d}= 0.707014,~\gamma_{d}=-0.99 ,~ \omega_{eff}=-0.996907,~q=-0.99536$). With another set of the model parameter: $\alpha=-0.791,~\omega=1.2,~\Gamma_{0}=2.209,~x_c=0.7$ it is shown that the critical point $P_6$ supports the observational data at late-times with cosmological parameters: ($\Omega_{m}= 0.3,~\Omega_{d}= 0.7,~\gamma_{d}=-1.113 ,~ \omega_{eff}=-1,~q=-1$).
Scaling solution $P_5$ shows exciting nature in its evolution. It corresponds to an accelerated scaling attractor at late-time attracted in quintessence era having the property of solving coincidence problem.
In fig.(\ref{P5}), for $\alpha=0.2,~\omega=0.661,~\Gamma_{0}=2.275$ the point $P_5$ corresponds to the universe's evolution having parameters values: $\Omega_{m}\approx0.31,~\Omega_{d}\approx0.69,~\omega_{eff}\approx-0.692,~q\approx-0.537$ and $\gamma_{d}=-0.661$. This phenomenon is in well agreement with present observation \cite{Planck,Harko2021} where the Umami Chaplygin gas (as DE) behaves as quintessence fluid. This solution can also solve the coincidence problem.
Finally, the scaling solution, namely, the normally hyperbolic set $P_6$ can represent the stable attractor in late phase of the universe which is plotted in fig. (\ref{P6stable}). For a particular choice of $x_c=1$, the solution can mimic the accelerated de Sitter solution which corresponds to late-phase and early phase of the universe. For this choice of coordinate the critical point becomes non-hyperbolic point with one vanishing eigenvalue and its stability can be analysed by center manifold theory. However, in the present work we employ numerical investigation to check the stability and the stability analysis through center manifold theory is left for a future work. Numerical investigations are presented through the figs. (\ref{P6earlydS}) and (\ref{P6latedS}).
In particular, the figure (\ref{P6latedS}) shows the phase space projection of perturbation of the system (\ref{autonomous1}) along the x-axis and y-axis for parameter values $\alpha=7.1$, $\omega=1.1$ and $\Gamma_{0}=10.1$. Sub-fig. \ref{Stable_Nx} and sub-fig. \ref{Stable_Ny} exhibit that the perturbations come back to $x=1$ and $y=-1$ axes as $N\longrightarrow \infty$. This indicates that the critical point with coordinate ($1,~-1$) is stable for the values of model parameters: $\alpha=7.1$, $\omega=1.1$ and $\Gamma_{0}=10.1$. Here, cosmological parameters for this point will take the values: $\Omega_{m}=0,~~\Omega_{d}=1,~~\omega_{eff}=q=-1$ and the DE behaves as cosmological constant ($\gamma_{d}=-1$) like fluid. Therefore, the point will describe a late-time de Sitter evolution of the universe for $|\omega|>1$. On the other hand, the point on the set may represent the early accelerated de Sitter solution for $|\omega|<1$, because there is an non-empty unstable sub-manifold in the positive eigendirection. The fig.(\ref{P6earlydS}) with $\alpha=7.1,~\omega=0.6,~\Gamma_{0}=10.1$ exhibits the evolution of perturbations near the critical point (with coordinate ($1,~-1$)) showing that perturbations increase gradually from $x=1$ and $y=-1$ as $N\longrightarrow \infty$. Thus, we can conclude that the point is unstable source. In fact, the point describes the accelerated de Sitter evolution at early times.
\\

We can summarize the results of this model in the following:
Our analysis provides some critical points exhibiting the early expansion of the universe as well as late-time acceleration by achieving the de Sitter solutions in phase plane. Some of critical points correspond to scaling attractors alleviating the coincidence problem successfully. These scenarios are also favored by observational data. Dust dominated decelerated intermediate phase of the universe is also achieved by some points. A numerical investigation is shown in fig.(\ref{evolutionLCDM}) which shows cosmologically viable solutions by identifying trajectory connecting a matter dominated critical point to DE dominated era. The future evolution of the model can mimic the $\Lambda$CDM. These solutions are achieved by scaling attractors and the cosmic evolution of relevant quantities is in agreement with observations. Despite these, we obtained a  region of parameters space: [$\omega \in (-1,~1),~\omega\neq0$ and $3+\alpha<\Gamma_{0}<2\alpha$, where $\alpha> 3$] (see in fig. (\ref{pr})) in which a series of critical points $P_2\longrightarrow P_3 \longrightarrow P_4$ (in fig. \ref{Complete Trajectory}) is obtained and they represent cosmologically interesting solutions, where a clear trajectory is identified connecting an early inflationary phase (represented by point $P_2$) to a late-time DE dominated solution (described by the point $P_4$) through a matter dominated era (described by the point $P_3$). The viability of our model is also confirmed by plotting the figure (in fig.\ref{Complete evolution}) numerically for physically relevant quantities ($\omega_{eff},~q$). The numerical investigation shows that the solution describes the cosmological evolutionary scheme of the universe where the effective equation of state parameter $\omega_{eff}$ evolves starting from an inflationary phase (i.e., $\omega_{eff}=-1$), passing through a sufficiently long amount of time in a matter era ($\omega_{eff}\approx 0$) and finally being attracted by late-time dark energy dominated accelerated phase (i.e., $\omega_{eff}<-1$). Therefore, a complete cosmic scenario is achieved. Note that phantom crossing behavior is also achieved in this scenario. Therefore, one can conclude that a unified cosmic evolution is achieved within this cosmological model of interacting Umami chaplygin gas with the same creation rate and this statement is validated by the figures \ref{Complete Stream}, \ref{Complete Trajectory} and \ref{Complete evolution}. One final remark of our study is that a  non-singular bouncing universe is also achieved for this model investigated numerically in fig.(\ref{bouncing_scenario}). The figures are plotted for parameter value $A=0.1$ and for different values of parameters $\alpha$, $\Gamma_{0}$ and $\omega$.

Thus an overall conclusion can be drawn as follows:
if we consider a cosmological model of matter creation with rate $\Gamma=\Gamma_{0} H$ and the universe is filled with pressureless dark matter(created matter) only. Then this model does not show phase transition. Next, if we consider the universe is filled with pressureless dark matter (created matter)  and perfect fluid as an additional fluid, then for non-interacting case, it would not be possible to achieve late time acceleration. But, for interacting case, the two-fluid model can be capable of providing the late -time acceleration. Moreover, if we consider a two-fluid model in context of particle creation when Umami chaplygin gas can play an exotic fluid, cosmological viable solutions are obtained. A complete cosmic scenario from early inflation to late-time acceleration connecting through matter dominated decelerated phase is achieved in this model. This type of scenario can not be obtained when interacting umami chaplygin fluid is considered without particle creation ($\Gamma=0$)\cite{Biswas2021} and including particle creation (with the same rate) in interacting DE (with EoS: $p_d =\omega_d \rho_{d}$) \cite{Biswas2017}. Interestingly, the creation rate has direct effects on cosmological dynamics. It may result differently when the matter contents of the universe has different equation of states.

Thus, investigations through dynamical analysis of interacting Umami Chaplygin fluid with different creation rates will be of future research interest.

\section*{Acknowledgments}

The authors are extremely thankful to the anonymous referees for their valuable comments, suggestions which help us a lot to improve our manuscript significantly. The authors are grateful to Prof. Subenoy Chakraborty for his valuable comments and insights on this work. We are also thankful to Dr. Supriya Pan for providing us his support and valuable comments while modifying the previous version of our work. We would like to thank Inter University Centre for Astronomy and Astrophysics (IUCAA) for providing research facilities and local hospitalities in an academic visit during which some modifications of this work have been done. The author Goutam Mandal acknowledges UGC, Government of India for providing Senior Research Fellowship [Award Letter No. F.82-1/2018(SA-III)] for Ph.D.



\end{document}